\documentclass[11pt]{article}
\usepackage{epsfig}
\usepackage{amssymb}
\newcommand{\be}{\begin{equation}}
\newcommand{\ee}{\end{equation}}
\newcommand{\nn}{\nonumber}
\newcommand{\bea}{\begin{eqnarray}}
\newcommand{\eea}{\end{eqnarray}}
\newcommand{\bfig}{\begin{figure}}
\newcommand{\efig}{\end{figure}}
\newcommand{\bc}{\begin{center}}
\newcommand{\ec}{\end{center}}
\long\def\symbolfootnote[#1]#2{\begingroup%
\def\thefootnote{\fnsymbol{footnote}}\footnote[#1]{#2}\endgroup}

\def\sq2{\sqrt{2}}
\def\Lum{{\cal L}}
\def\tb{\tan\beta}
\def\mb{m_b}
\newcommand\Rad{\Phi_{\rm rad}}
\newcommand{\smallz}{{\scriptscriptstyle Z}} %
\newcommand{\mz}{m_\smallz}
\newcommand{\smallw}{{\scriptscriptstyle W}}

\newcommand{\smalla}{{\scriptscriptstyle A}}
\newcommand{\ma}{m_\smalla}
\newcommand{\mt}{m_t}

\newcommand{\text}[1]{#1}

\newcommand{\PHB}{{\tt POWHEG BOX}}
\newcommand{\PH}{{\tt POWHEG}}

\newcommand{\hsm}{H}
\newcommand{\drbar}{\overline{\rm DR}}
\newcommand{\msbar}{\overline{\rm MS}}

\begin{document}

\begin{titlepage}
\nopagebreak
{\flushright{
        \begin{minipage}{5cm}
         RM3-TH/11-15 \\
         IFUM-990/FT
        \end{minipage}        }

}
\renewcommand{\thefootnote}{\fnsymbol{footnote}}
\vskip 1.5cm
\begin{center}
{\LARGE\bf Higgs production via gluon fusion in the \PH\ \\[7pt]
approach in the SM and in the MSSM}
\vskip 1.0cm
{\Large  E.~Bagnaschi$^{a,\,b}$, G.~Degrassi$^{c}$, P.~Slavich$^{b}$
and A.~Vicini$^{a}$}
\vspace*{8mm} \\
{\sl ${}^a$
    Dipartimento di Fisica, Universit\`a di Milano and
    INFN, Sezione di Milano,\\
    Via Celoria 16, I--20133 Milano, Italy}
\vspace*{2.5mm}\\
{\sl ${}^b$  LPTHE, 4, Place Jussieu, F-75252 Paris,  France}
\vspace*{2.5mm}\\
{\sl ${}^c$
  Dipartimento di Fisica, Universit\`a di Roma Tre and  INFN, Sezione di
  Roma Tre \\
  Via della Vasca Navale~84, I-00146 Rome, Italy}
\end{center}
\symbolfootnote[0]{{\tt e-mail:}}
\symbolfootnote[0]{{\tt bagnaschi@lpthe.jussieu.fr}}
\symbolfootnote[0]{{\tt degrassi@fis.uniroma3.it}}
\symbolfootnote[0]{{\tt slavich@lpthe.jussieu.fr}}
\symbolfootnote[0]{{\tt alessandro.vicini@mi.infn.it}}

\vskip 0.7cm

\begin{abstract}

We consider the gluon fusion production cross section of a scalar
Higgs boson at NLO QCD in the SM and in the MSSM. We implement the
calculation in the \PH\ approach, and match the NLO-QCD results with
the {\tt PYTHIA} and {\tt HERWIG} QCD parton showers. We discuss a few
representative scenarios in the SM and MSSM parameter spaces, with
emphasis on the fermion and squark mass effects on the Higgs boson
distributions.

\end{abstract}
\vfill
\end{titlepage}    
\setcounter{footnote}{0}
%

\section{Introduction}

Understanding the mechanism that leads to the breaking of the
electroweak symmetry and that is responsible for the generation of the
mass of the elementary particles is one of the major challenges of
high energy physics. The search for the Higgs boson(s) is currently
under way at the Tevatron and at the LHC, and limits on the Higgs mass
spectrum have already been set \cite{TevatronExp,LHCpapers}.  This
search requires an accurate control of all the Higgs production and
decay mechanisms, including the effects due to radiative corrections
\cite{Dittmaier:2011ti}.

In the Standard Model (SM) the gluon fusion process \cite{H2gQCD0} is
the dominant Higgs production mechanism both at the Tevatron and at
the LHC. The total cross section receives very large next-to-leading
order (NLO) QCD corrections, which were first computed in
ref.~\cite{H2gQCD1} in the so-called heavy-quark effective theory
(HQET), i.e.~including only the top-quark contributions in the limit
$\mt\to\infty$.  Later calculations~\cite{SDGZ,HK,babis1,ABDV,BDV}
retained the exact dependence on the masses of the top and bottom
quarks running in the loops.  The next-to-next-to-leading order (NNLO)
QCD corrections are also large, and have been computed in the HQET in
ref.~\cite{H2gQCD2}. The finite-top-mass effects at NNLO QCD have been
studied in ref.~\cite{H2gQCD3} and found to be small.  The resummation
to all orders of soft gluon radiation has been studied in
refs.~\cite{KLS,H2gQCD4}.  Leading third-order (NNNLO) QCD terms have
been discussed in ref.~\cite{H2gQCD5}.  The role of electroweak (EW)
corrections has been discussed in
refs.~\cite{H2gEW0,H2gEW1,APSU,BDVcpx}. The impact of mixed QCD-EW
corrections has been discussed in ref.~\cite{H2gQCDEW}.  The residual
uncertainty on the total cross section depends mainly on the
uncomputed higher-order QCD effects and on the uncertainties that
affect the parton distribution functions (PDF) of the proton
\cite{Demartin:2010er,Alekhin:2011sk,Dittmaier:2011ti}.

The Higgs sector of the Minimal Supersymmetric Standard Model (MSSM)
consists of two $SU(2)$ doublets, $H_1$ and $H_2$, whose relative
contribution to electroweak symmetry breaking is determined by the
ratio of vacuum expectation values of their neutral components,
$\tb\equiv v_2/v_1$. The spectrum of physical Higgs bosons is richer
than in the SM, consisting of two neutral CP-even bosons, $h$ and $H$,
one neutral CP-odd boson, $A$, and two charged bosons, $H^\pm$. The
couplings of the MSSM Higgs bosons to matter fermions differ from
those of the SM Higgs, and they can be considerably enhanced (or
suppressed) depending on $\tb$. As in the SM, gluon fusion is one of
the most important production mechanisms for the neutral Higgs bosons,
whose couplings to the gluons are mediated by top and bottom quarks
and their supersymmetric partners, the stop and sbottom squarks.

In the MSSM, the cross section for Higgs boson production in gluon
fusion is currently known at the NLO. The contributions arising from
diagrams with quarks and gluons can be obtained from the corresponding
SM results \cite{SDGZ,HK,babis1,ABDV,BDV} with an appropriate
rescaling of the Higgs-quark couplings. The contributions arising from
diagrams with squarks and gluons were first computed under the
approximation of vanishing Higgs mass in ref.~\cite{Dawson:1996xz},
and the full Higgs-mass dependence was included in later calculations
\cite{babis1,ABDV,BDV,MS}.  The contributions of two-loop diagrams
involving top, stop and gluino to both scalar and pseudoscalar Higgs
production were computed under the approximation of vanishing Higgs
mass in refs.~\cite{HS,Franziska}, whose results were later confirmed
and cast in a compact analytic form in refs.~\cite{DS1,DDVS}.

The approximation of vanishing Higgs mass can provide reasonably
accurate results as long as the Higgs mass is well below the threshold
for creation of the massive particles running in the loops. For the
production of the lightest scalar Higgs, this condition does apply to
the two-loop diagrams involving top, stop and gluino, but it obviously
does not apply to the corresponding diagrams involving the bottom
quark, whose contribution can be relevant for large values of
$\tb$. In turn, the masses of the heaviest scalar and of the
pseudoscalar might very well approach (or exceed) the threshold for
creation of top quarks or even of squarks. Unfortunately, retaining
the full dependence on the Higgs mass in the quark-squark-gluino
contributions has proved a rather daunting task. A calculation based
on a combination of analytic and numerical methods was presented in
ref.~\cite{babis2} (see also ref.~\cite{spiraDb}), but neither
explicit analytic results nor a public computer code have been made
available so far. However, ref.~\cite{DS2} presented an approximate
evaluation of the bottom-sbottom-gluino contributions to scalar
production, based on an asymptotic expansion in the large
supersymmetric masses that is valid up to and including terms of
${\cal O}(\mb^2/m_\phi^2)$, ${\cal O}(\mb/M_{SUSY})$ and ${\cal
  O}(\mz^2/M_{SUSY}^2)$, where $m_\phi$ denotes a Higgs boson mass and
$M_{SUSY}$ denotes a generic superparticle mass. An independent
calculation of the bottom-sbottom-gluino contributions, restricted to
the limit of a degenerate superparticle mass spectrum, was also
presented in ref.~\cite{hhm}, confirming the results of
ref.~\cite{DS2}. More recently, ref.~\cite{DDVS} presented an
evaluation of the quark-squark-gluino contributions to pseudoscalar
production that is also based on an asymptotic expansion in the large
supersymmetric masses, but does not assume any hierarchy between the
pseudoscalar mass and the quark mass, thus covering both the
top-stop-gluino and bottom-sbottom gluino cases.

The total cross section, without acceptance cuts, provides important
information about the Higgs boson production rate.  On the other hand,
especially at the LHC, it is very likely that a Higgs boson is
produced in association with a jet, generating a transverse momentum
$p_T^H$ of the Higgs boson.  The first studies on Higgs+jet final
states in the SM were performed in ref.~\cite{Hj0}, using the
real-parton emission amplitudes that enter the calculation of the
NLO-QCD corrections to the inclusive Higgs production cross
section. The NLO corrections to the Higgs+jet final state at large
transverse momentum of the Higgs boson were subsequently studied in
ref.~\cite{Hj1}, and the resummation of all the logarithmically
enhanced terms, matched with the NLO calculation of the Higgs+jet
final state, was discussed in ref.~\cite{DFGK}.  The impact on the
Higgs+jet final state of the NLO-EW corrections and of the finite
masses of the particles running in the loops was discussed in
ref.~\cite{realtb}.  In the MSSM, the production of a neutral Higgs
boson in association with one jet was discussed in
refs.~\cite{Brein,MSSMH+j}.

A different set of observables are the differential distributions of
the Higgs boson, inclusive over QCD radiation.  For the SM case,
results were presented in ref.~\cite{Hdiff1} at NLO-QCD and in
ref.~\cite{Hdiff2} at NNLO-QCD.  The Higgs boson transverse momentum
spectrum, including the NLO-QCD corrections matched with the
resummation of next-to-next-to-leading logarithmic (NNLL) enhanced
terms, has been studied in
refs.~\cite{Balazs:2000wv,HqT,deFlorian:2011xf}.  In the MSSM a study
of the Higgs distributions, at NLO-QCD, was discussed in
ref.~\cite{hhm}.

If a new scalar particle is discovered at the Tevatron or at the LHC,
a major question will be to determine whether it is a Higgs boson and,
in that case, whether it belongs to the particle spectrum of the SM,
of the MSSM or of any other model.  An example could be represented by
a MSSM Higgs boson whose production cross section is close to the
production cross section for a SM Higgs boson of equal mass.  In this
case, an accurate study of the differential distributions involving
the Higgs boson might shed some light on the underlying model.

A precise analysis of the experimental data requires the use of
NLO-QCD results merged with the description of initial-state multiple
gluon emission via a QCD parton shower (PS) Monte Carlo (MC) such as
{\tt HERWIG} \cite{herwig} or {\tt PYTHIA} \cite{pythia}.  However,
the merging of NLO-QCD matrix elements with PS faces the problem of
avoiding double counting, as addressed in
refs.~\cite{mcnlo,Nason:2004rx}.
The \PH\ method \cite{Frixione:2007vw} allows to systematically merge
NLO calculations with vetoed PS, avoiding double counting and
preserving the NLO accuracy of the calculation.  The procedure can be
implemented using a set of tools and results available in the
so-called \PHB~\cite{powhegggh}. The latter provides a general
framework that exploits the universal nature of initial-state
collinear divergences and the factorization property of soft radiation
to automatize the subtraction of all the soft and/or collinear
divergent terms from the NLO matrix elements of an arbitrary process.
The \PH\ method does not rely on the details of the shower MC and, by
construction, guarantees an accuracy at NLO + leading logarithmic (LL)
QCD. In ref.~\cite{Alioli:2008tz} it has been shown that, with an
appropriate choice of scale for the strong coupling constant, the
merging procedure can also reproduce the next-to-leading logarithmic
(NLL) terms.

At present, no code exists that merges the NLO-QCD results for the
gluon fusion process with a QCD PS, retaining the exact dependence on
the Higgs mass and on the masses of the particles running in the
loops.
Two implementations of the NLO-QCD results merged with a PS are
available\footnote{ These implementations are also available as a
  subprocess of {\tt HERWIG++} \cite{herwig++ggh}.  }: the one in {\tt
  MC@NLO}~\cite{mcnloggh} and the one in the
\PHB\ framework~\cite{Alioli:2008tz}.  However, both implementations
are limited to the SM case, and beyond LO they only include results
computed in the HQET.
Conversely, the codes {\tt HIGLU}~\cite{HIGLU} and {\tt
  iHixs}~\cite{iHixs} contain the full dependence on the Higgs and
quark masses up to NLO, and {\tt HIGLU} also allows to include the
full squark-gluon contributions from ref.~\cite{MS}, but neither code
is matched to a shower MC.
Recently, a step toward the inclusion of the finite-quark-mass effects
in PS was taken in ref.~\cite{Alwall:2011cy}, where parton-level
events for Higgs production accompanied by zero, one or two partons
are generated with matrix elements computed in the HQET, and then,
before being passed to the PS, they are re-weighted by the ratio of
the exact one-loop amplitudes over the approximate ones. This
procedure is equivalent to generating events directly with the exact
one-loop amplitudes, yet it is much faster.

We aim to provide a code that fills the remaining gap, using matrix
elements that include the dependence on the masses, both in the SM and
in the MSSM, properly matched to an external shower MC.  For the SM
case, we use NLO matrix elements with full dependence on the Higgs,
top and bottom masses. For the MSSM case, we use matrix elements with
exact dependence on the quark, squark and Higgs masses in the
contributions of real-parton emission. For the two-loop virtual
contributions, the approximation of vanishing Higgs mass is employed
in the diagrams involving superpartners, while the rest is computed
exactly.

The plan of the paper is as follows: in section \ref{sec:POWHEG} we
describe the basic features of the \PH\ implementation of $gg \to
\phi$; in section \ref{sec:SM} we discuss our SM implementation with
exact dependence on the fermion masses, presenting a numerical
analysis valid for an on-shell Higgs; section \ref{sec:MSSM} is
devoted to analyzing the MSSM case; finally, in section
\ref{sec:concl} we draw our conclusions.



\section{\PH\ implementation of $gg \to \phi$ \label{sec:POWHEG}}
\noindent In this section we briefly discuss the implementation of the
gluon-fusion Higgs production process in the \PHB\ framework,
following closely ref.~\cite{Alioli:2008tz} (see also
ref.~\cite{Hamilton:2009za}).  We fix the notation keeping the
discussion at a general level, without referring to a specific
model. In the next sections the formulae presented below will be
specialized to the SM and MSSM cases.

The generation of the hardest emission is done in \PH\ according to the
following formula:
\begin{eqnarray}
\label{eq:POWHEG}
d\sigma &=& \bar{B}(\bar{\Phi}_1)\, d \bar{\Phi}_1
 \left\{ \Delta\left(\bar{\Phi}_1,p_T^{min}\right)
+
\Delta\left(\bar{\Phi}_1, p_T \right)\, \frac{
 R\left(\bar{\Phi}_1,\Phi_{\rm rad} \right)}{ B\left(\bar{
    \Phi}_1\right)} \, d\Phi_{\rm rad} \right\} 
\nn\\ 
&+& \sum_q R_{q \bar q}\left(\bar{
  \Phi}_1,\Phi_{\rm rad} \right) d \bar{\Phi}_1 d\Phi_{\rm rad} ~.
\end{eqnarray}
In the equation above the variables $\bar{\Phi}_1 \equiv (M^2,Y)$
denote the invariant mass squared and the rapidity of the Higgs boson,
which describe the kinematics of the Born (i.e., lowest-order) process
$gg\rightarrow\phi$. The variables $\Rad\equiv (\xi, y,\phi)$ describe
the kinematics of the additional final-state parton in the real
emission processes. In particular, denoting by $k_2^\prime$ the
momentum of the final-state parton in the partonic center-of-mass
frame, or
\be
\label{eq:k2}
  k^\prime_2= k_2^{\prime\, 0}\ (1, \sin\theta \sin\phi, \sin\theta \cos\phi,
  \cos\theta), 
\ee
we have
\be
\label{eq:radvar}
k_2^{\prime\, 0} = \frac{\sqrt{s}}{2} \xi, \qquad y = \cos\theta\,,
\ee
where $s$ is the partonic center-of-mass energy squared.

The factor $\bar{B}(\bar{\Phi}_1)$ in eq.~(\ref{eq:POWHEG}) is related
to the total cross section computed at NLO in QCD.  It contains the
value of the differential cross section, for a given configuration of
the Born final state variables, integrated over the radiation
variables.  The integral of this quantity on $d\bar{\Phi}_1$ without
acceptance cuts yields the total cross section.  This factor is
responsible for the correct NLO-QCD normalization of the result, and
is computed in the initialization phase using the real and virtual
NLO-QCD corrections.

The terms within curly brackets in eq.~(\ref{eq:POWHEG}) describe the
real emission spectrum of an extra parton: the first term is the
probability of not emitting any parton with transverse momentum larger
than a cutoff $p_T^{min}$, while the second term is the probability of
not emitting any parton with transverse momentum larger than a given
value $p_T$ times the probability of emitting a parton with transverse
momentum equal to $p_T$. The sum of the two terms fully describes the
probability of having either zero or one additional parton in the
final state. The probability of non-emission of a parton with
transverse momentum $k_T$ larger than $p_T$ is obtained using the
\PH\ Sudakov form factor
\be
\Delta(\bar \Phi_1,p_T)=
\exp
\left\{
-\int d\Phi_{\rm rad}
\frac{R(\bar{\Phi}_1,\Phi_{\rm rad})}{B(\bar{\Phi}_1)}
\theta(k_T-p_T)
\right\}~.
\label{hgen}
\ee

Finally, the last term in eq.~(\ref{eq:POWHEG}) describes the effect
of the $q \bar q \to \phi g$ channel, which has been kept apart in the
generation of the first hard emission because it does not factorize
into the Born cross section times an emission factor.

We now discuss the various terms appearing in eq.~(\ref{eq:POWHEG}) in
more detail.  We have:
\bea
\bar{B}(\bar{\Phi}_1)
&=&
B_{gg}(\bar{\Phi}_1)
~+~
V_{gg}(\bar{\Phi}_1)
 \nn \\
&+&\int d\Phi_{\rm rad}
\left\{
\hat R_{gg}\left(\bar\Phi_1,\Phi_{\rm rad}\right)
+
\sum_q \left[
\hat R_{gq}\left(\bar\Phi_1,\Phi_{\rm rad}\right)+
\hat R_{qg}\left(\bar\Phi_1,\Phi_{\rm rad}\right) \right]
\right\}
~+~  {\rm c. \: r.}~,
\label{Bexp}
\eea
where
\begin{eqnarray}
B_{gg} (\bar{\Phi}_1) &=& \mathcal{B}_{gg} (\bar{\Phi}_1) 
\, \Lum_{gg}\,, \label{BggL} \\
V_{gg} (\bar{\Phi}_1) &=& \mathcal{V}_{gg} (\bar{\Phi}_1)
\, \Lum_{gg}\,, \label{VggL}\\
\hat{R}_{gg}(\bar{\Phi}_1,\Rad) &=&
  \hat{\mathcal{R}}_{gg}(\bar{\Phi}_1,\Rad)  
\, \Lum_{gg}\,, \label{RggL} \\
\hat{R}_{gq}(\bar{\Phi}_1,\Rad)  &=&
\hat{\mathcal{R}}_{gq}(\bar{\Phi}_1,\Rad)  
\, \Lum_{gq}\,, \label{RgqL}\\
\hat{R}_{qg}(\bar{\Phi}_1,\Rad) &=& \hat{\mathcal{R}}_{qg}(\bar{
  \Phi}_1,\Rad) 
\, \Lum_{qg}\,, \label{RqgL}
\end{eqnarray}
with $\Lum_{ab}$ the luminosity for the partons $a$ and $b$.  
In eq.~(\ref{Bexp}) ``\,c.~r.'' denotes the collinear remnants
multiplied by the relevant parton luminosity.  The remnants are the
finite leftovers after the subtraction of the initial-state collinear
singularities into the parton distribution function is performed, and
their explicit expressions are given in eqs.~(2.36), (2.37) and
(3.7)--(3.10) of ref.~\cite{Alioli:2008tz}.

The function $\mathcal{B}_{gg} (\bar{\Phi}_1) $ in
eq.~(\ref{BggL}) represents the squared matrix element of the Born
contribution to the process, averaged over colors and helicities of
the incoming gluons, and multiplied by the flux factor $1/(2M^2)$. It
is given by
\be
\mathcal{B}_{gg} (\bar{\Phi}_1) = \frac{G_\mu\, \alpha_s^2
  (\mu_R^2\,) M^2}{256\, \sqrt{2} \, \pi^2}\, \left| \mathcal{H}^{1\ell}
\right|^2 ~,
\label{Bgg}
\ee
where $\mathcal{H}$ is the form factor for the coupling of the Higgs
boson with two gluons, whose explicit form depends on the particle
content of the model considered and will be detailed in the following
sections.  It is decomposed in one- and two loop parts as
\be
\mathcal{H}  = \mathcal{H}^{1\ell} 
          + \frac{\alpha_s}{\pi} \, \mathcal{H}^{2\ell} 
          + {\cal O}(\alpha_s^2)~.
\label{hhexpansion}
\ee

The regularized two-loop virtual contributions are contained in
\be
\mathcal{V}_{gg} (\bar{\Phi}_1)
= {\alpha_s \over \pi} \left[ C_A \, \frac{\pi^2}3  
	  + \beta_0 \ln \left( \frac{\mu_R^2}{\mu_F^2} \right)
          +  2\, \mathrm{Re}
           \left(\frac{{\mathcal H}^{2\ell}}{{\mathcal H}^{1\ell}} \right) 
             \right] \mathcal{B}_{gg}(\bar{\Phi}_1)  ~.
\label{Vgg}
\ee
In the equation above, $\mu_R$ and $\mu_F$ are the renormalization and
factorization scale, respectively, $C_A =N_c$ ($N_c$ being the number
of colors), and $\beta_0 = (11\, C_A - 2\, N_f)/6 $ ($N_f$ being the
number of active flavors) is the one-loop beta function of the strong
coupling.

The hatted functions $\hat{\mathcal{R}}_{ij}$ in
eqs.~(\ref{RggL})--(\ref{RqgL}) are the Frixione, Kunst and
Signer~\cite{Frixione:1995ms,Frixione:1997np} infrared-subtracted
counterparts of $\mathcal{R}_{ij}$
\be
\label{eq:FKSrin}
\hat{\mathcal R}_{ij}(\bar{\Phi}_1,\Rad) = \frac{1}{\xi} 
\left\{ \frac{1}{2} \left( \frac{1}{\xi} \right)_{\!\!+}
\left[\left( \frac{1}{1-y}\right)_{\!\!+} \!\!+ \left(
\frac{1}{1+y} \right)_{\!\!+} \right] \right\}
\left[ (1-y^2) \, \xi^2 \, \mathcal{R}_{ij}(\bar{\Phi}_1,\Rad) \right]~,
\ee
where $\mathcal{R}_{ij}$ are the squared amplitudes, averaged over the
incoming helicities and colors and multiplied by the flux factor
$1/(2s)$, for the NLO partonic subprocesses ($gg \to \phi g, gq\to
\phi q, qg\to q \phi$):
\bea
\label{eq:Rgg}
\mathcal{R}_{gg} (\bar{\Phi}_1,\Rad)
&=& \frac{3\, G_\mu \,\alpha_s^3\, M^8}{\sqrt{2} \, \pi\, 2\, s}
\frac{ \left| {\cal A}_{gg}(s,t,u)\right|^2}{s t u}\,,\\
\mathcal{R}_{gq}(\bar{\Phi}_1,\Rad) 
&=& - \frac{G_\mu \,\alpha_s^3\,  M^4}{\sqrt{2} \, \pi\, 6 \,s}
\,\frac{s^2+u^2}{(s+u)^2\,t}\,\left| {\cal A}_{qg}(s,t,u)\right|^2,\\
\label{eq:Rqg}
\mathcal{R}_{qg}(\bar{\Phi}_1,\Rad) 
&=& - \frac{G_\mu \,\alpha_s^3\,  M^4}{\sqrt{2} \, \pi\, 6\, s}
\,\frac{s^2+t^2}{(s+t)^2\,u}\,\left| {\cal A}_{qg}(s,u,t)\right|^2,
\eea
where $s=M^2/(1-\xi), \: t= - (s/2)\, \xi\, (1+y)$ and $u=- (s/2)\,
\xi\,(1-y)$.

The complete real matrix elements that enter the \PH\ Sudakov form
factor, eq.~(\ref{hgen}), read
\bea
R(\bar{\Phi}_1,\Phi_{\rm rad}) &=&
R_{gg}(\bar{\Phi}_1,\Phi_{\rm rad}) + \sum_q \left[
R_{gq}(\bar{\Phi}_1,\Phi_{\rm rad}) + R_{qg}(\bar{\Phi}_1,\Phi_{\rm rad}) \right]~,\\
B\! \left(\bar{\Phi}_1\right) &=& B_{gg}\! \left(\bar{\Phi}_1\right)~, 
\eea
where the functions $R_{ab}$ are the non-infrared-subtracted
counterparts of eqs.~(\ref{RggL})--(\ref{RqgL}).  The probability for
the emission of the first and hardest parton is described with the
exact matrix element in all the phase space regions.

Finally, the contribution of the  $q \bar q \to  \phi g$ channel is
\be
R_{q \bar q}(\bar{\Phi}_1,\Rad) = \mathcal{R}_{q \bar q}(\bar{
  \Phi}_1,\Rad) \, \Lum_{q \bar q},
\ee
with
\be
\label{eq:Rqq}
\mathcal{R}_{q\bar{q}}(\bar{\Phi}_1,\Rad) = 
\frac{4\,G_\mu \,\alpha_s^3\, M^4}{\sqrt{2} \, \pi\, 9\, s}
\,\frac{t^2+u^2}{(t+u)^2\,s}\,\left| {\cal A}_{q \bar q}(s,t,u)\right|^2 \, .
\ee

The functions ${\cal A}_{gg},\,{\cal A}_{q g}$ in
eqs.~(\ref{eq:Rgg})--(\ref{eq:Rqg}) and ${\cal A}_{q \bar q}$ in
eq.~(\ref{eq:Rqq}) depend on the particle content of the model
considered, and will be defined in the following sections.

\vfill
\newpage

\section{SM results \label{sec:SM}}

The current public release of \PH\ \cite{powhegggh} contains matrix
elements evaluated in the HQET. It also gives the user the possibility
of rescaling the term $\bar{B}(\bar{ \Phi}_1)$ in
eq.~(\ref{eq:POWHEG}) by a normalization factor defined as the ratio
between the exact Born contribution where the full dependence from the
top and bottom masses is kept into account and the Born contribution
evaluated in the HQET. In the following we describe the modifications
we have introduced in the code to include the full fermion-mass
dependence at the NLO and the effect of the two-loop EW corrections.

\subsection{Modifications in \PH}
\label{sec:SMmod}

The inclusion of the fermion-mass effects is achieved using for the
functions ${\mathcal H}^{1\ell},\,{\mathcal H}^{2\ell},\,{\cal
  A}_{gg},\,{\cal A}_{q g},\, {\cal A}_{q \bar q}$ the exact results
instead of those computed in the HQET. For the Born term we have
\be
{\mathcal H}^{1\ell} = - 4 \,T_F\,\sum_{q=t,b}
\lambda_q\,y_{q}
 \left[ 2 - \left( 1 -4 y_{q} \right)  \, \frac{1}{2} 
\ln^2 \left(x_q \right) \right]\, , 
\label{eq:3} 
\ee
where 
\be
y_q \equiv \frac{m^2_q}{M^2}~ , ~~~~~~~~~~~~~~~
x_q \equiv \frac{\sqrt{1- 4 y_q} - 1}{\sqrt{1- 4 y_q} + 1}~, 
\label{eq:xy}
\ee 
$T_F= 1/2$ is the matrix normalization factor of the fundamental
representation of $SU(N_c)$, and $\lambda_q$ is a normalization factor
for the Higgs-quark coupling. In the SM case $\lambda_q=1$ for both
the top and the bottom quark.

The form factor ${\mathcal H}^{2\ell}$ contains the mass-dependent
contribution of the two-loop virtual corrections, and can be cast in
the following form:
\be
{\mathcal H}^{2\ell}  =  T_F\,\sum_{q=t,b}
        \, \lambda_q\, \Biggl( C_F\, {\mathcal G}^{(2\ell, C_R)}_{1/2} (x_q)+
                    C_A \,{\mathcal G}^{(2\ell, C_{A})}_{1/2} (x_q)  \Biggr) 
~+~ {\rm h.c.}~,
\label{G2}
\ee
where $C_F = (N_c^2-1)/(2N_c)$.  Explicit analytic expressions for
${\mathcal G}^{(2\ell,C_R (C_A))}_{1/2}$ given in terms of harmonic
polylogarithms can be found in ref.~\cite{ABDV}.  It should be noticed
that ${\mathcal G}^{(2\ell,C_R)}_{1/2}$ depends on the choice of
renormalization scheme for the quark mass entering the one-loop part
of the form factor. In ref.~\cite{ABDV} expressions for ${\mathcal
  G}^{(2\ell,C_R)}_{1/2}$ with on-shell (OS) or $\msbar$ parameters
are presented.  In our implementation we allow the choice among the
OS, $\msbar$ or $\drbar$ renormalization schemes.

Concerning the real emission contributions, we have for the $g g \to
\hsm g$ channel
\be
\left| {\cal A}_{gg}(s,t,u) \right|^2  =
 |A_2 (s,t,u)|^2 + |A_2 (u,s,t)|^2 + |A_2 (t,u,s)|^2 +
      |A_4 (s,t,u)|^2 ~,
\label{Agg}
\ee
where the functions $A_2$ and $A_4$ can be cast in the following form:
\bea
A_2 (s,t,u) & = & T_F\,\sum_{q=t,b}  \lambda_q\,y_q^2 \,
\left[ b_{1/2} (s_q,t_q,u_q) + b_{1/2} (s_q,u_q,t_q) \right] , \\
A_4 (s,t,u) & = & T_F\,\sum_{q=t,b}  \lambda_q\,y_q^2 \,
\left[ c_{1/2} (s_q,t_q,u_q) + c_{1/2} (t_q,u_q,s_q) 
  +\,  c_{1/2} (u_q,s_q,t_q) \right] ,
\label{A24fun}
\eea
with 
\be
s_q \equiv \frac{s}{m_q^2},~~~~~t_q \equiv \frac{t}{m_q^2}, ~~~~~~~
u_q \equiv \frac{u}{m_q^2}~.
\label{eq:defstu}
\ee
Explicit expressions for the functions $b_{1/2} (s_q,t_q,u_q)$ and
$c_{1/2} (s_q,t_q,u_q)$ are given in ref.~\cite{BDV}.

The function ${\cal A}_{q \bar q}(s,t,u)$ relevant for the $q \bar q
\to \hsm g$ channel is given by
\be
{\cal A}_{q \bar q} ( s,t,u)  =  T_F\, \sum_{q=t,b}  
       \lambda_q\,y_q \, d_{1/2} (s_q,t_q,u_q)~ , 
\ee
and $d_{1/2} (s_q,t_q,u_q)$ can be found in ref.~\cite{BDV}. Finally
${\cal A}_{q g}(s,t,u)$ relevant for the $q g \to \hsm g$ channel
can be obtained from ${\cal A}_{q \bar q}(s,t,u)$ via
\be
{\cal A}_{qg}(s, t, u) =
{\cal A}_{q \bar q}(t,s,u)~.
\ee

The two-loop EW corrections are included as a factor $(1 + \delta_{\rm
  {\scriptscriptstyle EW}})$ which multiplies the term
$\bar{B}(\bar{\Phi}_1)$ in the first line of
eq.~(\ref{eq:POWHEG}). This choice follows from the current structure
of \PH\ where the $q \bar q \to \hsm g$ channel is kept apart, because
it is not proportional to the Born cross section in the collinear
limit. In the SM case, the values of the correction $ \delta_{\rm
  {\scriptscriptstyle EW}}$ as a function of the Higgs boson mass can
be obtained from ref.~\cite{APSU}.

\subsection{SM: numerical results \label{sec:sm-numerical}}

In this section we present numerical results for the production of an
on-shell Higgs boson in the SM. We focus our analysis on the inclusion
of the exact quark-mass dependence in the NLO corrections and on the
effect of the EW corrections. We also consider the effect of merging
\PH\ with a PS. The results have been obtained for the LHC with
center-of-mass energy of 7 TeV, using the following numerical values
for the physical input parameters: $G_\mu=1.16637\cdot 10^{-5}$
GeV$^{-2}$, $m_t=172.5$ GeV and $m_b=4.75$ GeV
\cite{Dittmaier:2011ti}.  We have used the MSTW2008 \cite{mstw2008}
NLO set of PDF to describe to partonic content of the proton.  In the
code the value of $\alpha_s(m_Z)$ is set accordingly to the choice
made in the PDF set: in our case $\alpha_s(m_Z)=0.12018$. When
discussing the distributions in the Higgs transverse momentum $p_T^H$,
a cut $p_T^H > 0.8$ GeV has been enforced.  The renormalization and
the factorization scales have been set equal to the Higgs boson
mass: $\mu_R=\mu_F=m_H$.

\begin{figure}[t]
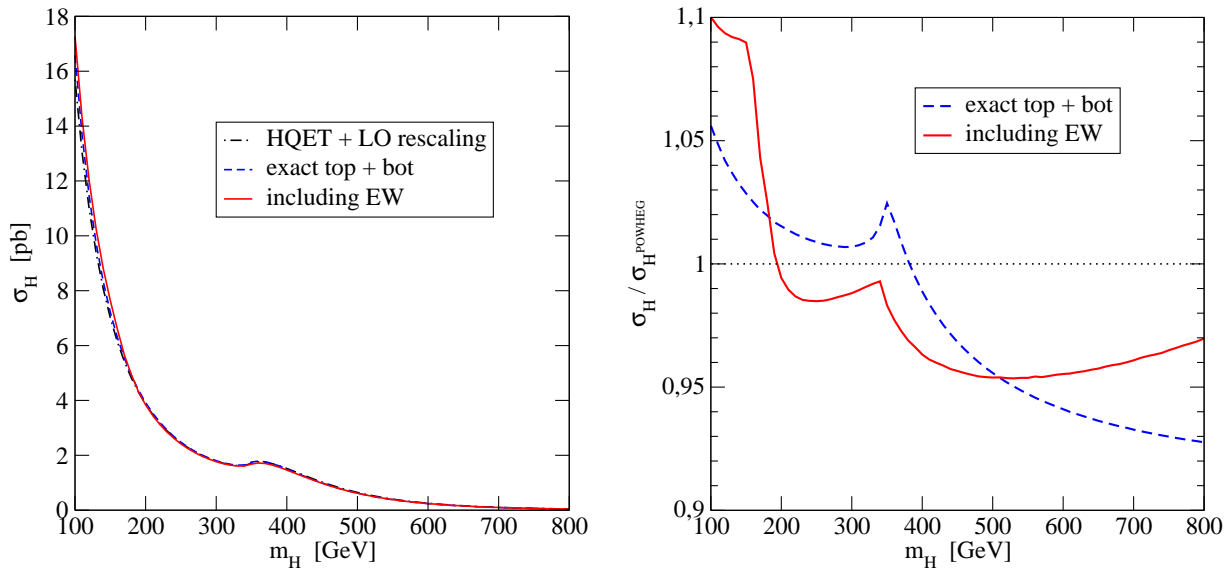

\begin{center}
\includegraphics[height=75mm]{figures-sm/xseclin}~~~
\includegraphics[height=75mm]{figures-sm/ratios}\\
\caption{\small Total cross section for SM Higgs production in gluon
  fusion at the LHC (7 TeV), as a function of the Higgs mass,
  including different subsets of radiative corrections.  We show the
  absolute predictions (left panel) and their ratio with respect to
  the current \PH\ implementation (right panel).
\label{fig:SMxsec}
} 
\end{center}
\end{figure}

In the left panel of figure \ref{fig:SMxsec} we plot the total Higgs
production cross section, without acceptance cuts, in three different
approximations: the dot-dashed line corresponds to the current public
\PH\ implementation, in which the NLO-QCD corrections are computed in
the HQET and are then rescaled with the exact Born cross section
(which includes the full dependence on the top and bottom masses); the
dashed line corresponds to the \PH\ implementation presented in this
paper, where the complete NLO-QCD calculation is employed, i.e.~the
top and bottom contributions are treated exactly in the NLO
corrections; the solid line also includes the effect of the EW
corrections.

In the right panel of figure \ref{fig:SMxsec} we plot the ratio
between the Higgs production cross section obtained using our version
of \PH\ and the one obtained using the current public version. The
dashed line omits the effect of the EW corrections, while the solid
line includes it.  For $m_H\lesssim 160$ GeV the exact treatment of
the quark masses results in an increase up to $\sim 6\%$ in the cross
section, with a further increase (up to a combined $\sim 10\%$) when
the EW corrections are included. This effect is mainly due to the
bottom-quark contribution, which is not negligible when the Higgs
boson is light. For $m_H \gtrsim 180$ GeV the quark-mass effects and
the EW corrections have opposite sign, resulting in a $\sim - 2 \%$
correction in the Higgs mass range up to $2 \, \mt$. Above the
threshold for real top-quark production, where the approximation $\mt
\to \infty$ is not valid, the large corrections due to the quark-mass
effects in the QCD contribution are partially screened by the EW
corrections. 
In the rest of the section we discuss the kinematic distributions of a
SM Higgs boson at NLO QCD. Since the effect of the EW corrections is
very close to an overall rescaling of the total cross section, we
neglect them in the following and focus on the effect of the QCD
corrections.

\begin{figure}[!ht]
\begin{center}
\includegraphics[height=80mm,angle=270]
{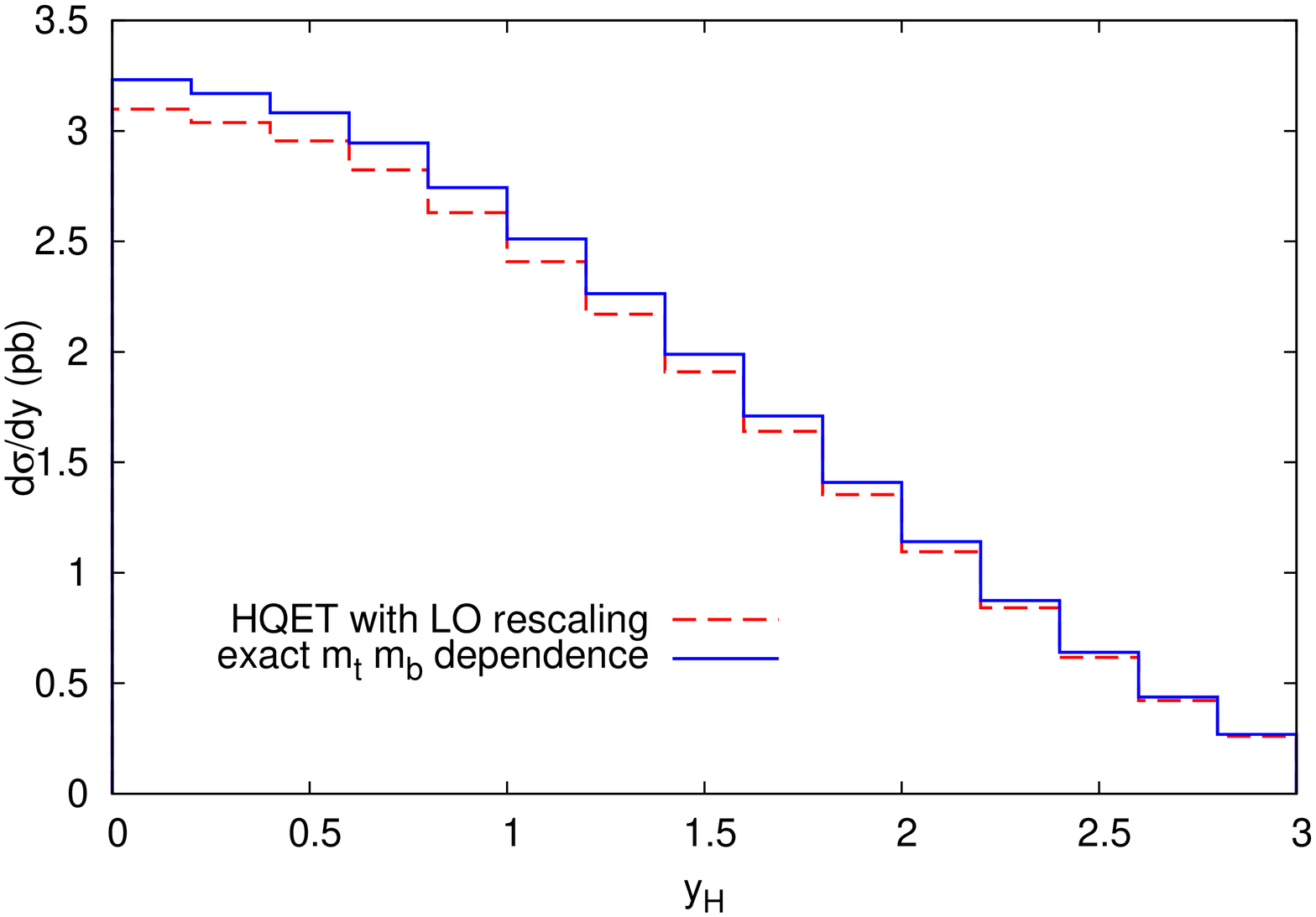}~
\includegraphics[height=80mm,angle=270]
{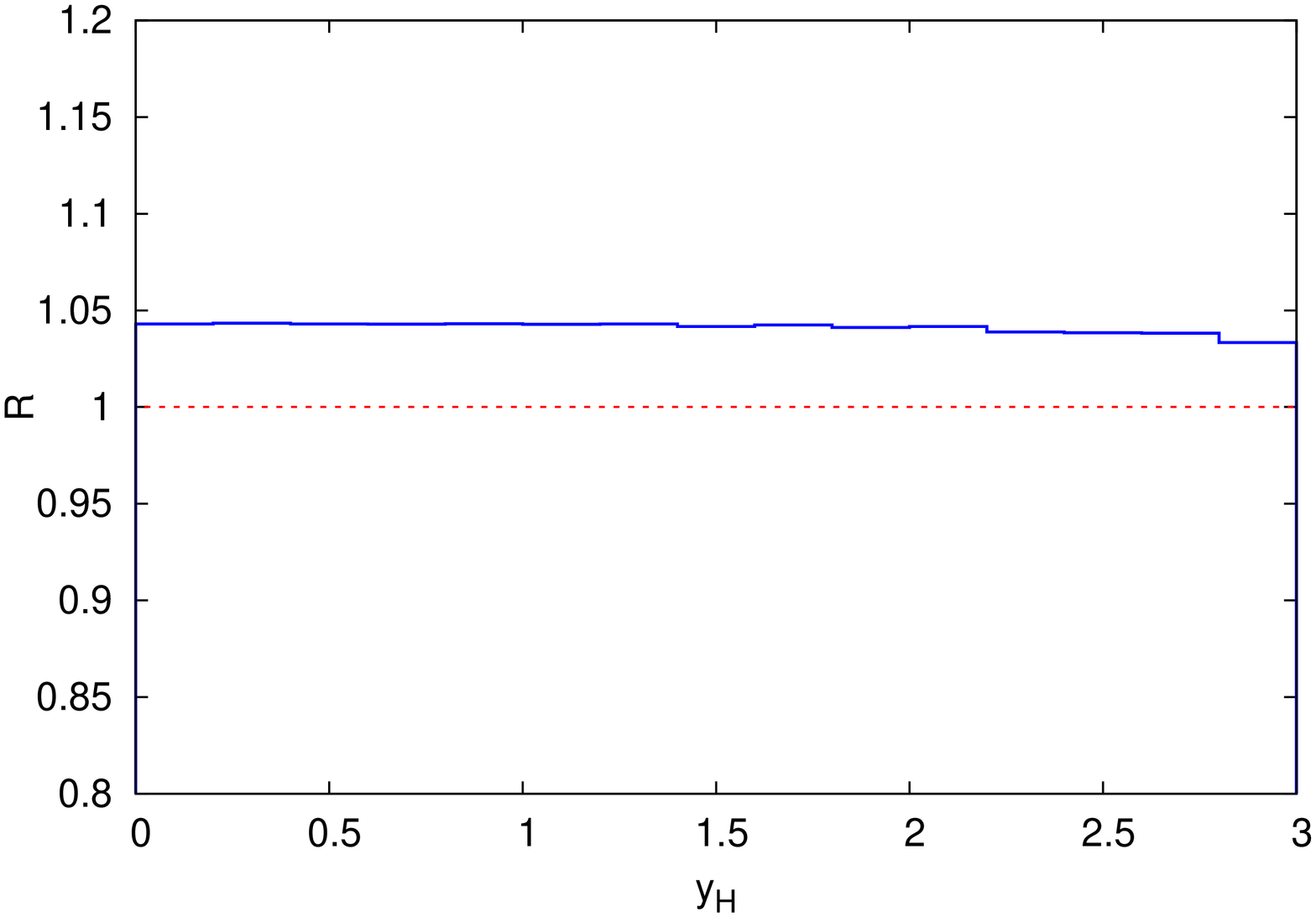}\\~\\
\includegraphics[height=80mm,angle=270]
{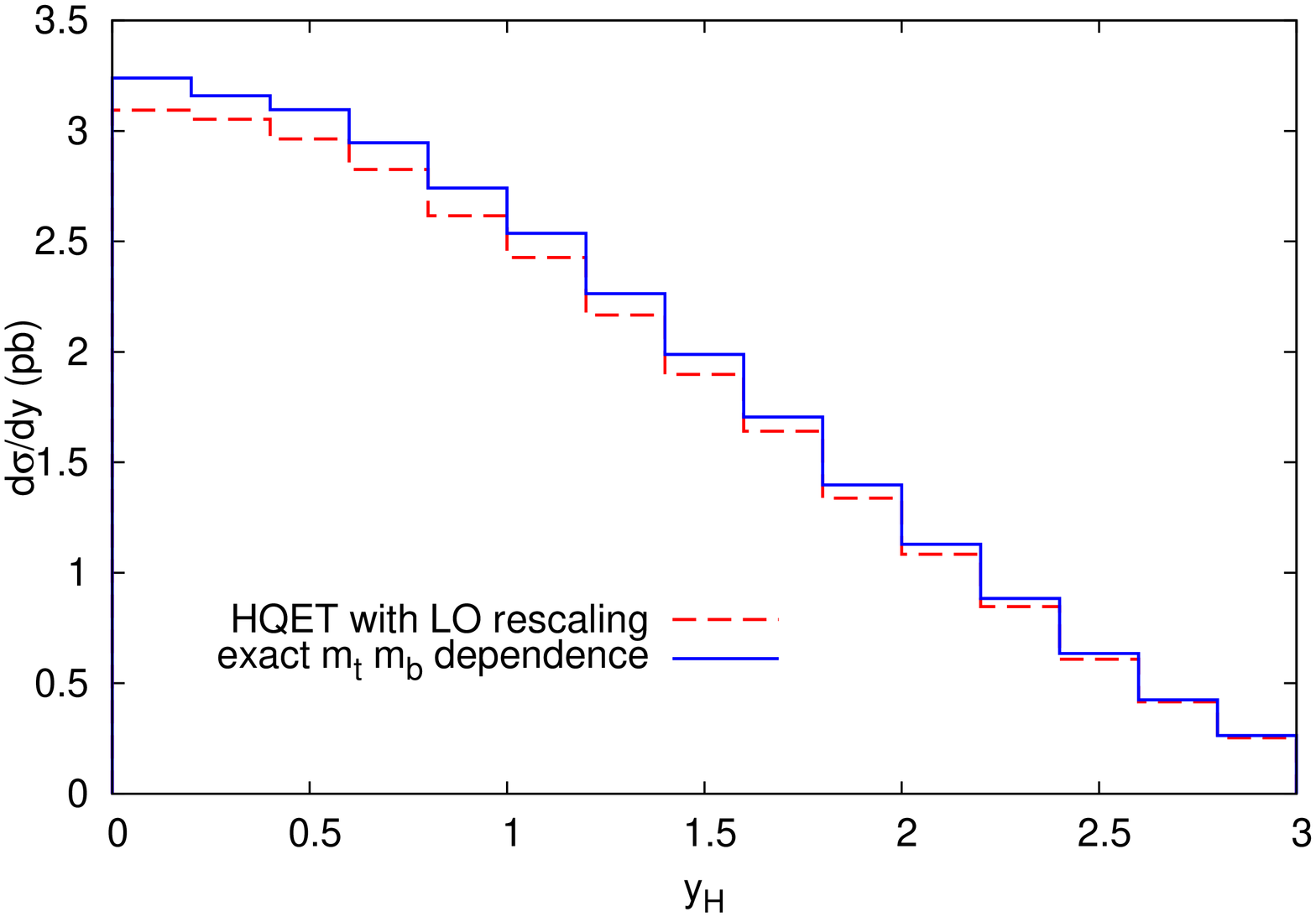}~
\includegraphics[height=80mm,angle=270]
{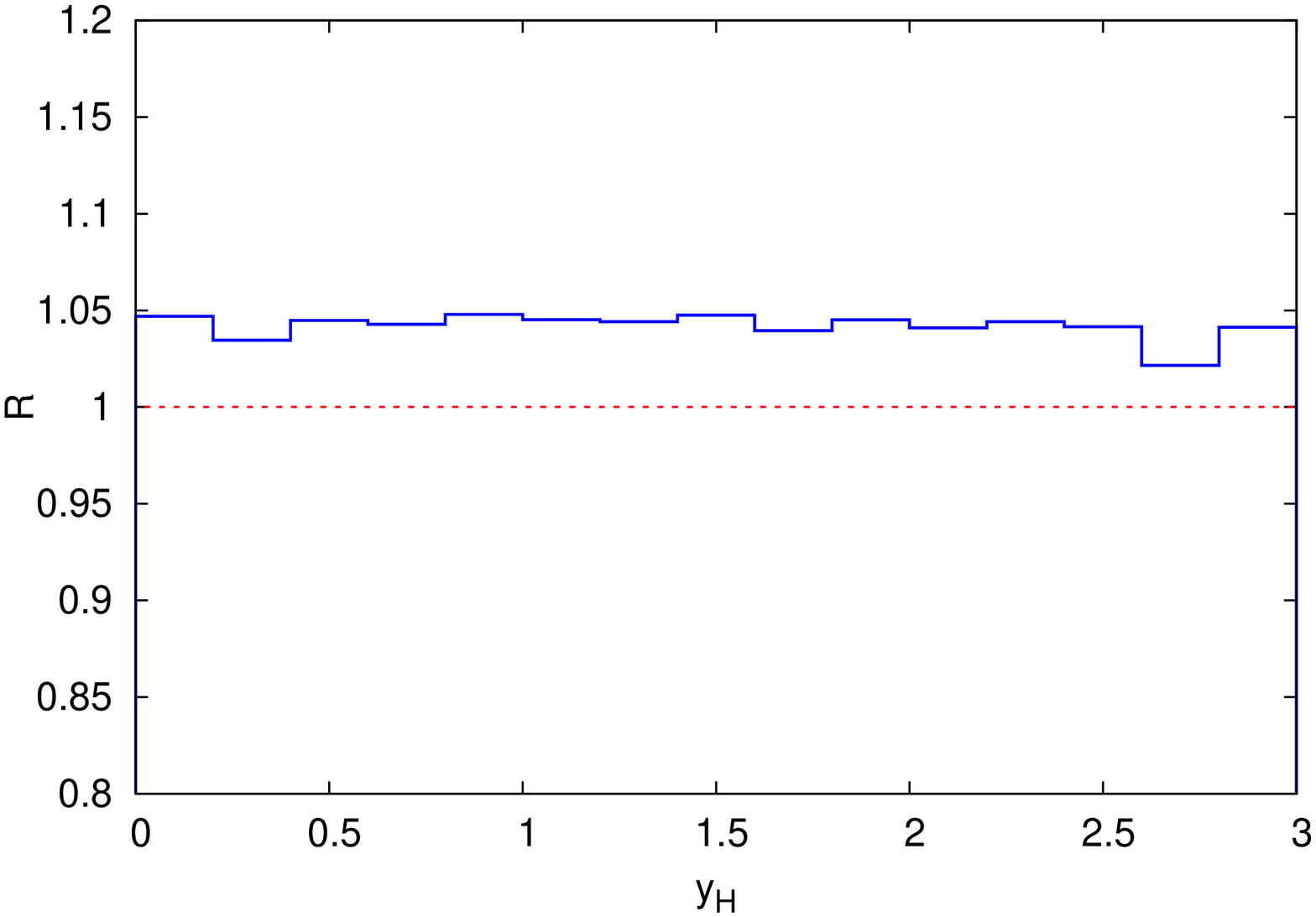}\\~\\
\caption{\small Rapidity distribution for a SM Higgs with $m_H=120$
  GeV. Left plots: in red (dashed) the current implementation of \PH,
  in which the NLO-QCD corrections are computed in the HQET and are
  rescaled by the LO cross section with exact top and bottom mass
  dependence; in blue (solid) the exact NLO-QCD corrections with full
  top and bottom mass dependence. The results are obtained at
  fixed-order NLO QCD (upper plots), or including the effects of the
  Sudakov form factor and of the {\tt PYTHIA} PS (lower plots). Right
  plots: ratio of the result of exact calculation over the result of
  the current \PH\ implementation.
\label{fig:SMy120}
} 
\end{center}
\end{figure}

In figure \ref{fig:SMy120} we show the rapidity distribution for a
Higgs boson with mass $m_H=120$ GeV. In the left panels, the dashed
red lines correspond to the distributions obtained using the current
implementation of \PH, and the solid blue lines correspond to our
implementation. In the right panels we plot the ratio between the
distributions obtained with the two implementations. We compare two
different approximations: in the upper panels we show the pure (i.e.,
fixed-order) NLO-QCD calculation, while in the lower panels we show
the event distributions according to the basic formula of \PH,
eq.~(\ref{eq:POWHEG}), which includes the effects of the Sudakov form
factors, merged also with the {\tt PYTHIA} QCD PS.  As appears from
the plots, the exact treatment of the quark masses results in a $\sim
5\%$ enhancement, uniformly distributed over the whole rapidity
range. In the absence of acceptance cuts, the results obtained by
combining \PH\ and {\tt PYTHIA} do not differ significantly from the
pure NLO-QCD results.

\begin{figure}[p]
\begin{center}
\includegraphics[height=80mm,angle=270]
{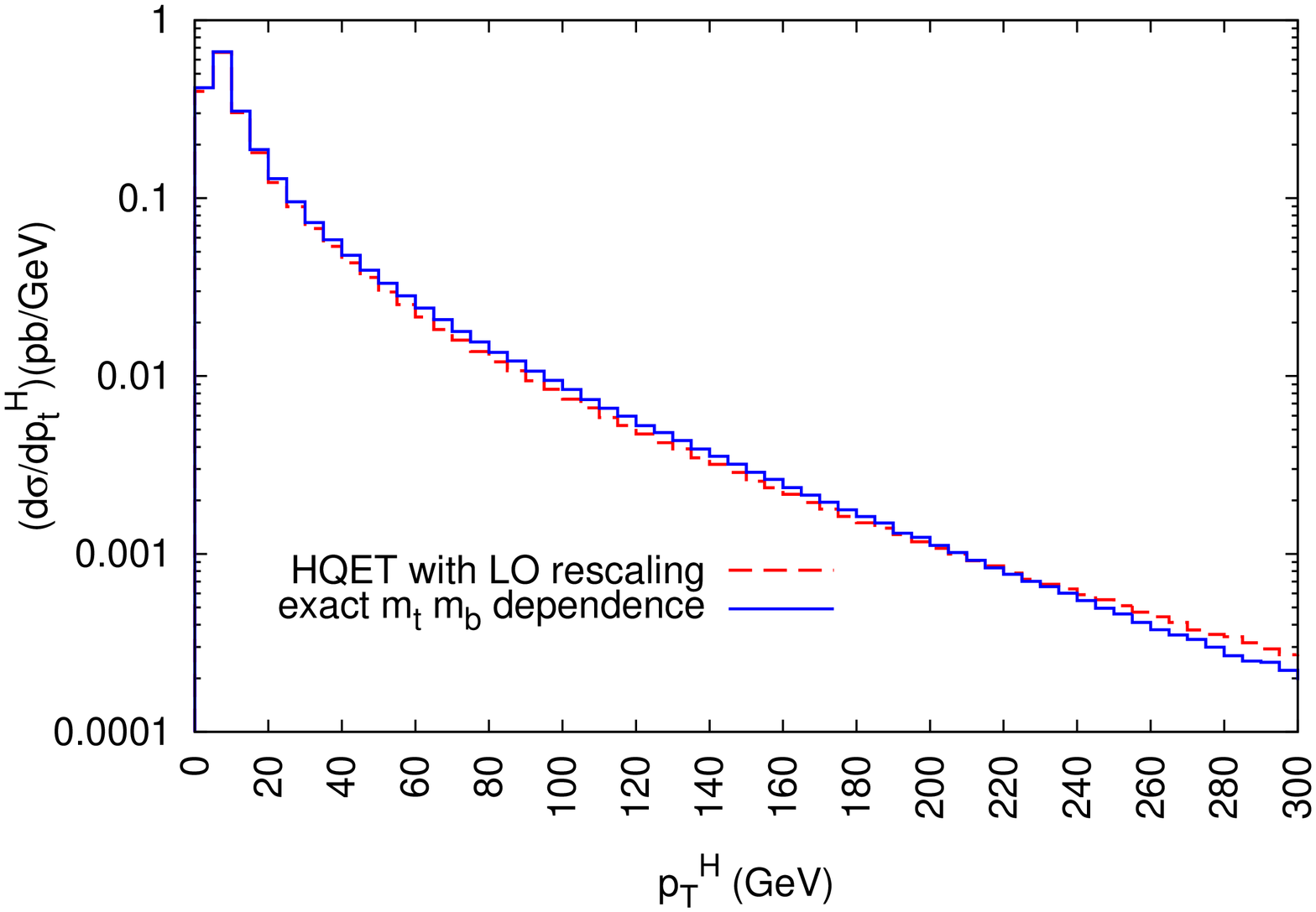}~
\includegraphics[height=80mm,angle=270]
{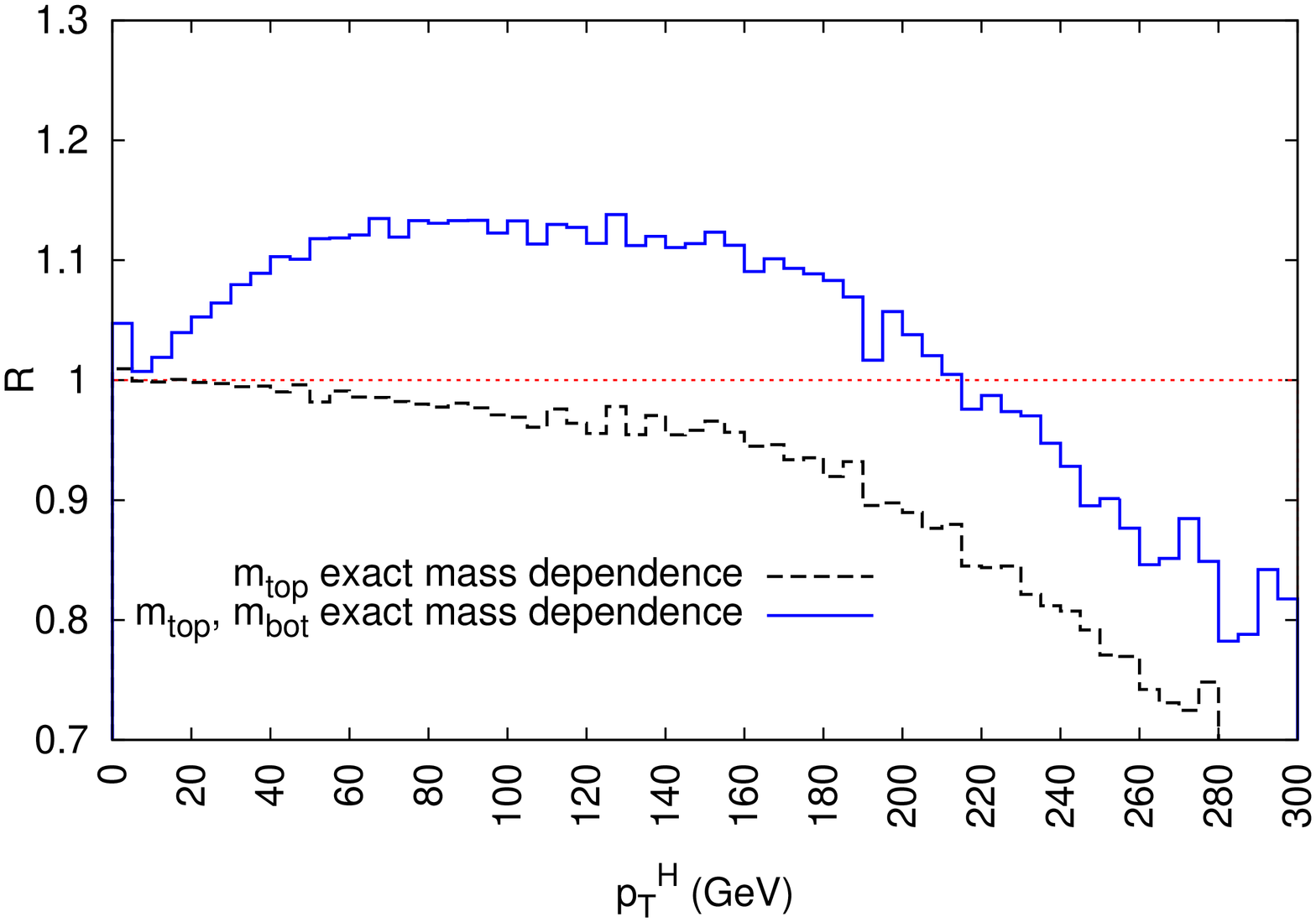}\\
\includegraphics[height=80mm,angle=270]
{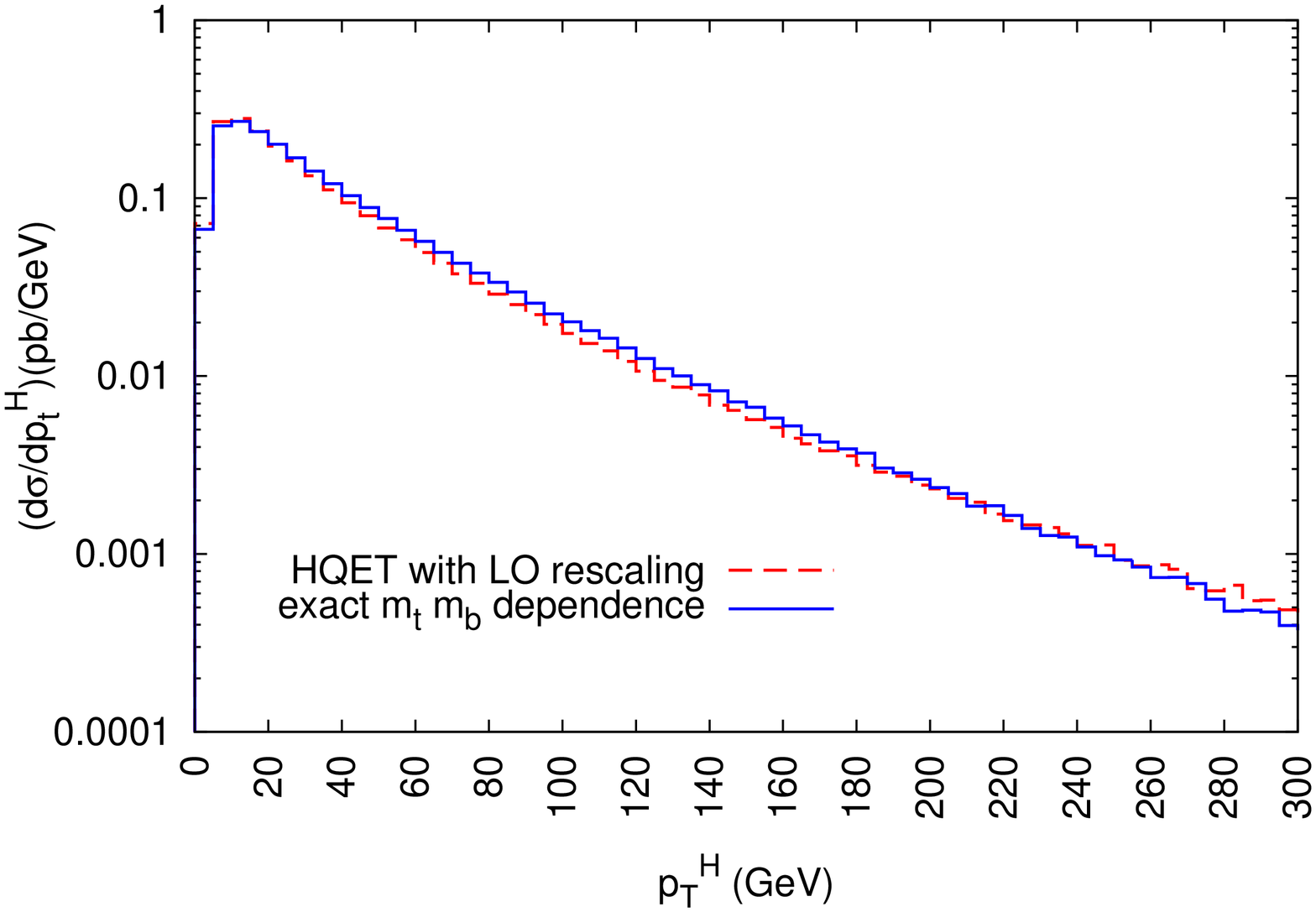}~
\includegraphics[height=80mm,angle=270]
{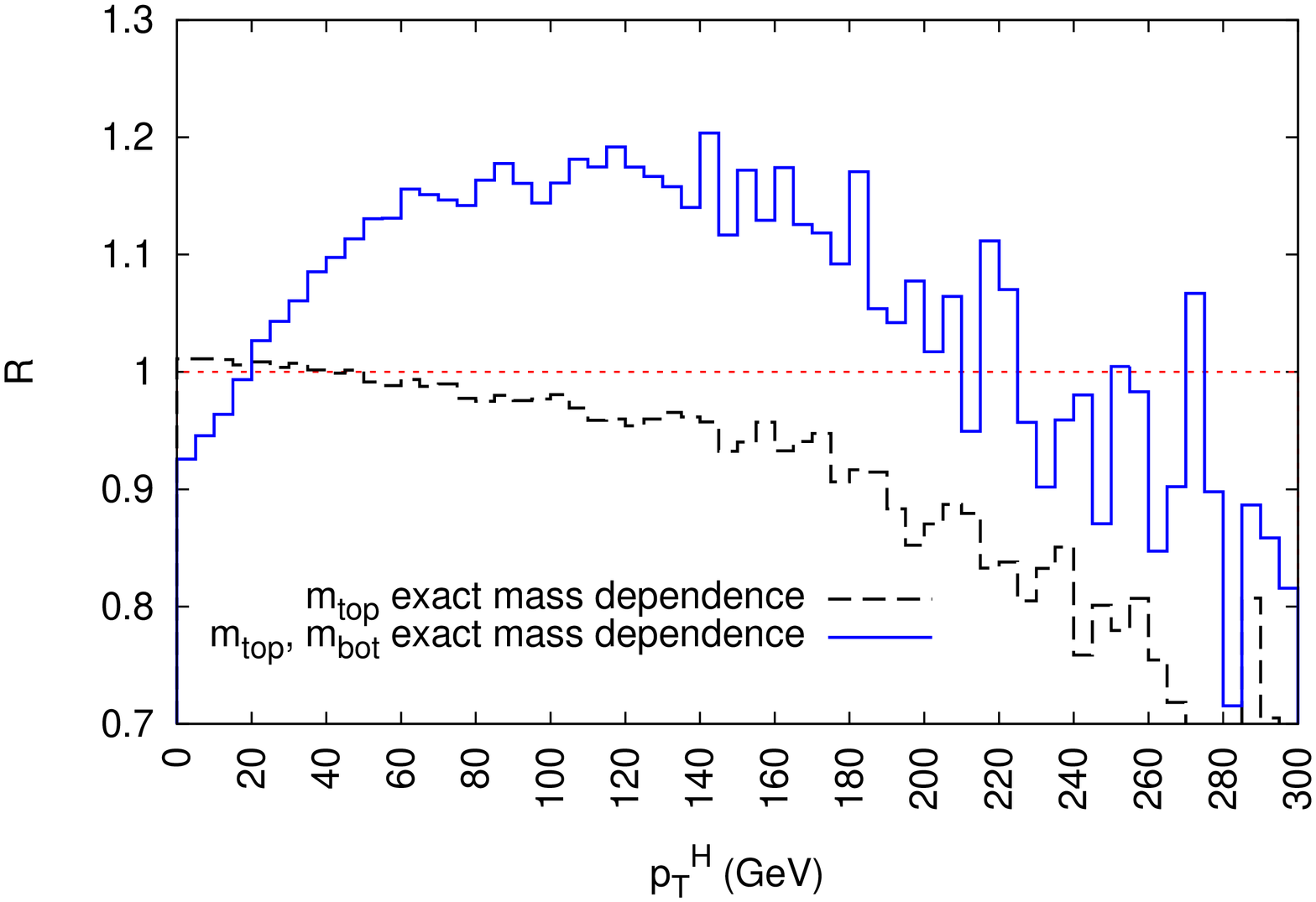}\\
\includegraphics[height=80mm,angle=270]
{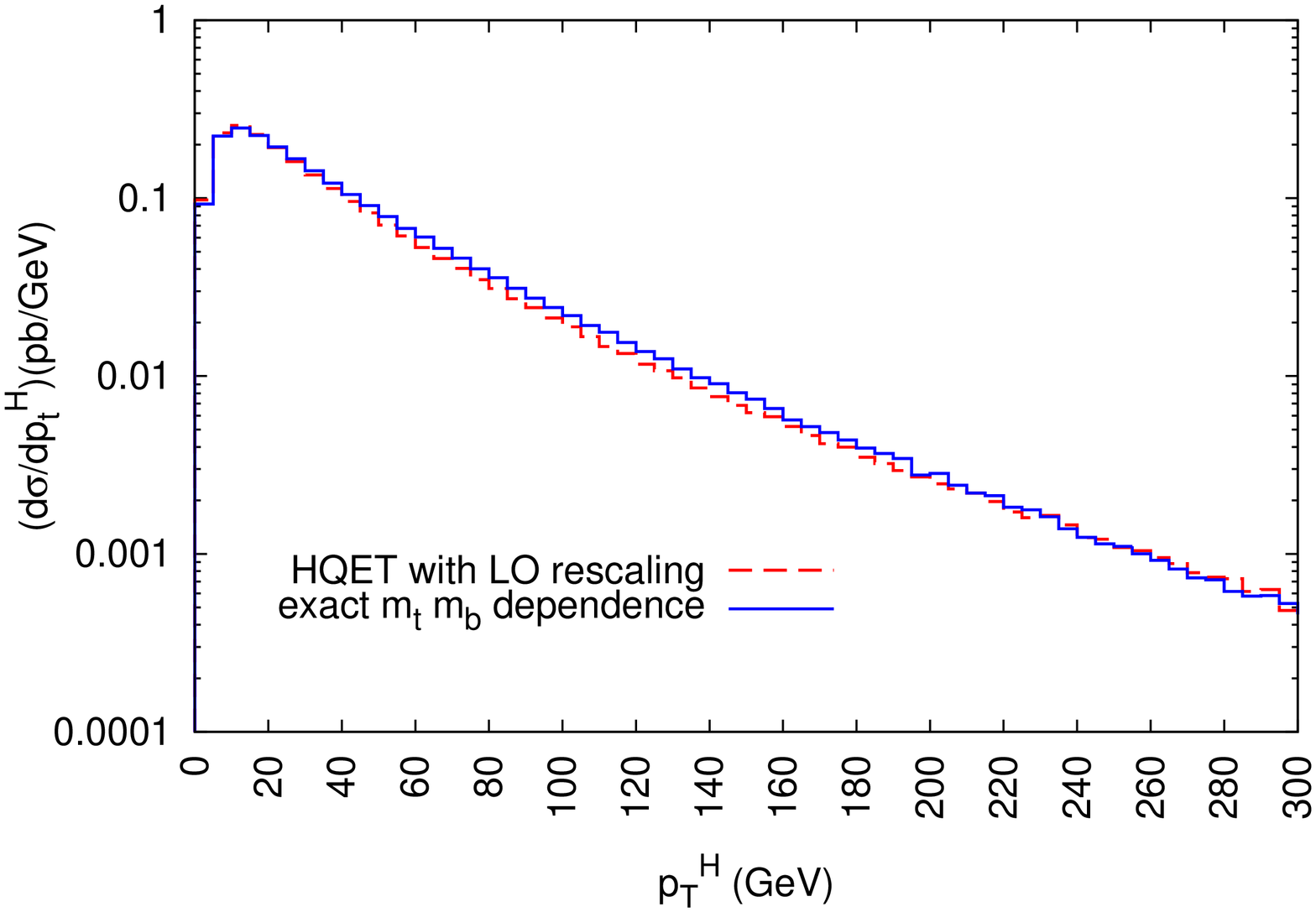}~
\includegraphics[height=80mm,angle=270]
{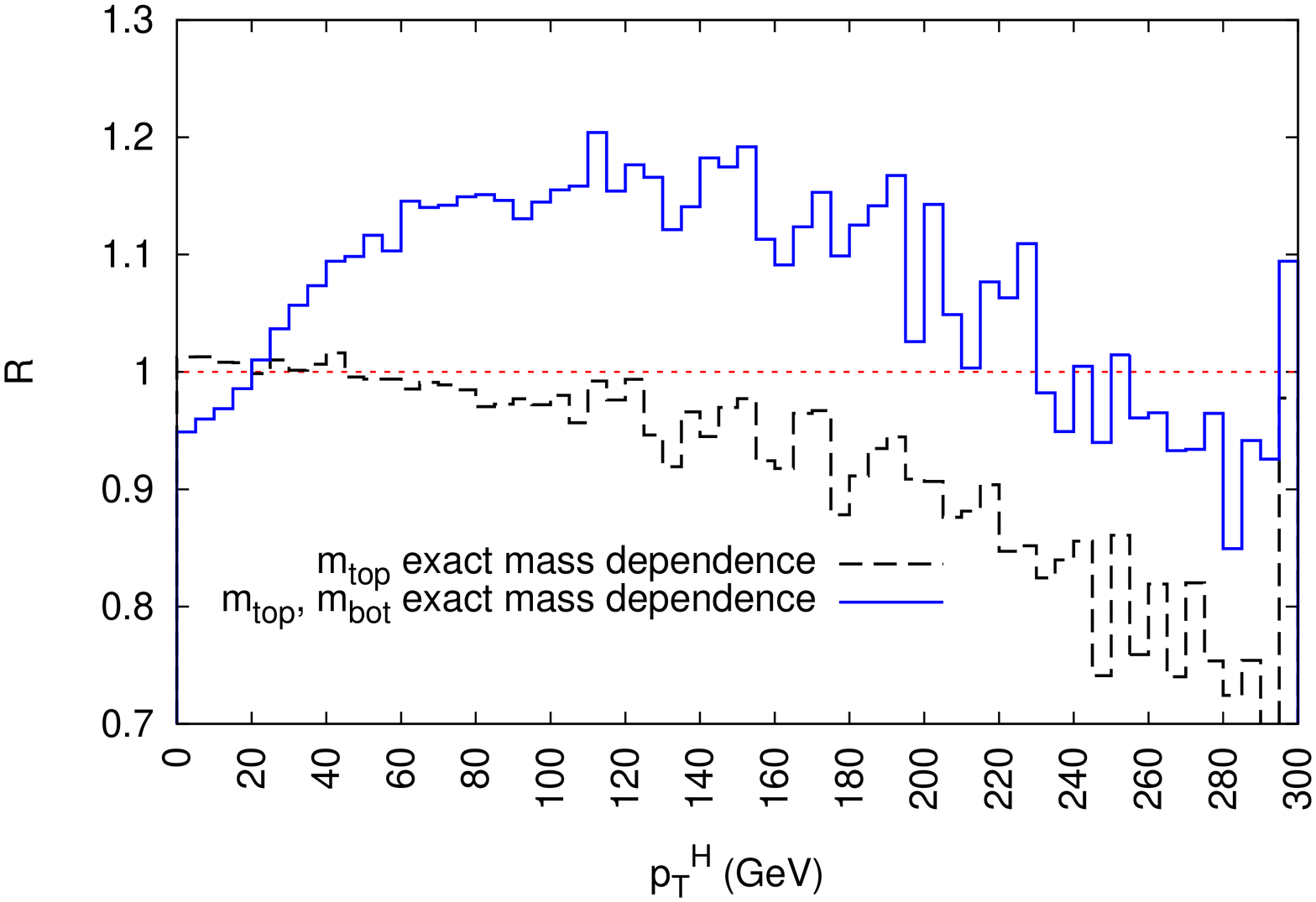}\\
\caption{\small Transverse momentum distribution for a SM Higgs with
  $m_H=120$ GeV.  Left plots: in red (dashed) the current
  \PH\ implementation, in which the NLO-QCD corrections are computed
  in the HQET and are rescaled by the LO cross section with full top
  and bottom mass dependence; in blue (solid) the exact NLO-QCD
  corrections with full top and bottom mass dependence. The results
  are obtained at NLO QCD (upper plots), including the effects of the
  Sudakov form factor (middle plots), including also the effects of
  the {\tt PYTHIA} QCD PS (lower plots). Right plots: the full NLO-QCD
  results (blue, solid) and the ones obtained by introducing in
  \PH\ only the exact top-mass dependence (black, dashed), both
  normalized to the results of the current \PH\ implementation.
\label{fig:SMpt120}
} 
\end{center}
\end{figure}

In figure \ref{fig:SMpt120} we show the transverse momentum
distribution for a Higgs boson of mass $m_H=120$ GeV.  In the left
panels we compare the current \PH\ implementation with ours but,
differently from figure \ref{fig:SMy120}, we show separately the pure
NLO-QCD, \PH\ and \PH\ + {\tt PYTHIA} calculations. In the right
panels we plot the full results (blue, solid lines) and the ones
obtained by introducing in \PH\ only the exact top-mass dependence
(black, dashed lines), both normalized to the results of the current
\PH\ implementation.

The pure NLO-QCD calculation (upper panels) diverges for vanishing
Higgs transverse momentum, although in figure \ref{fig:SMpt120} the
divergence is masked by the lower cut on $p_T^H$.  The plot on the
right shows that for $p_T^H \lesssim m_t$ it is the inclusion of the
bottom-quark contribution in the NLO corrections that gives rise to a
positive correction with respect to the result obtained with the
current \PH\ version, while for $p_T^H\gtrsim m_t$ there is a
substantial modification of the distribution, with a large negative
correction driven by the use of the exact top-quark contribution.

The inclusion of the effect of the Sudakov form factor (central
panels) modifies the distribution in such a way that the latter
vanishes in the limit $p_T^H\to 0$.  As can be read from
eq.~(\ref{eq:POWHEG}), the probability of emitting the Higgs in
association with a parton depends on the product $\Delta\times R/B$,
where $R$ is the squared matrix element for real-parton emission, $B$
is the Born amplitude, and $\Delta$ is the Sudakov factor, which in
turn is exponentially suppressed by $R/B$, see eq.~(\ref{hgen}). In
the current \PH\ implementation, the emission probability is computed
in terms of the ratio $R(t,\infty)/B(t,\infty)$, where both $R$ and
$B$ are evaluated in the HQET.\footnote{In the current
  \PH\ implementation only the total cross section is rescaled by the
  exact top+bottom LO result.}  For small $p_T^H$, the Sudakov factor
with exact top and bottom mass dependence, $\Delta(t+b,exact)$, used
in our implementation is smaller than the corresponding factor
$\Delta(t,\infty)$ used in the current \PH\ implementation, because
$R(t+b,exact)/B(t+b,exact)>R(t,\infty)/B(t,\infty)$. This inequality
holds for two reasons: first, the $p_T^H$ distribution is proportional
to $R$, and $R(t+b,exact)>R(t,\infty)$ for $p_T^H<200$
GeV~\cite{realtb}; second, the inclusion of the bottom contribution
reduces the LO cross section with respect to the result obtained in
the HQET~\cite{SDGZ}. Thus, as shown in the right plot, for small
$p_T^H$ the Sudakov factor suppresses the $p_T^H$ distribution by
almost 10\% with respect to the result obtained in the current
\PH\ implementation.  Since the emission probability is also directly
proportional to the ratio $R/B$, starting from $p_T^H \simeq 30$ GeV
this factor prevails over the Sudakov factor, and the distribution
with exact dependence on the quark masses becomes larger than the one
in the current \PH\ implementation by up to $\sim 15\%$. Finally, for
$p_T^H\gtrsim \mt$ the inclusion of the full top-mass dependence leads
to a negative correction, similar to the one already observed in the
pure NLO-QCD calculation. The inclusion of multiple gluon emission
with the {\tt PYTHIA} QCD-PS (lower panels) does not change
dramatically -- in the absence of acceptance cuts -- the results
obtained including only the hardest emission.

%
\begin{figure}[p]
\begin{center}
\includegraphics[height=80mm,angle=270]
{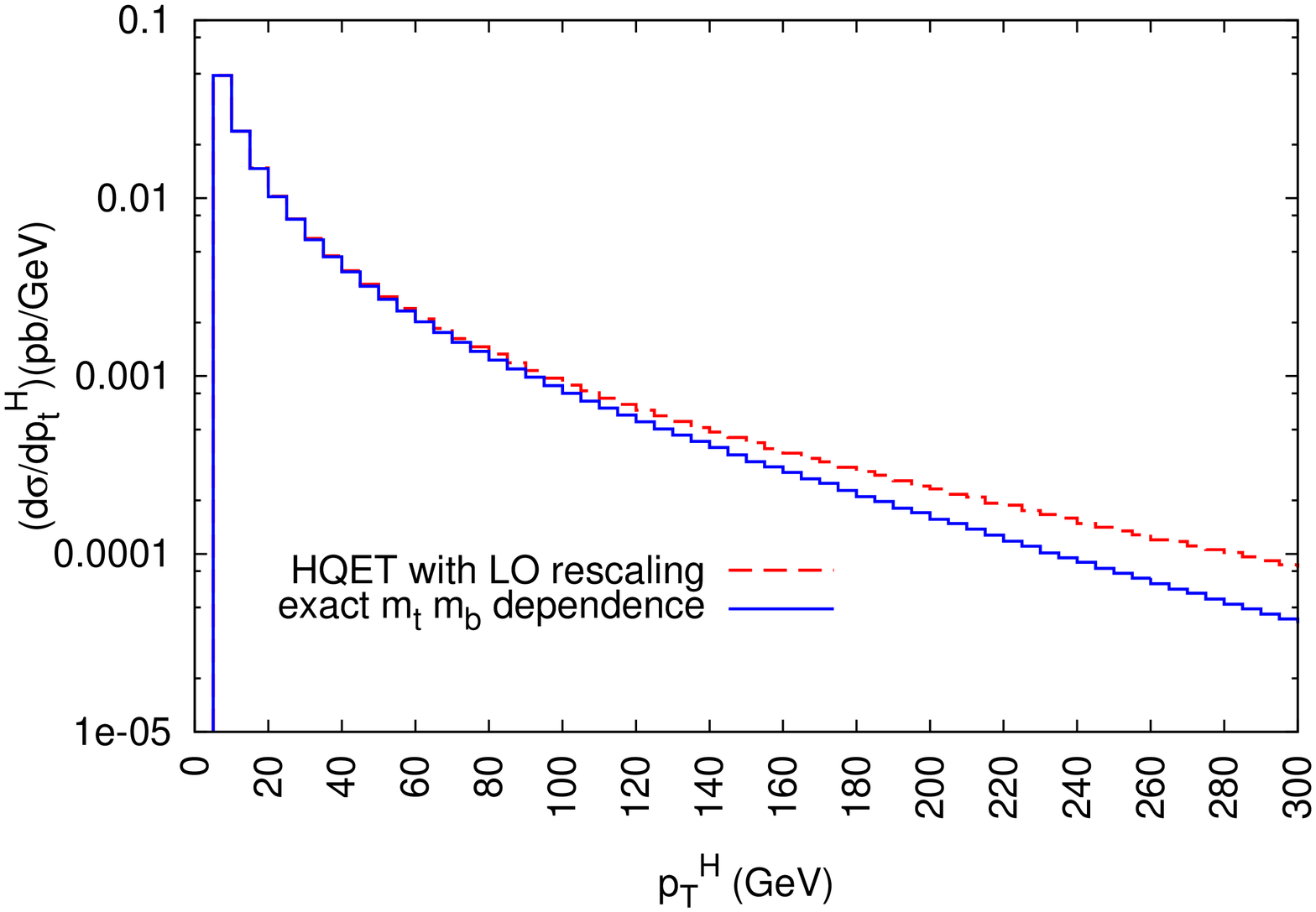}~
\includegraphics[height=80mm,angle=270]
{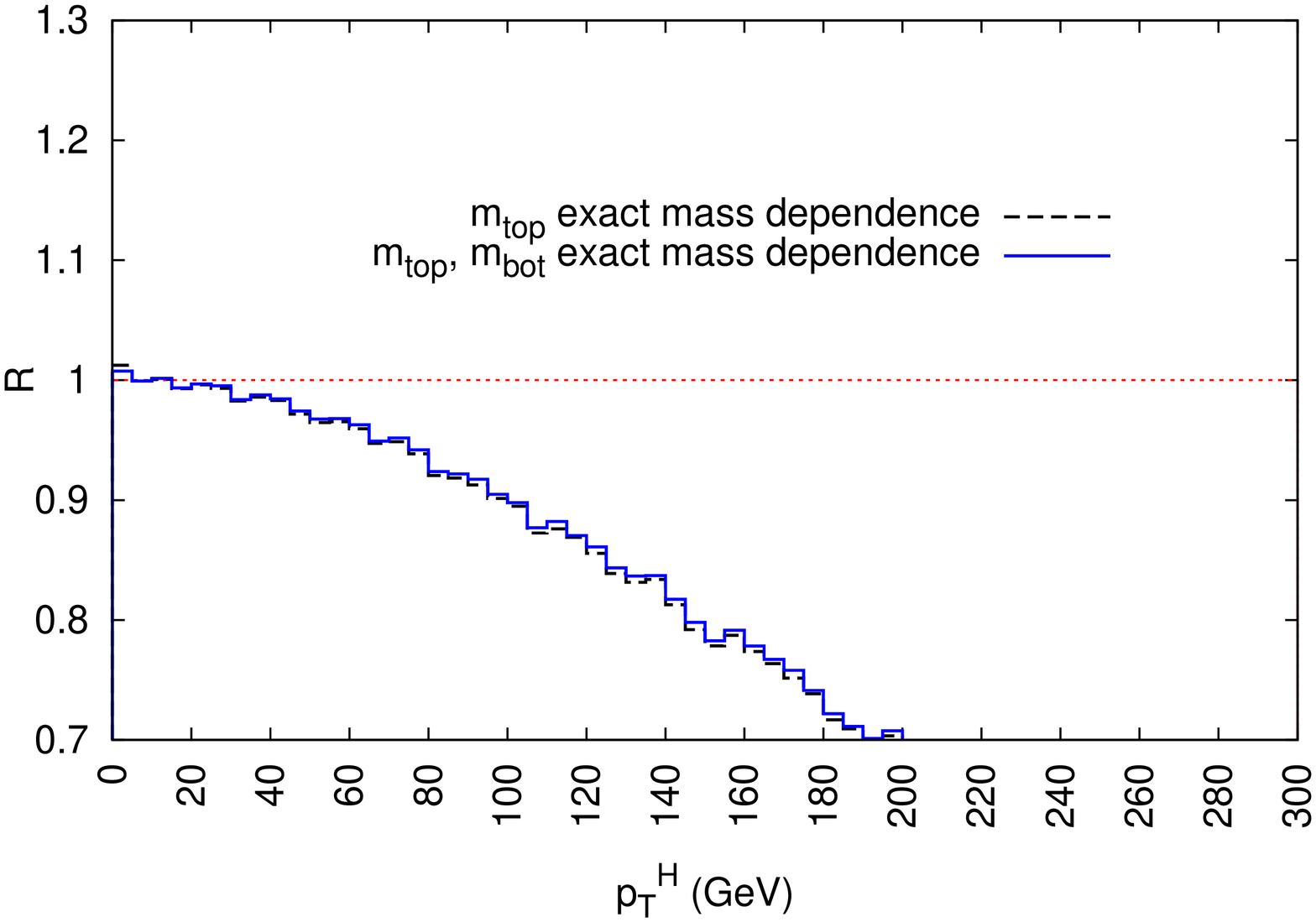}\\
\includegraphics[height=80mm,angle=270]
{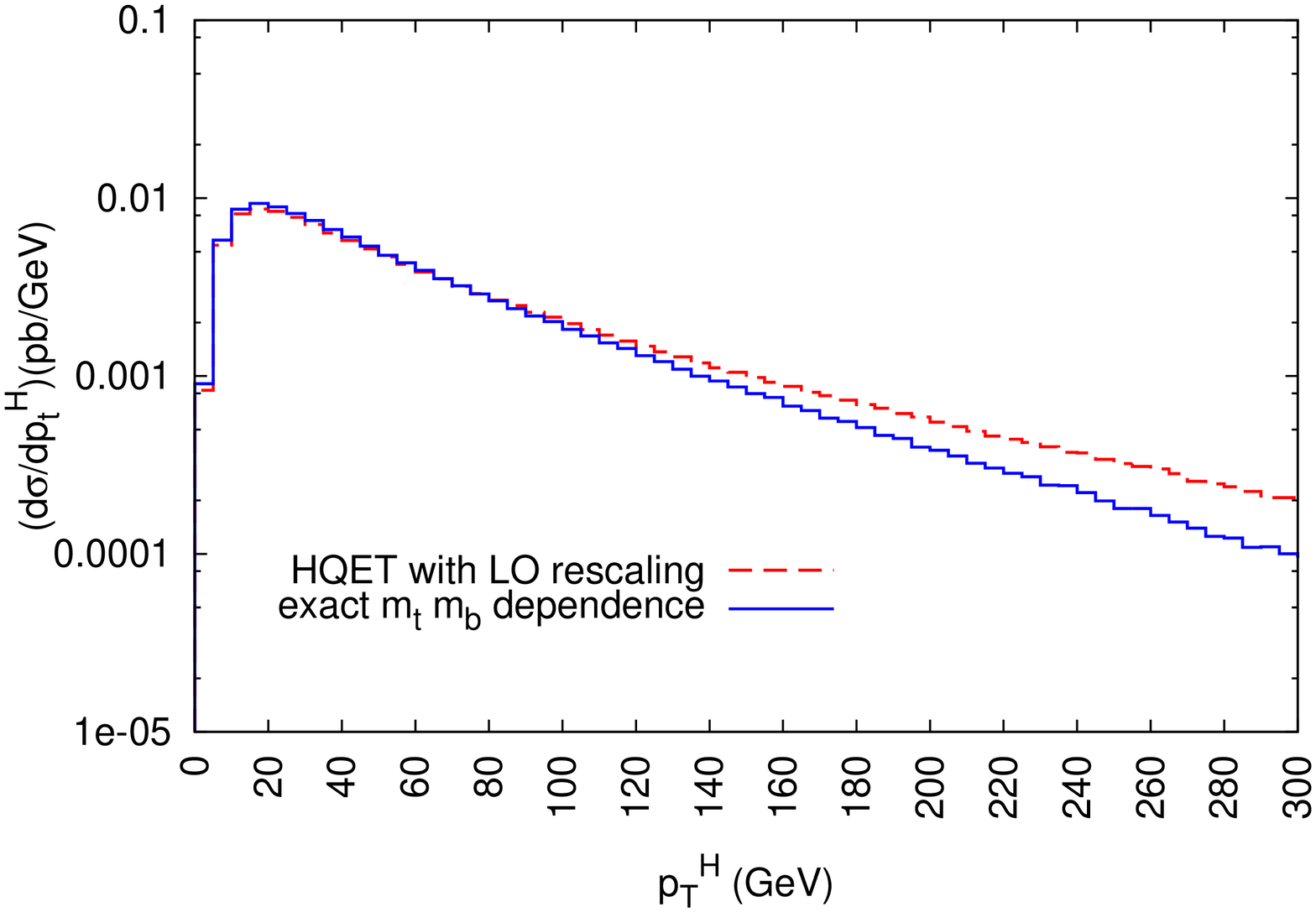}~
\includegraphics[height=80mm,angle=270]
{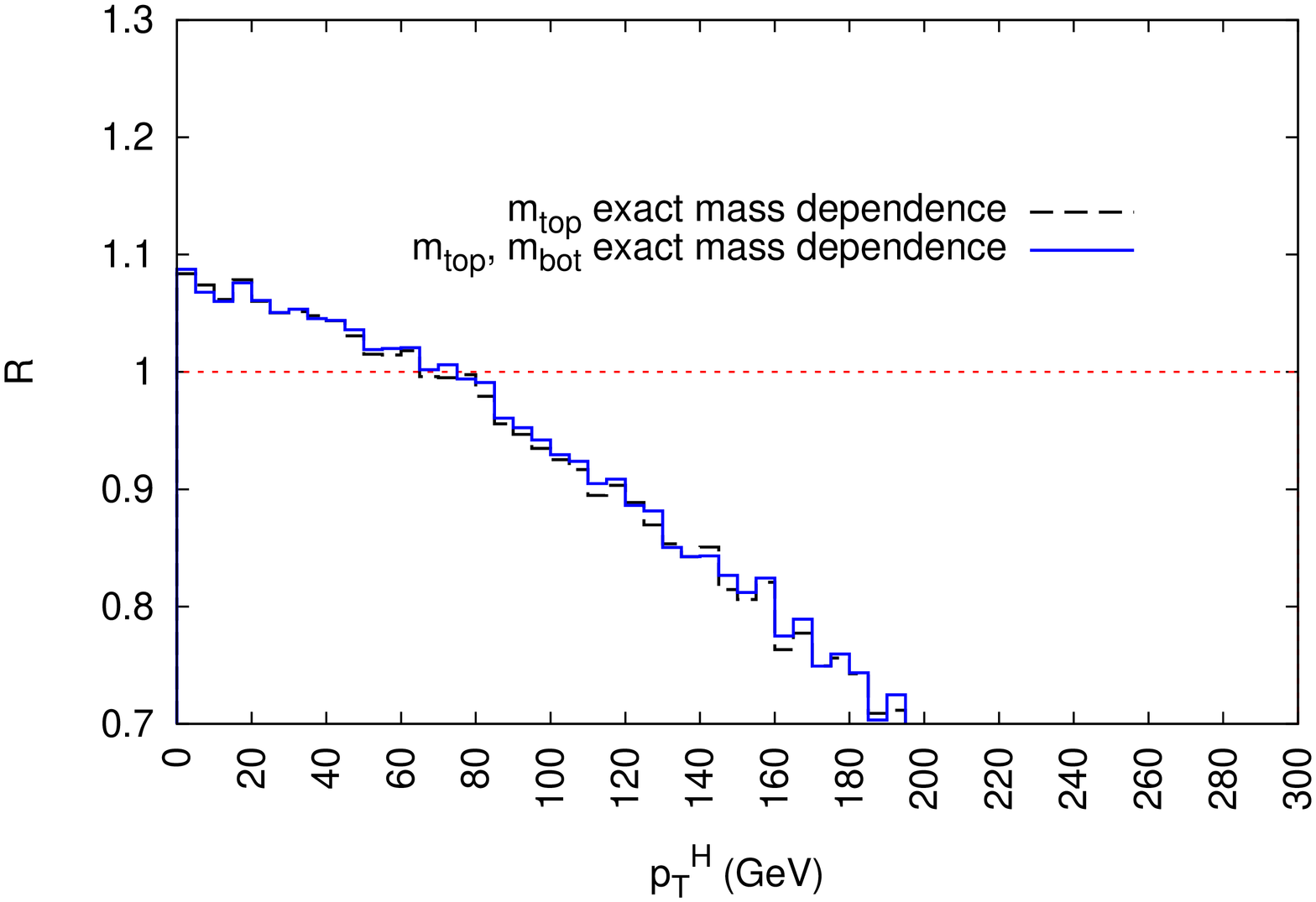}\\
\includegraphics[height=80mm,angle=270]
{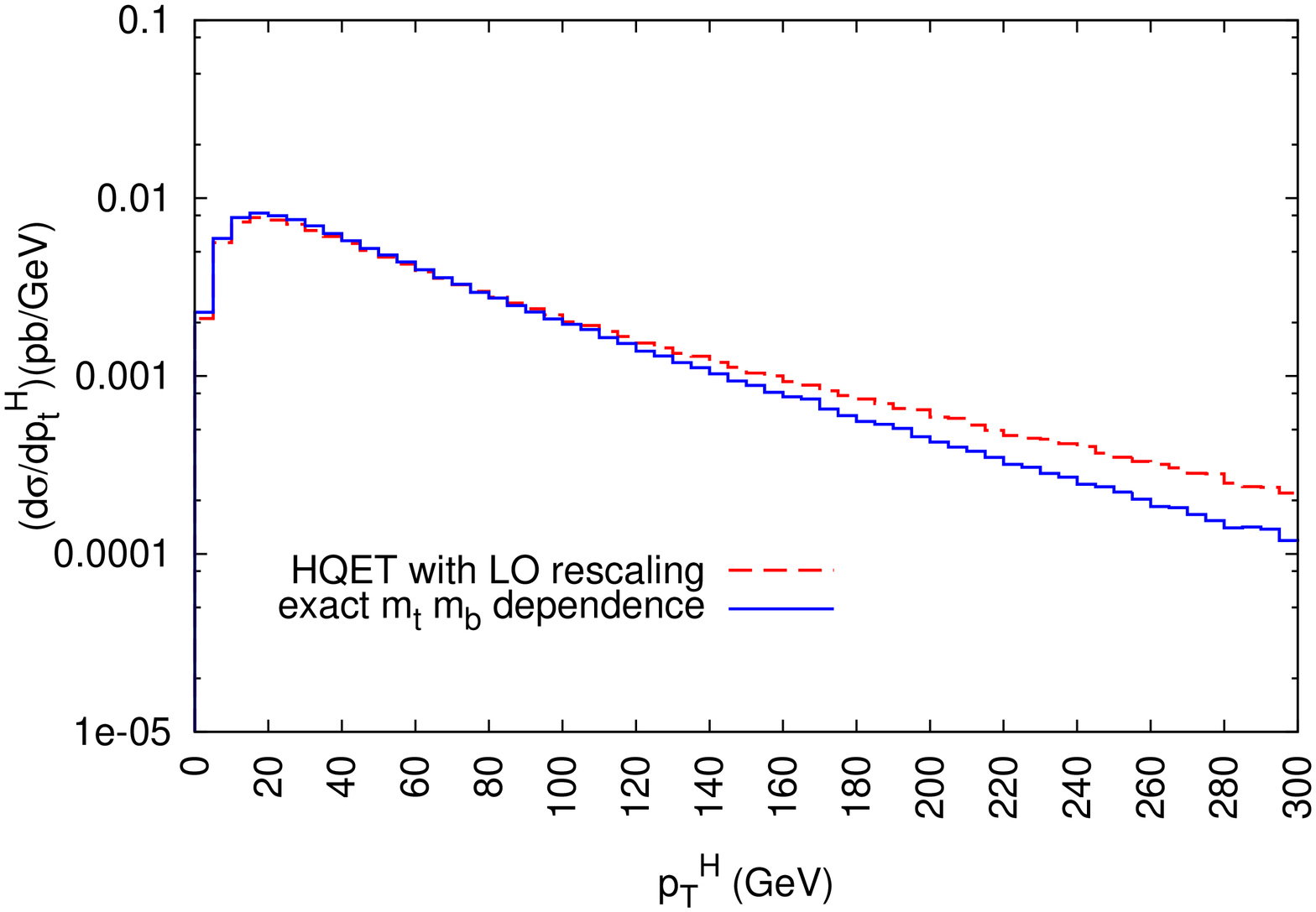}~
\includegraphics[height=80mm,angle=270]
{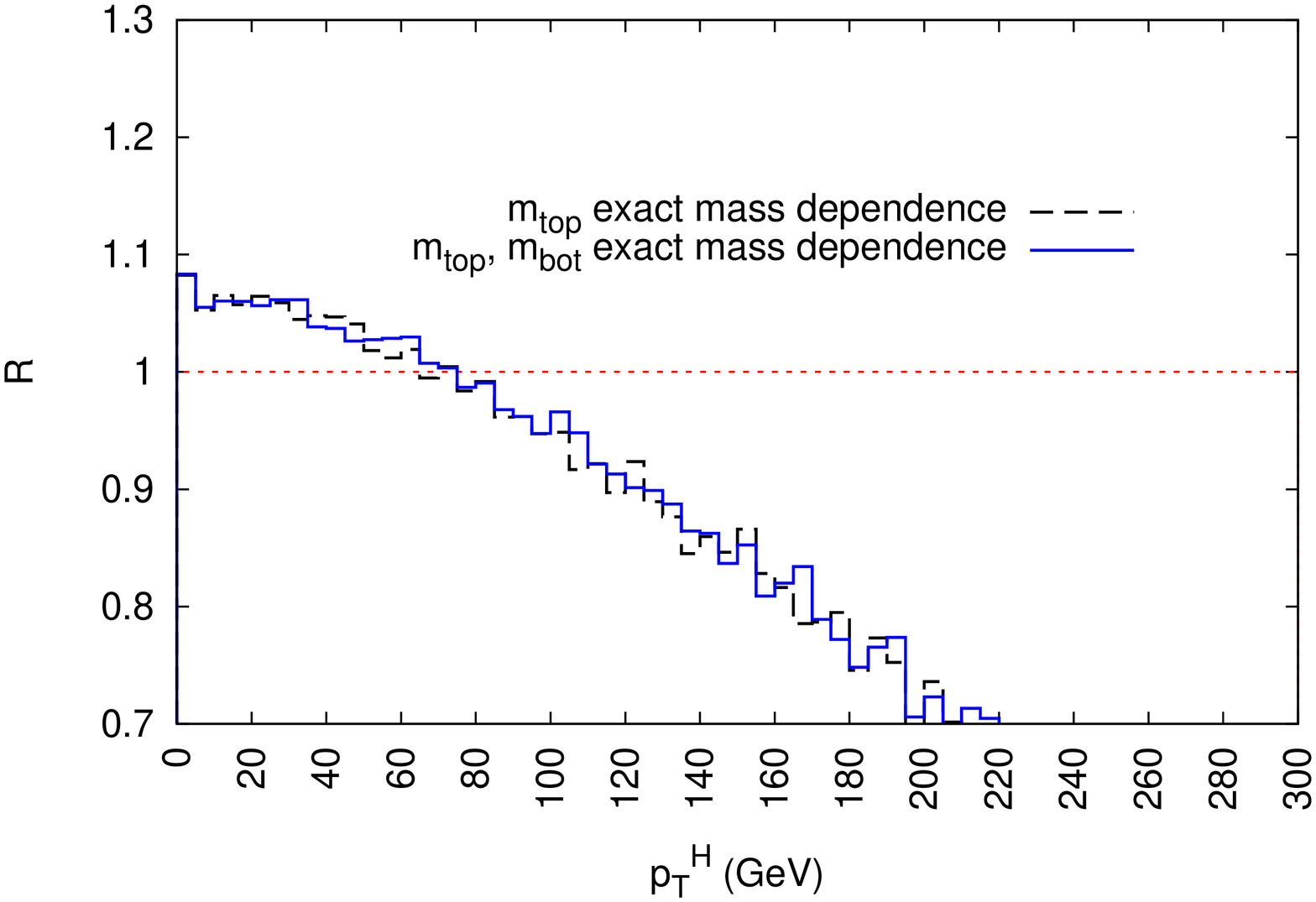}\\
\caption{\small Transverse momentum distribution for a SM Higgs with
  $m_H=500$ GeV.  Left plots: in red (dashed) the current
  \PH\ implementation, in which the NLO-QCD corrections are computed
  in the HQET and are rescaled by the LO cross section with full top
  and bottom mass dependence; in blue (solid) the exact NLO-QCD
  corrections with full top and bottom mass dependence. The results
  are obtained at NLO QCD (upper plots), including the effects of the
  Sudakov form factor (middle plots), including also the effects of
  the {\tt PYTHIA} QCD PS (lower plots). Right plots: the full NLO-QCD
  results (blue, solid) and the ones obtained by introducing in
  \PH\ only the exact top-mass dependence (black, dashed), both
  normalized to the results of the current \PH\ implementation.
\label{fig:SMpt500}
} 
\end{center}
\end{figure}

Figure \ref{fig:SMpt500} shows the transverse momentum distribution
for a Higgs boson of mass $m_H=500$ GeV. The meaning of the different
curves is the same as in figure \ref{fig:SMpt120}.  In the pure
NLO-QCD calculation (upper panels), the effect of using a finite top
mass in the NLO corrections is always negative and monotonically
decreasing for increasing transverse momentum.  Contrary to the
light-Higgs case, the use of a finite top mass in the Sudakov form
factor (central panels) yields an enhancement of the transverse
momentum distribution at small values of $p_T^H$.  For $p_T^H\gtrsim
80$ GeV the mass effects induce instead a dampening of the spectrum.
As for the case of $m_H = 120$ GeV, the introduction of the QCD PS
(lower panels) does not modify substantially the distributions.  As
can be seen from the plots on the right, for such a large value of the
Higgs mass the bottom contribution is highly suppressed.

\begin{figure}[t]
\begin{center}
\includegraphics[height=80mm,angle=270]
{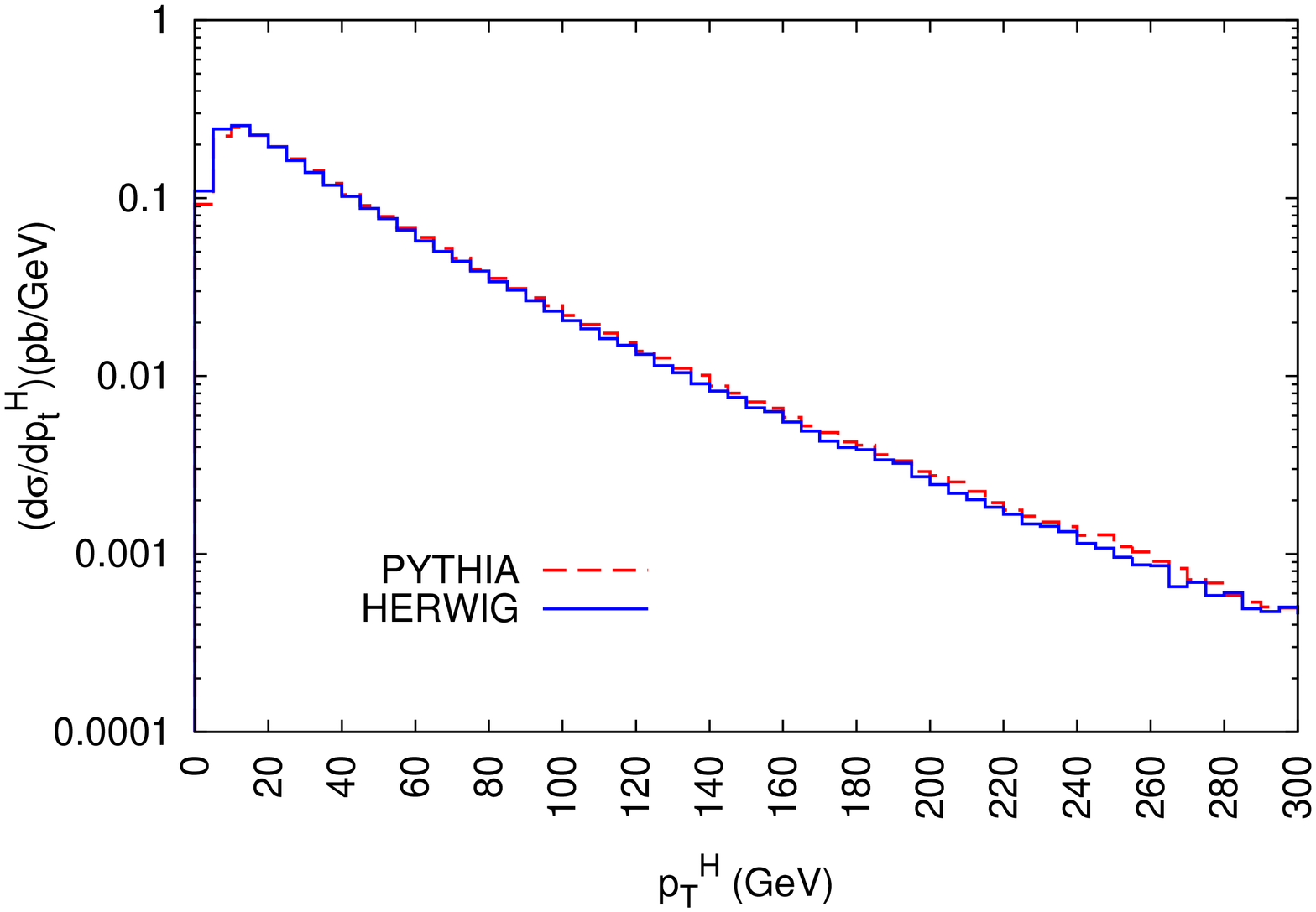}~
\includegraphics[height=80mm,angle=270]
{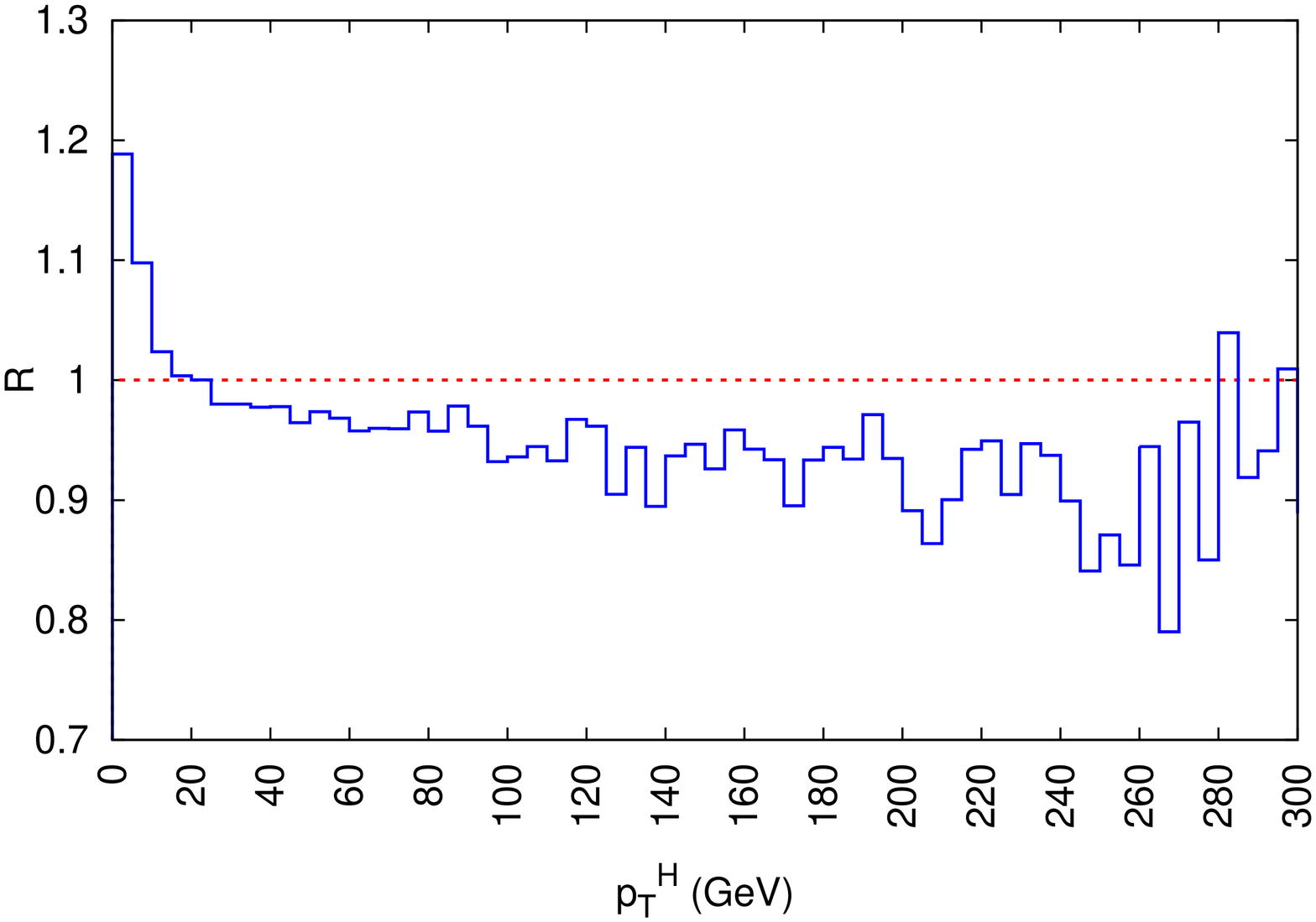}
\caption{\small Transverse momentum distribution for a SM Higgs with
  $m_H=120$ GeV.  Left: comparison of the results obtained with {\tt
    POWHEG+PYTHIA} (red dashed line) and with {\tt POWHEG+HERWIG}
  (blue solid line).  Right: ratio {\tt POWHEG+HERWIG} over {\tt
    POWHEG+PYTHIA}.
\label{fig:HEvsPY}
} 
\end{center}
\end{figure}

In order to illustrate the uncertainty arising from the use of
different PS algorithms, in the left panel of figure \ref{fig:HEvsPY}
we compare the transverse momentum distributions obtained with our
implementation of \PH\ matched with either {\tt PYTHIA} (red dashed
line) or {\tt HERWIG} (blue solid line), for a Higgs boson of mass
$m_H=120$ GeV.  In the right panel of figure \ref{fig:HEvsPY} we plot
the ratio of the distribution obtained with {\tt POWHEG+HERWIG} over
the one obtained with {\tt POWHEG+PYTHIA}.  The {\tt HERWIG} showering
algorithm yields a differential cross section larger by up to 10-20\%
with respect to {\tt PYTHIA} in the region of small transverse momenta
($p_T^H\leq 20$ GeV), and smaller by 5-10\% for large values of
$p_T^H$. We remark that this behavior is not specific to our
implementation of \PH. Indeed, the current implementation yields
differences of similar size when matched with {\tt HERWIG} instead of
{\tt PYTHIA}.

In our SM analysis the top and bottom masses are renormalized in the
on-shell scheme. Therefore, the dependence of the differential cross
section on the renormalization and factorization scales does not
differ significantly from the case of the HQET. As an example, for
$m_H=120$ GeV and $p_T^H=20$ GeV, we find for our \PH\ implementation
an uncertainty band of $[-16\%,+21\%]$ around the central value when
the factorization and renormalization scales are varied in a range
between $0.5\, m_H$ and $2\,m_H$ while keeping the ratio of the two
scales in the range $[0.5,2]$. In the case of the HQET the
corresponding uncertainty band is $[-16\%,+20\%]$.

The phenomenological relevance of the finite quark mass effects can be
established by comparing the size of the latter to the most accurate
estimate of the theoretical uncertainty band for the Higgs transverse
momentum distribution.  This band can be found e.g.~in
ref.~\cite{deFlorian:2011xf}, where NNLO-QCD results in the HQET have
been matched analytically with the effects of soft-gluon
resummation. For $m_H=165$ GeV and 10 GeV $\leq p_T^H\leq$ 60 GeV, the
band width, for a variation of the factorization, renormalization and
resummation scales in a range between $0.5\, m_H$ and $2\,m_H$ (while
keeping the ratio of any two scales in the range $[0.5,2]$), amounts
to $[-8\%,+12\%]$ with respect to the central value computed with all
the scales set equal to $m_H$.  For the same value of $m_H$ and range
of $p_T^H$, we find that the mass effects on the shape of the Higgs
transverse momentum distribution range between $-8\%$ and $+8\%$,
i.e.~they are of the same size as the uncertainty band of
ref.~\cite{deFlorian:2011xf}.

\vfill

\section{MSSM results \label{sec:MSSM}}
We extended our implementation of the gluon-fusion Higgs production
process in \PH\ to describe also the production of the neutral CP-even
bosons of the MSSM, $h$ and $H$. Differently from the case of the SM,
the CP-even Higgs boson masses of the MSSM can be predicted in terms
of the other parameters of the model. At the tree level it is
customary to consider the pseudoscalar mass $\ma$ and $\tb$ as free
parameters which, together with $\mz$, determine the masses $m_h$ and
$m_H$ of the CP-even Higgs bosons and their couplings to the quarks
(as well as to the squarks). Radiative corrections, however, induce in
the Higgs masses and couplings a dependence on all of the MSSM
parameters (see, e.g., ref.~\cite{radcor}).  When studying Higgs boson
production in the MSSM, therefore, it is necessary to compute the
entire spectrum of masses and couplings of the model in a consistent
way, starting from a given set of input parameters. In our numerical
analysis, we use the code {\tt SoftSusy} \cite{softsusy} to compute
the MSSM spectrum starting from a set of running parameters expressed
in the $\drbar$ renormalization scheme. However, it is in principle
possible to interface our calculation of the Higgs production cross
section with other spectrum calculators (such as, e.g., {\tt
  FeynHiggs} \cite{feynhiggs}) that adopt different choices of
renormalization scheme for the input parameters.

\subsection{Modifications in \PH}
In order to describe the production of the neutral CP-even bosons of
the MSSM, three modifications to the SM implementation of \PH\ are
required: a rescaling of the Higgs-quark couplings in the top and
bottom quark contributions, the introduction of all the contributions
from diagrams involving superpartners, and, finally, a rearrangement
of the EW corrections (which have been computed only in the SM case).

For the production of the lightest scalar, $h$, the normalization
factors for the Higgs-quark couplings entering the functions
${\mathcal H}^{1\ell},\,{\mathcal H}^{2\ell},\,{\cal A}_{gg},\,{\cal
  A}_{q g},\, {\cal A}_{q \bar q}$ in section \ref{sec:SMmod} become
\be
\label{lambdalight}
\lambda_t = \frac{\cos\alpha}{\sin\beta}~,~~~~~~~
\lambda_b = -\frac{\sin\alpha}{\cos\beta}~,
\ee
where $\alpha$ is the effective mixing angle that diagonalizes the
radiatively corrected mass matrix in the CP-even Higgs sector. The
corresponding normalization factors for the production of the heaviest
scalar, $H$, can be obtained from eq.~(\ref{lambdalight}) through the
replacements $\cos\alpha\rightarrow\sin\alpha$ in $\lambda_t$ and
$-\sin\alpha\rightarrow \cos\alpha$ in $\lambda_b$.

Diagrams with a squark running in the loop give an additional
contribution to the one-loop form factor ${\mathcal H}^{1\ell}$:
\be
\Delta{\mathcal H}^{1\ell} = 4 \,T_F\, \sum_{\tilde q_i}
\frac{\lambda_{\tilde q_i}}{m^2_{\tilde q_i}}\,y_{\tilde q_i}\,
 \left[1+y_{\tilde q_i} \ln^2 \left(x_{\tilde q_i} \right)\right]\, , 
\label{1lsquark} 
\ee
where the sum runs over the six flavors for the squarks $\tilde q_i$
and the two mass eigenstates $(i=1,2)$ for each flavor. The variables
$y_{\tilde q_i}$ and $x_{\tilde q_i}$ are defined in analogy to $y_q$
and $x_q$ in eq.~(\ref{eq:xy}), and the couplings for the stop and
sbottom mass eigenstates $\tilde t_1$ and $\tilde b_1$ to the lightest
scalar $h$ read
\bea
\lambda_{\tilde t_1} &=& -
\frac{\sin\alpha}{\sin\beta}\,\left\{
-\frac12 \,\sin2\theta_t\,\mu\,\mt + \frac18\,\mz^2\sin2\beta
\left[1 + \cos2\theta_t\,\left(1 - \frac83\sin^2\theta_{\smallw}
\right)\right]\right\}\nn\\
&&+ 
\frac{\cos\alpha}{\sin\beta}\,\left\{\mt^2 + 
\frac12 \,\sin2\theta_t\,A_t\,\mt - \frac14\,\mz^2\sin^2\beta
\left[1 + \cos2\theta_t\,\left(1 - \frac83\sin^2\theta_{\smallw}
\right)\right]\right\}~,\label{lamst}\\
&&\nn\\
\lambda_{\tilde b_1} &=& -
\frac{\sin\alpha}{\cos\beta}\,\left\{ \mb^2
+ \frac12 \,\sin2\theta_b\,A_b\,\mb - \frac14\,\mz^2\cos^2\beta
\left[1 + \cos2\theta_b\,\left(1 - \frac43\sin^2\theta_{\smallw}
\right)\right]\right\}\nn\\
&&+ 
\frac{\cos\alpha}{\cos\beta}\,\left\{
-\frac12 \,\sin2\theta_b\,\mu\,\mb + \frac18\,\mz^2\sin2\beta
\left[1 + \cos2\theta_b\,\left(1 - \frac43\sin^2\theta_{\smallw}
\right)\right]\right\}~.\label{lamsb}
\eea
In the equations above $\mu$ is the higgsino mass parameter in the
MSSM superpotential, $A_q$ (for $q=t,b$) are the soft SUSY-breaking
Higgs-squark couplings, $\theta_q$ are the left-right squark mixing
angles and $\theta_\smallw$ is the Weinberg angle. The couplings for
the squark mass eigenstates $\tilde t_2$ and $\tilde b_2$ can be
obtained from the corresponding couplings for $\tilde t_1$ and $\tilde
b_1$ through the replacements $\sin2\theta_q\rightarrow
-\sin2\theta_q$ and $\cos2\theta_q\rightarrow -\cos2\theta_q$. The
squark couplings to the heaviest scalar $H$ can be obtained from the
squark couplings to $h$ through the replacements
$-\sin\alpha\rightarrow \cos\alpha$ and
$\cos\alpha\rightarrow\sin\alpha$.

The couplings for the up-type and down-type squarks of the first two
generations can be obtained from the stop and sbottom couplings,
respectively, by setting the quark mass and the squark mixing angle to
zero. However, it can be seen from
eqs.~(\ref{1lsquark})--(\ref{lamsb}) that all contributions from the
first two generations of squarks are suppressed by the ratio
$\mz^2/{m^2_{\tilde q_i}}\,$. Furthermore, there are significant
cancellations among the contributions of the four squarks in each
generation (indeed, the total contribution vanishes for degenerate
squark masses). Therefore, in what follows we neglect the first two
generations, and focus on the stop and sbottom contributions.

Additional contributions to the two-loop form factor ${\cal
  H}^{2\ell}$ arise from diagrams with squarks and gluons, with four
squarks, and with quarks, squarks and gluinos. In our
\PH\ implementation we use the results of ref.~\cite{DS1} for the
stop contributions, obtained in the limit of vanishing Higgs mass, and
the results of ref.~\cite{DS2} for the sbottom contributions, obtained
via an asymptotic expansion in the superparticle masses.

The functions $A_2$, $A_4$ and ${\cal A}_{q\bar q}$ entering the real
emission contributions in section \ref{sec:SMmod} also receive additional
contributions from diagrams with a squark running in the loop:
\bea
\Delta A_2 (s,t,u) & = & T_F\,\sum_{\tilde q_i} 
\frac{\lambda_{\tilde q_i}}{m^2_{\tilde q_i}}\,y^2_{\tilde q_i}\,
\left[ b_0 (s_{\tilde q_i},t_{\tilde q_i},u_{\tilde q_i}) 
+ b_0 (s_{\tilde q_i},u_{\tilde q_i},t_{\tilde q_i}) \right]~, \\
\Delta A_4 (s,t,u) & = & T_F\,\sum_{\tilde q_i} 
\frac{\lambda_{\tilde q_i}}{m^2_{\tilde q_i}}\,y^2_{\tilde q_i}\,
\left[ c_0 (s_{\tilde q_i},t_{\tilde q_i},u_{\tilde q_i}) 
+ c_0 (t_{\tilde q_i},u_{\tilde q_i},s_{\tilde q_i}) 
  +\,  c_0 (u_{\tilde q_i},s_{\tilde q_i},t_{\tilde q_i}) \right] ~,\\
\Delta{\cal A}_{q \bar q} ( s,t,u)  & = & T_F\,\sum_{\tilde q_i}   
\frac{\lambda_{\tilde q_i}}{m^2_{\tilde q_i}}\,y_{\tilde q_i}\,
d_0 (s_{\tilde q_i},t_{\tilde q_i},u_{\tilde q_i})~,
\eea
where $s_{\tilde q_i},t_{\tilde q_i}$ and $u_{\tilde q_i}$ are defined
in analogy to $s_q, t_q$ and $u_q$ in eq.~(\ref{eq:defstu}). Explicit
expressions for the functions $b_0 (s_{\tilde q_i},t_{\tilde
  q_i},u_{\tilde q_i}), c_0 (s_{\tilde q_i},t_{\tilde q_i},u_{\tilde
  q_i})$ and $d_0 (s_{\tilde q_i},t_{\tilde q_i},u_{\tilde q_i})$ are
given in ref.~\cite{BDV}.

Finally, we need to adapt the electroweak correction $\delta_{\rm
  {\scriptscriptstyle EW}}$ to the case of the MSSM. Although a
calculation of the contributions to $\delta_{\rm {\scriptscriptstyle
    EW}}$ from diagrams involving superpartners is not currently
available, we can obtain a partial estimate of the EW corrections in
the MSSM by introducing in the SM result appropriate rescaling factors
for the couplings of the Higgs boson. In particular, when the Higgs
boson mass is below the threshold for real top production the EW
correction in the SM is dominated by the contribution of two-loop
diagrams involving light quarks, in which the Higgs boson couples to a
gauge boson \cite{H2gEW1,APSU,BDVcpx}. We can therefore approximate
the EW correction for the production of the lightest scalar $h$ as
\be
\label{EWMSSM}
\delta_{\rm {\scriptscriptstyle EW}} ~\approx~ \sin(\beta-\alpha)\,
\frac{\alpha_{\rm em}}{|{\cal H}^{1\ell}|^2}\,
\left[{\rm Re}\, {\cal H}^{1\ell}~{\rm Re}\, {\cal G}^{2\ell}_{\rm lf}
+{\rm Im}\, {\cal H}^{1\ell}~{\rm Im}\, {\cal G}^{2\ell}_{\rm lf}\,\right]~,
\ee
where the one-loop form factor ${\cal H}^{1\ell}$ is computed in the
MSSM (i.e., it contains both the quark and squark contributions) and
the explicit expression for the two-loop EW light-fermion contribution
${\cal G}^{2\ell}_{\rm lf}$ can be found in ref.~\cite{BDVcpx}. In the
case of the production of the heaviest scalar $H$ the factor
$\sin(\beta-\alpha)$, which rescales the Higgs-gauge boson couplings,
must be replaced by $\cos(\beta-\alpha)$. However, we recall that the
approximation of including only the light-fermion contributions
becomes less justified when $m_H \gtrsim 2\,\mt$.

\subsection{MSSM: numerical results \label{sec:MSSMmhmax}  }

In this section we present numerical results for the production of the
lightest CP-even Higgs boson, $h$, in a representative region of the
MSSM parameter space. Events are generated with our implementation of
\PH, then matched with the {\tt PYTHIA} PS. We compute the total
inclusive cross section, as well as the transverse momentum
distribution, for the production of a light Higgs in gluon fusion, and
we compare them with the corresponding quantities computed for a SM
Higgs boson with the same mass.

For the relevant soft SUSY-breaking parameters (and for $\mu$) we
choose
\be
\label{eq:MSSMinputs}
m_Q=m_U=m_D= 500~{\rm GeV}\,,~~~X_t = 1250~{\rm GeV}\,,
~~~M_3=2\,M_2=4\,M_1= 400~{\rm GeV},~~~\left|\mu\right| = 200~{\rm GeV},
\ee
where: $m_Q,\,m_U$ and $m_D$ are the soft SUSY-breaking mass terms for
stop and sbottom squarks; $X_t \equiv A_t - \mu\cot\beta$ is the
left-right mixing term in the stop mass matrix; $M_i$ (for $i=1,2,3$)
are the soft SUSY-breaking gaugino masses. We consider the input
parameters in eq.~(\ref{eq:MSSMinputs}) as expressed in the $\drbar$
renormalization scheme, at a reference scale $Q$ of the order of the
squark masses (in particular, we take $Q = 500$ GeV). Our choice for
$X_t$ is modeled on the so-called ``$m_h^{\rm max}$ scenario'', in
which the stop-induced radiative corrections maximize the mass of the
lightest scalar $h$, allowing it to satisfy the lower bounds from LEP
even for relatively low values of the stop masses (for our choices of
parameters the physical masses of the two stops are around 280 GeV and
660 GeV, respectively). We consider both signs for the parameter
$\mu$, keeping in mind that, in our conventions, the
$\tan\beta$-dependent corrections to the relation between the bottom
mass and the bottom Yukawa coupling \cite{hrs} enhance the Higgs
couplings to bottom and sbottoms for $\mu<0$ and suppress them for
$\mu>0$.

We perform a scan on the parameters that determine the Higgs boson
masses and mixing at tree level, $\ma$ and $\tan\beta$, varying them
in the ranges\footnote{We remark, however, that parts of the
  $(\ma,\tan\beta)$ plane considered in our study have recently been
  excluded by searches for Higgs bosons decaying into tau pairs at the
  LHC \cite{LHCsearches}, albeit for different choices of the SUSY
  parameters.} $90~{\rm GeV} \leq \ma \leq 200~{\rm GeV}$ and $2 \leq
\tb \leq 50$. For each value of $\tb$ we derive the soft SUSY-breaking
Higgs-stop coupling $A_t$ from the condition on $X_t$, then we fix the
corresponding Higgs-sbottom coupling as $A_b = A_t$.

For each point in the parameter space, we use the code {\tt SoftSusy}
\cite{softsusy} to compute the physical (i.e., radiatively corrected)
Higgs boson masses $m_h$ and $m_H$, and the effective Higgs mixing
angle $\alpha$. We obtain from {\tt SoftSusy}~\footnote{In the MSSM
  analysis we use $M_t = 173.1$ GeV, $m_b(m_b) = 4.16$ GeV and
  $\alpha_s(m_Z) = 0.1172$ as inputs for {\tt SoftSusy}.} also the
MSSM running quark masses $\widehat m_t$ and $\widehat m_b$, expressed
in the $\drbar$ scheme at the scale $Q=500$ GeV. The running quark
masses are used both in the calculation of the running stop and
sbottom masses and mixing angles, and in the calculation of the top
and bottom contributions to the form factors for Higgs boson
production (the latter are computed using the $\drbar$ results
presented in refs.~\cite{DS1,DS2}). As discussed in ref.~\cite{DS2},
the use of $\widehat m_b(Q)$ in the one-loop form factor for gluon
fusion, ${\cal H}^{1\ell}$, induces potentially large contributions,
enhanced either by $\tan\beta$ or by $\ln (\mb^2/Q^2)$, in the
two-loop form factor ${\cal H}^{2\ell}$. We checked that our results
are not significantly altered if we compute ${\cal H}^{1\ell}$ in
terms of the running bottom mass expressed at the lower scale $Q=m_h$.

\begin{figure}[p]
\begin{center}
\includegraphics[height=80mm,angle=0]{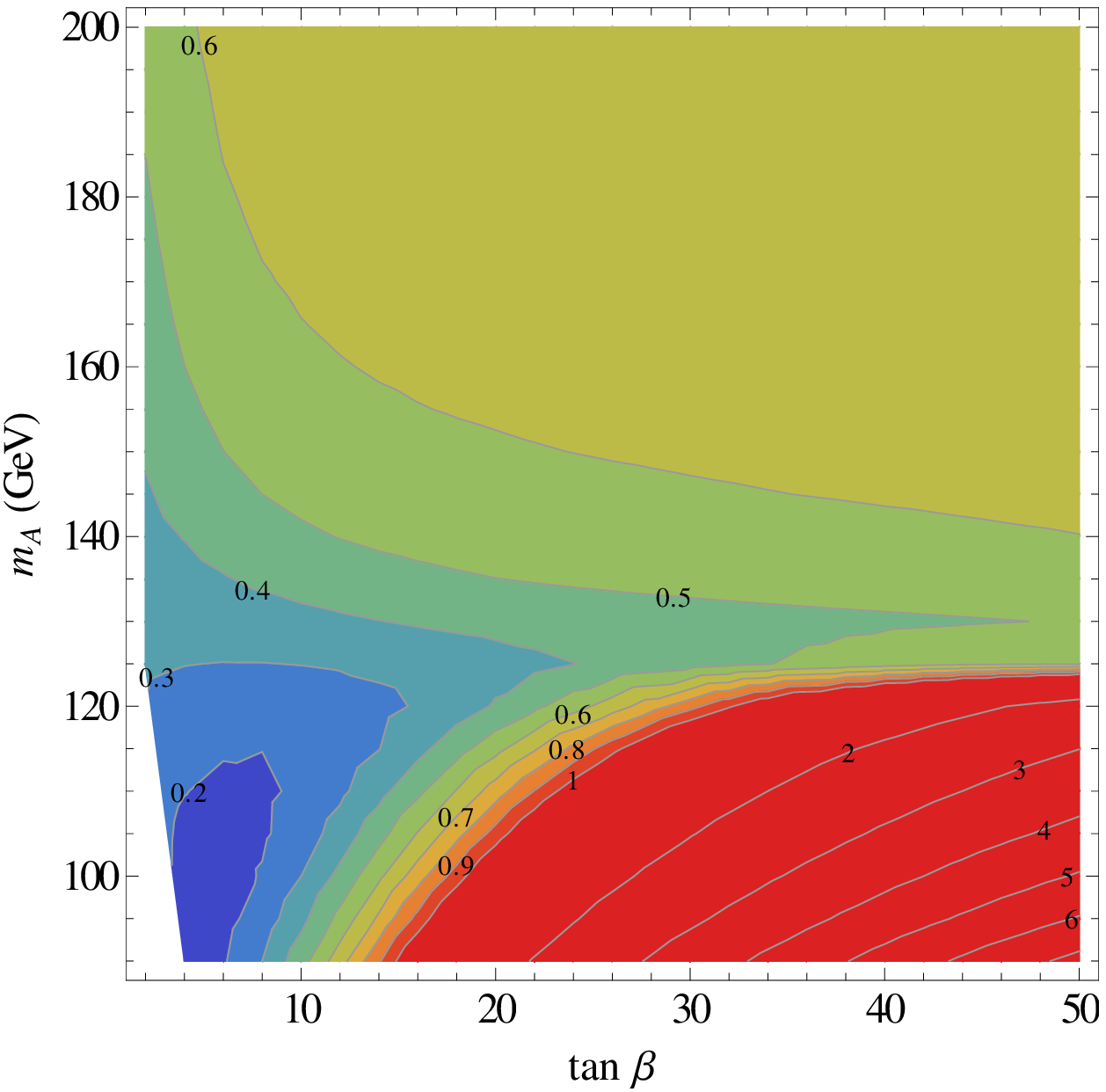}
~~~~
\includegraphics[height=80mm,angle=0]{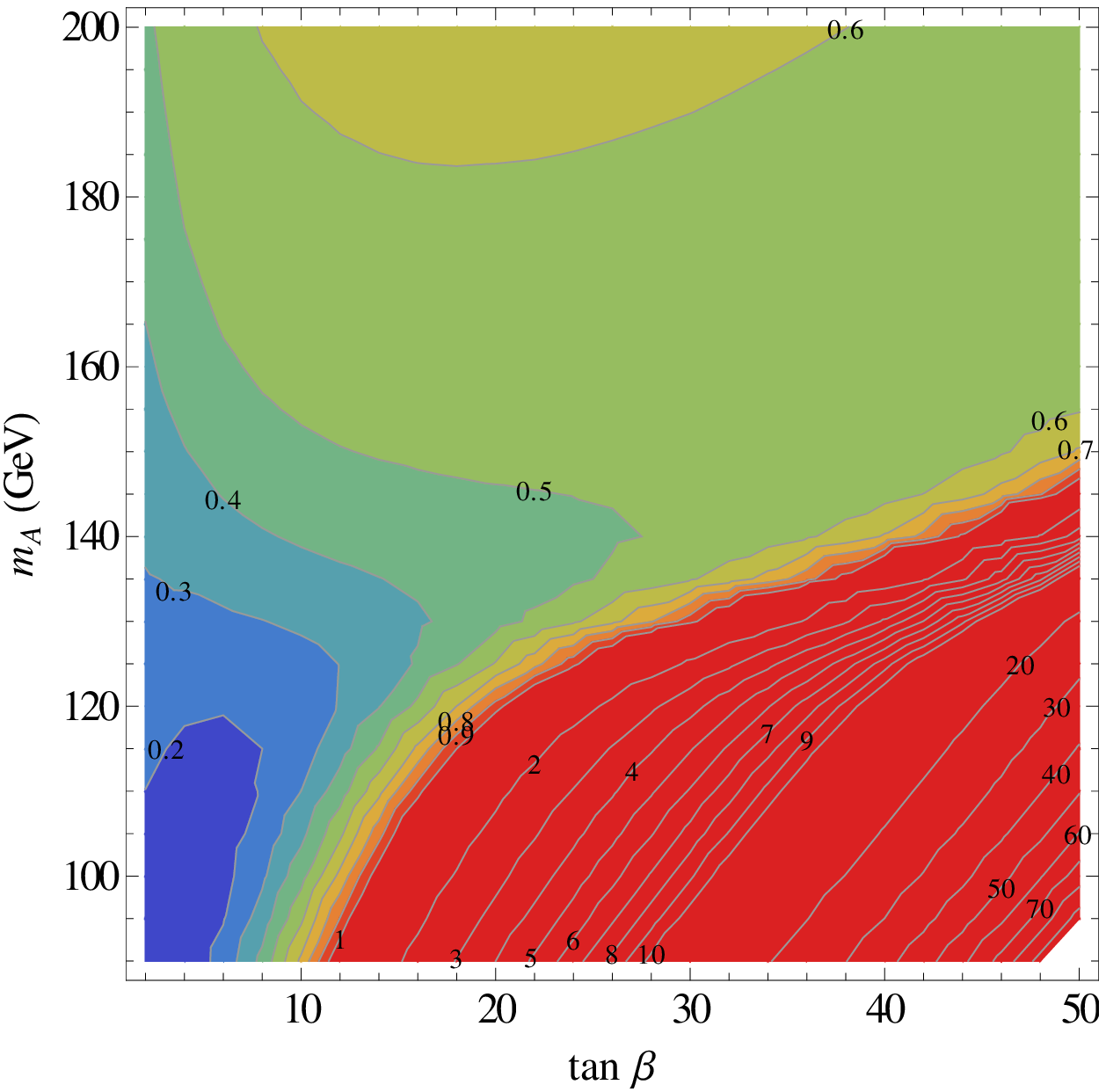}
\caption{\small Ratio of the total cross section for $h$ production in
  the MSSM over the cross section for the production of a SM Higgs
  boson with the same mass. The plot on the left is for $\mu>0$ while
  the plot on the right is for $\mu<0$.
\label{fig:MSSMvsSMxsec}
} 
\end{center}
\end{figure}

\begin{figure}[p]
\begin{center}
\includegraphics[height=80mm,angle=0]{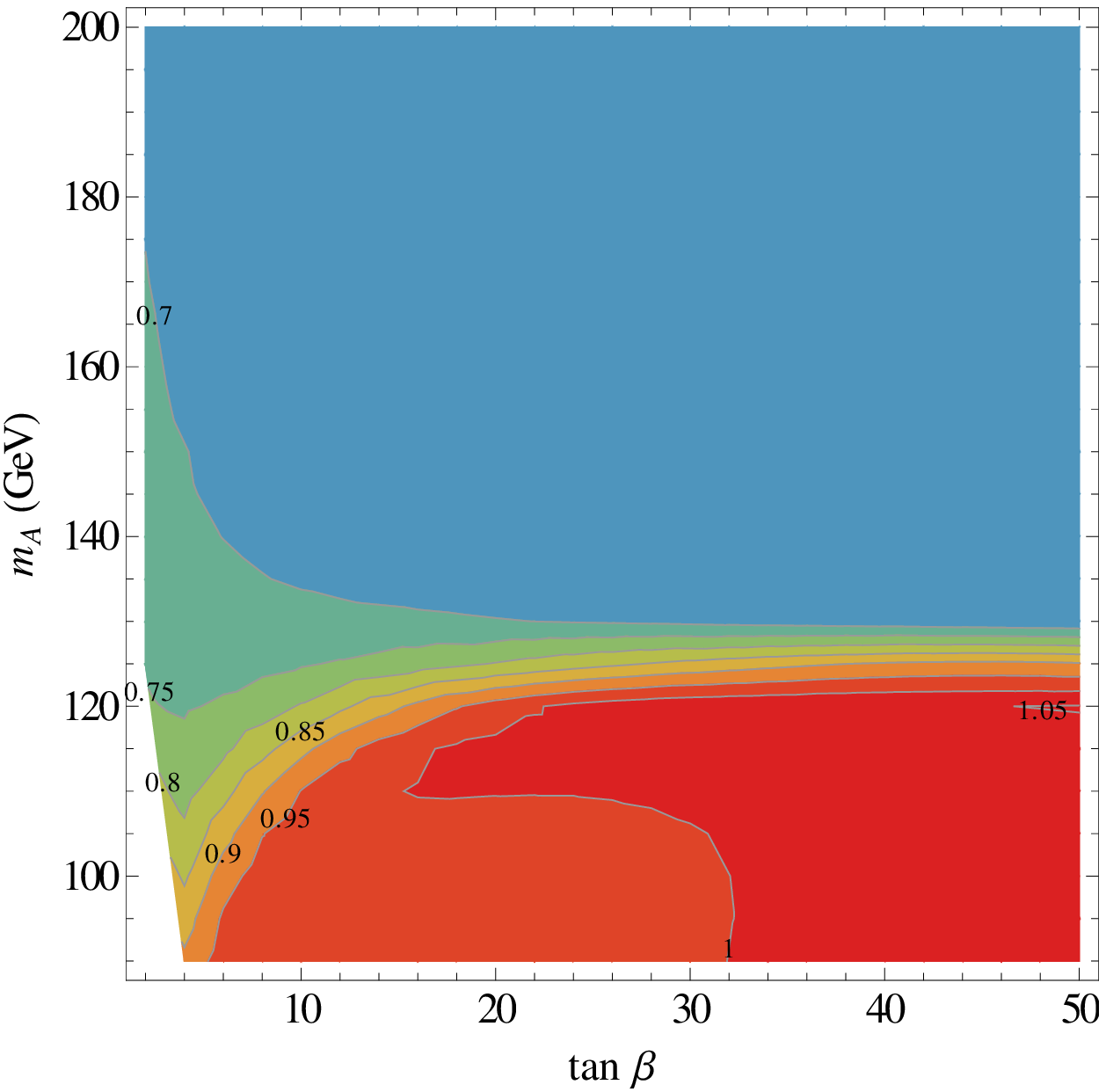}
~~~~
\includegraphics[height=80mm,angle=0]{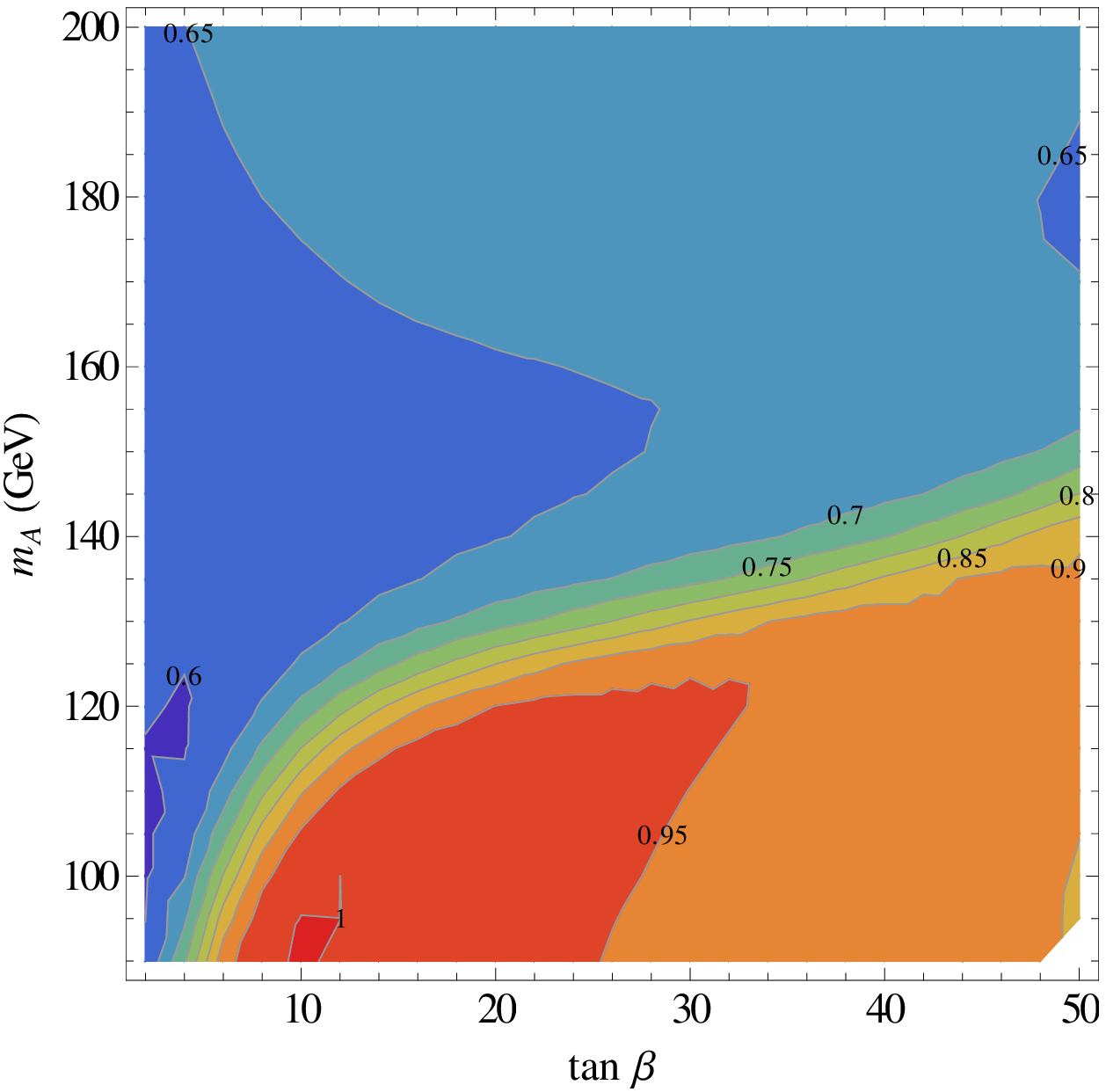}
\caption{\small Ratio of the full cross section for $h$ production in
  the MSSM over the approximated cross section computed with only
  quarks running in the loops. The plot on the left is for $\mu>0$
  while the plot on the right is for $\mu<0$.
\label{fig:MSSMvsONLYQUARKS}
} 
\end{center}
\end{figure}

In figure \ref{fig:MSSMvsSMxsec} we plot the ratio of the cross
section for the production of the lightest scalar $h$ in the MSSM over
the cross section for the production of a SM Higgs boson with the same
mass. For a consistent comparison, we adopt the $\drbar$ scheme in
both the MSSM and the SM calculations. The plot on the left is
obtained with $\mu>0$, while the plot on the right is obtained with
$\mu<0$. In order to interpret the plots, it is useful to recall that
for small values of $m_A$ it is the heaviest scalar $H$ that has
SM-like couplings to fermions, while the coupling of $h$ to top
(bottom) quarks is suppressed (enhanced) by $\tb$. In the lower-left
region of the plots, with small $m_A$ and moderate $\tb$, the
enhancement of the bottom contribution does not compensate for the
suppression of the top contribution, and the MSSM cross section is
smaller than the corresponding SM cross section. On the other hand,
for sufficiently large $\tan\beta$ (in the lower-right region of the
plots) the enhancement of the bottom contribution prevails, and the
MSSM cross section becomes larger than the corresponding SM cross
section.  For $\mu<0$ the coupling of $h$ to bottom quarks is further
enhanced by the $\tb$-dependent threshold corrections \cite{hrs}, and
the ratio between the MSSM and SM predictions can significantly exceed
a factor of ten. Finally, for sufficiently large $m_A$, i.e.~when the
couplings of $h$ to quarks approach their SM values, the MSSM cross
section is smaller than the SM cross section.

It is interesting to note that for intermediate values of $m_A$ there
is a band along which the two cross sections are similar to each
other.  Indeed, the observation of a scalar particle with cross
section in agreement with the SM prediction does not necessarily imply
that the Higgs boson is the SM one. However, as will be discussed
below, a more detailed study of the Higgs kinematic distributions can
help discriminate between the two models.


To assess the genuine effect of the squark contributions (as opposed
to the effect of the modifications in the Higgs-quark couplings), we
plot in figure \ref{fig:MSSMvsONLYQUARKS} the ratio of the full MSSM
cross section for $h$ production over the approximated MSSM cross
section computed with only quarks running in the loops. As in figure
\ref{fig:MSSMvsSMxsec}, the plot on the left is obtained with $\mu>0$,
while the plot on the right is obtained with $\mu<0$.  We observe
that, in most of the considered region of the MSSM parameter space,
the squark contributions reduce the total cross section.  We identify
three regions: $i)$ for sufficiently large $\tb$ and sufficiently
small $m_A$ the squark contribution is modest, ranging between $-10\%$
and $+5\%$; this region roughly coincides with the one in which the
total MSSM cross section is dominated by the $\tan\beta$-enhanced
bottom quark contribution, and is larger than the SM cross section;
$ii)$ a transition region, where the corrections rapidly become as
large as $-30\%$; this region coincides with the one in which the SM
and MSSM cross sections are similar to each other; $iii)$ for
sufficiently large $m_A$ the squark correction is almost constant,
ranging between $-40\%$ and $-30\%$; this region coincides with the
one in which the MSSM cross section is smaller than the corresponding
SM cross section.

\begin{figure}[t]
\begin{center}
\includegraphics[height=80mm,angle=-90]
{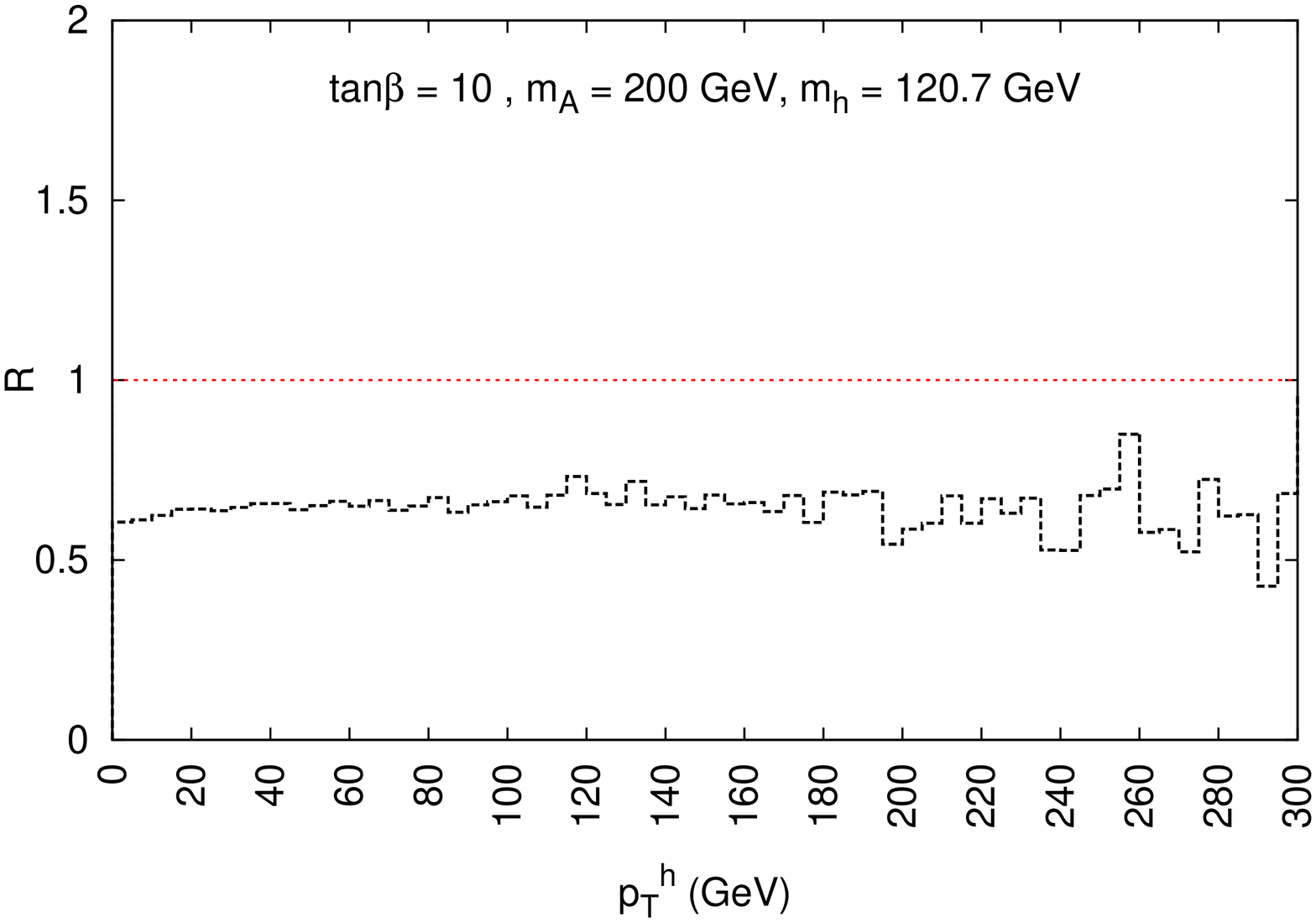}
~~~
\includegraphics[height=80mm,angle=-90]
{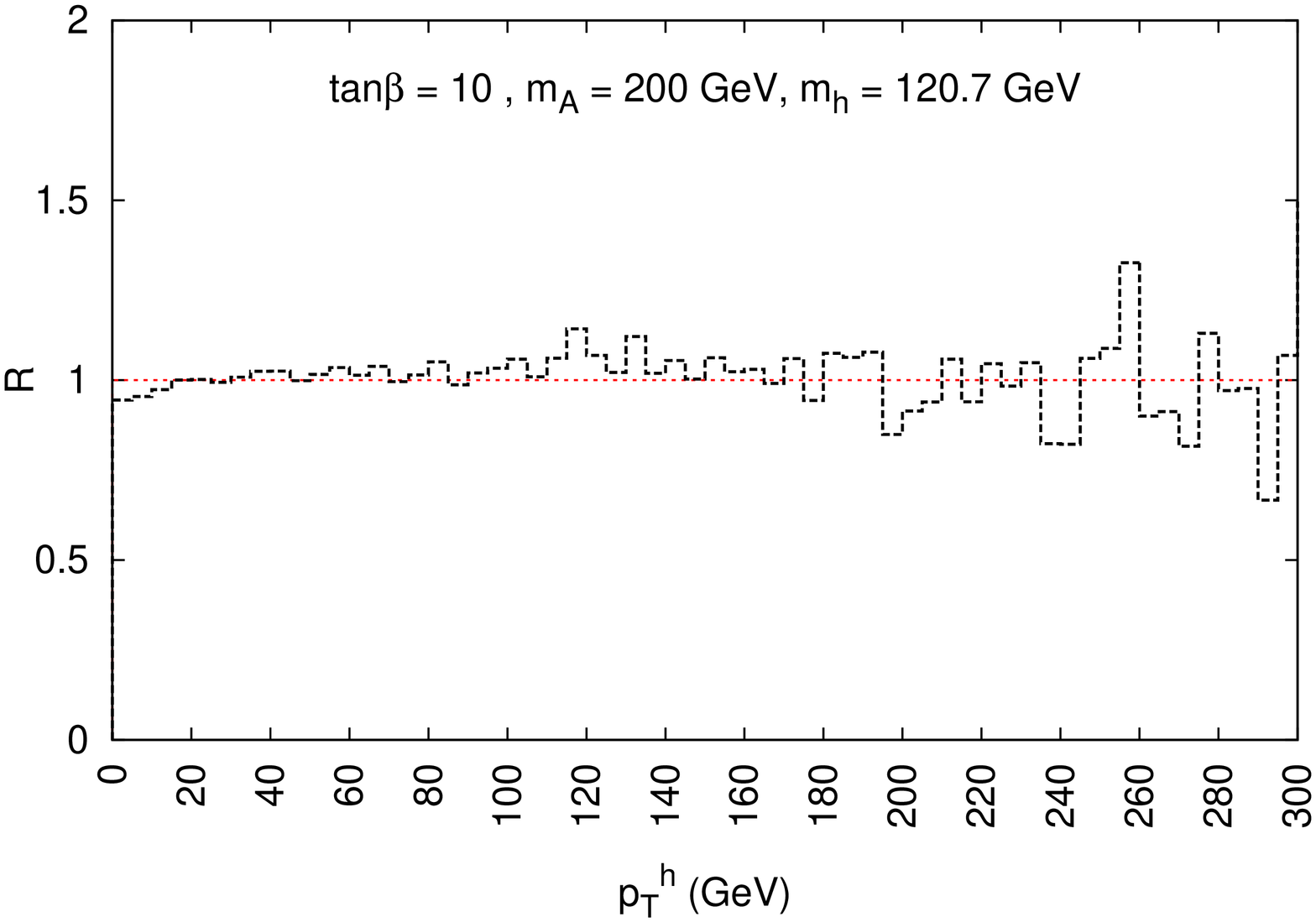}
\caption{\small Left: ratio of the transverse momentum distribution
  for the lightest scalar $h$ in the MSSM over the distribution for a
  SM Higgs boson with the same mass. Right: ratio of the corresponding
  shapes.
\label{fig:tb10}
} 
\end{center}
\end{figure}

We now discuss the distribution of the transverse momentum $p_T^h$ of
a light scalar $h$, considering two distinct scenarios.  First, we
take a point in the MSSM parameter space ($m_A = 200$ GeV, $\tan
\beta= 10$ and $\mu > 0$) in which the coupling of $h$ to the bottom
quark is not particularly enhanced with respect to the SM value, so
that the bottom contribution to the cross section is not particularly
relevant. Because a light Higgs boson cannot resolve the top and
squark vertices, unless we consider very large transverse momentum, we
expect the form of the $p_T^h$ distribution to be very similar to the
one for a SM Higgs boson of equal mass, the two distributions just
differing by a scaling factor related to the total cross section.
This is illustrated in the left plot of figure \ref{fig:tb10}, where
we show the ratio of the transverse momentum distribution for $h$ over
the transverse momentum distribution for a SM Higgs boson of equal
mass. In the right plot of figure \ref{fig:tb10} we show the ratio of
the corresponding shapes, i.e.~the distributions normalized to the
corresponding cross sections.  This ratio, as expected, is close to
one in most of the $p_T^h$ range.

\begin{figure}[p]
\begin{center}
\includegraphics[height=80mm,angle=270]{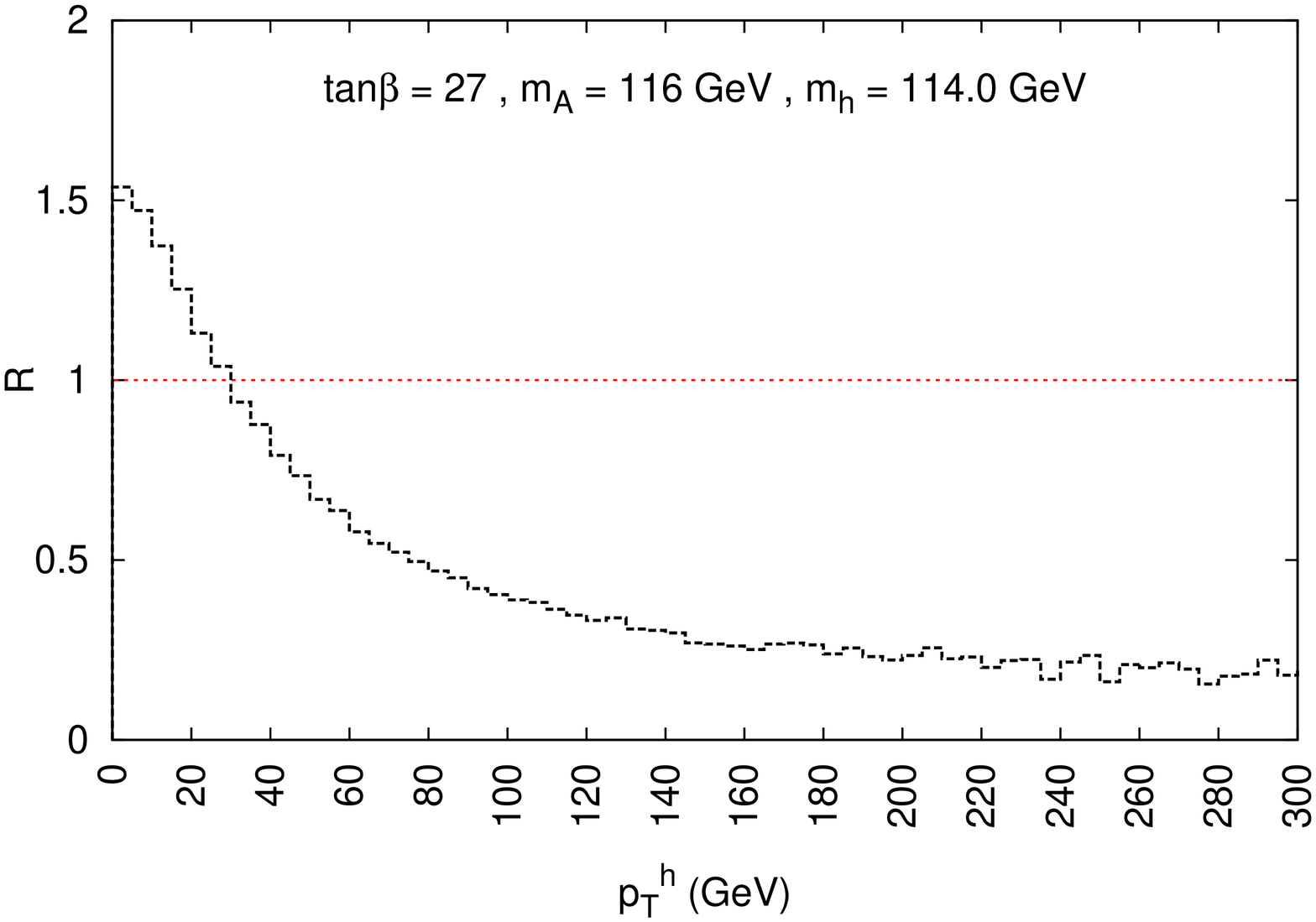}
~~~
\includegraphics[height=80mm,angle=270]{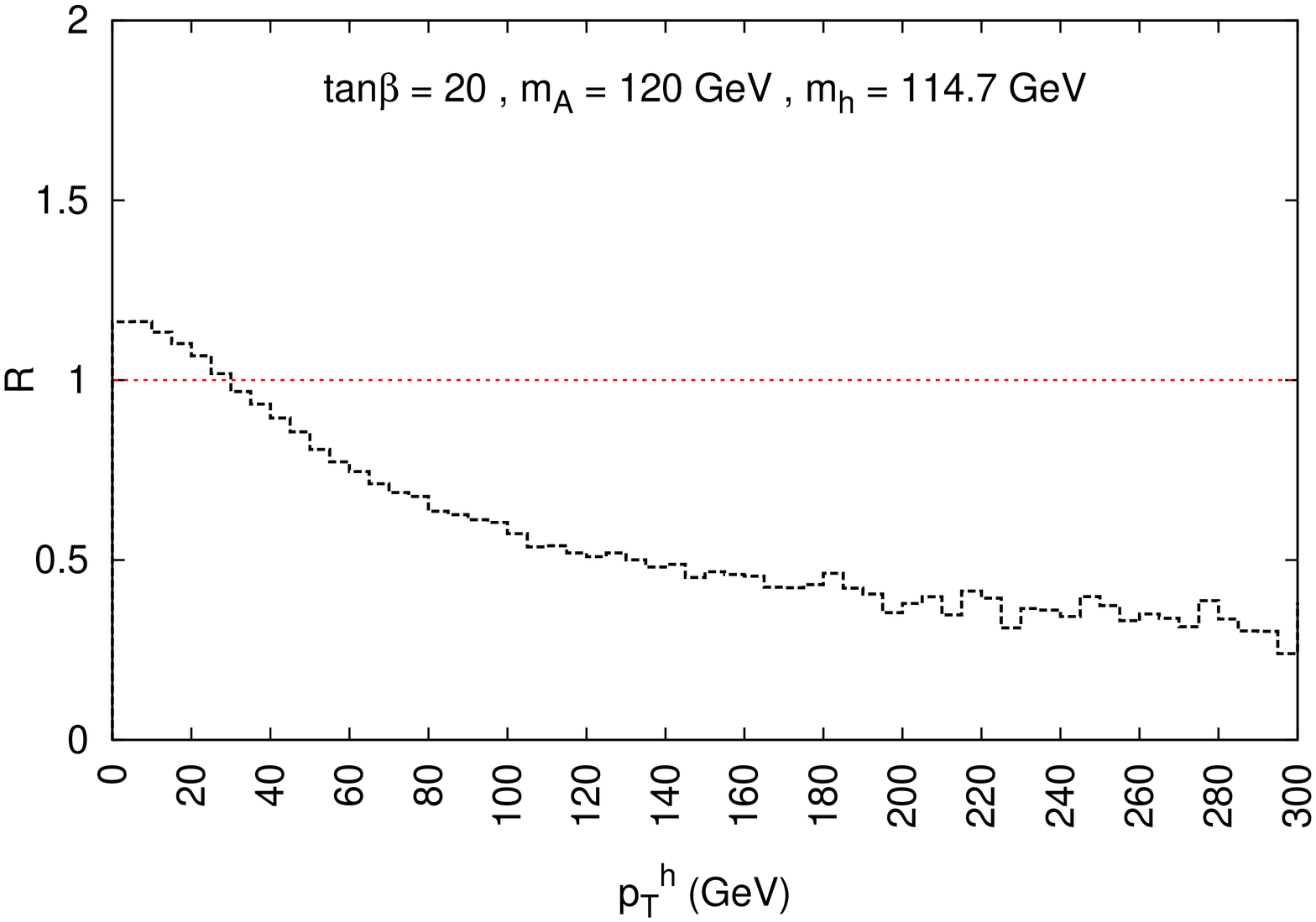}
\\~\\
\includegraphics[height=80mm,angle=270]{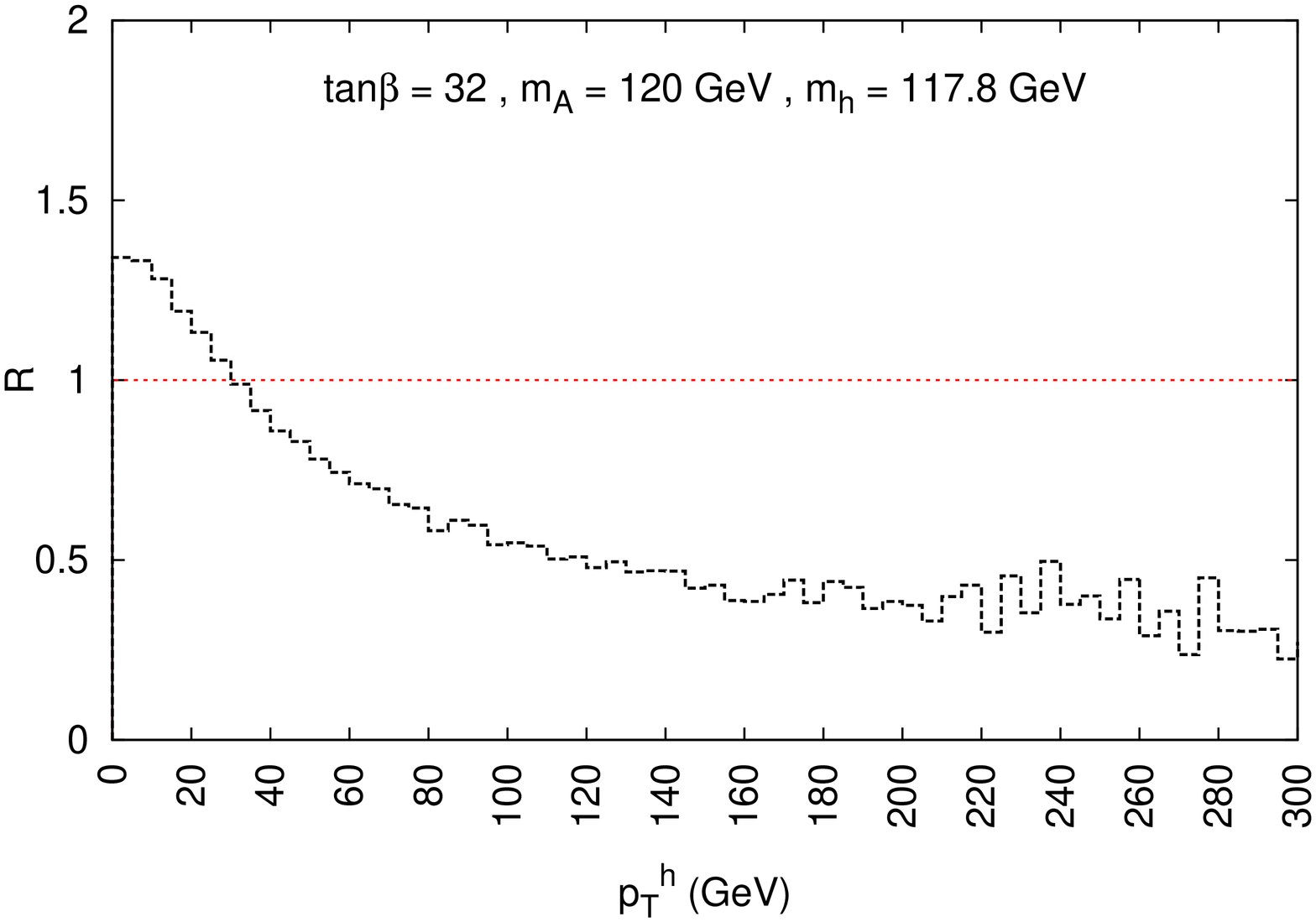}
~~~
\includegraphics[height=80mm,angle=270]{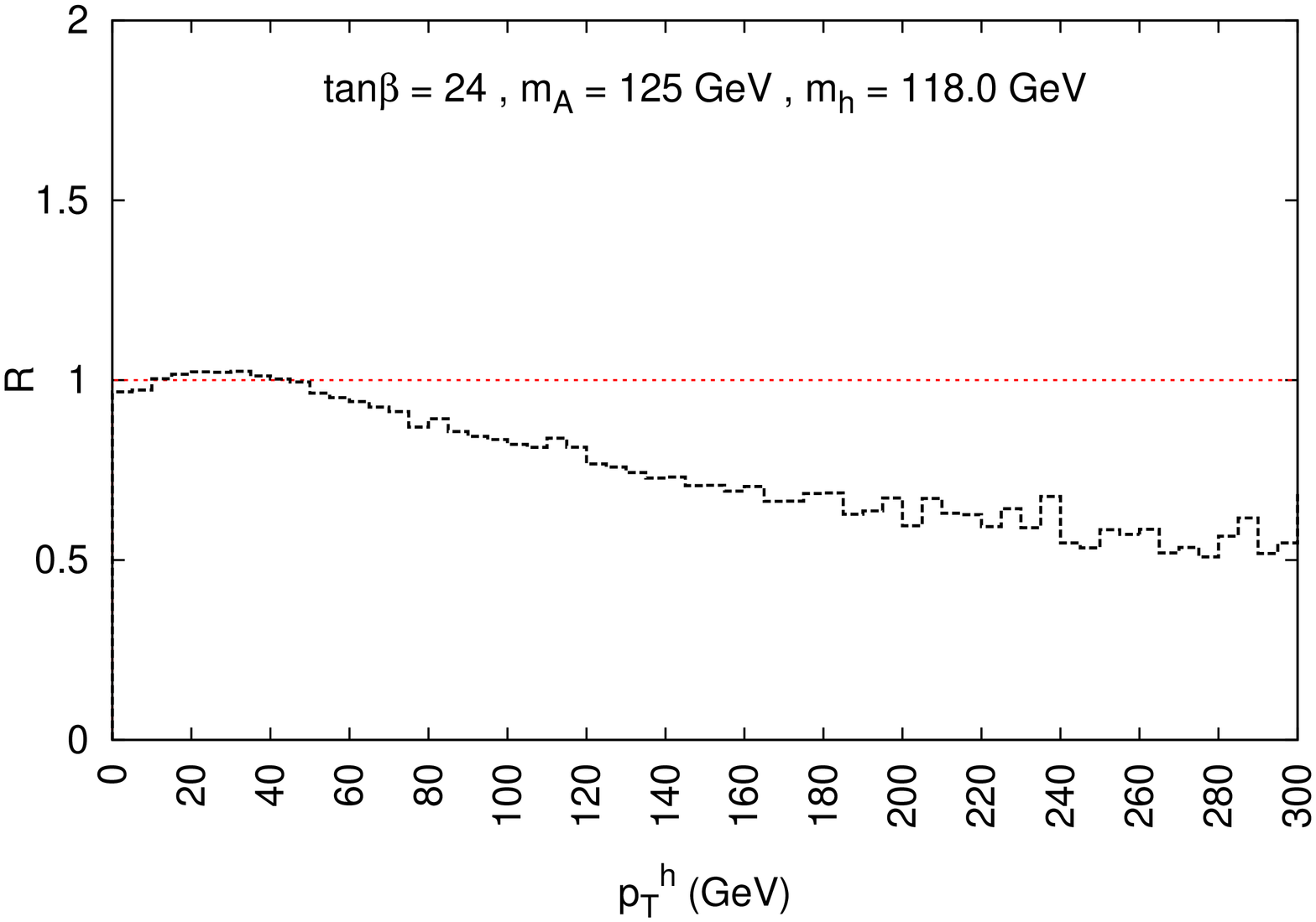}
\\~\\
\includegraphics[height=80mm,angle=270]{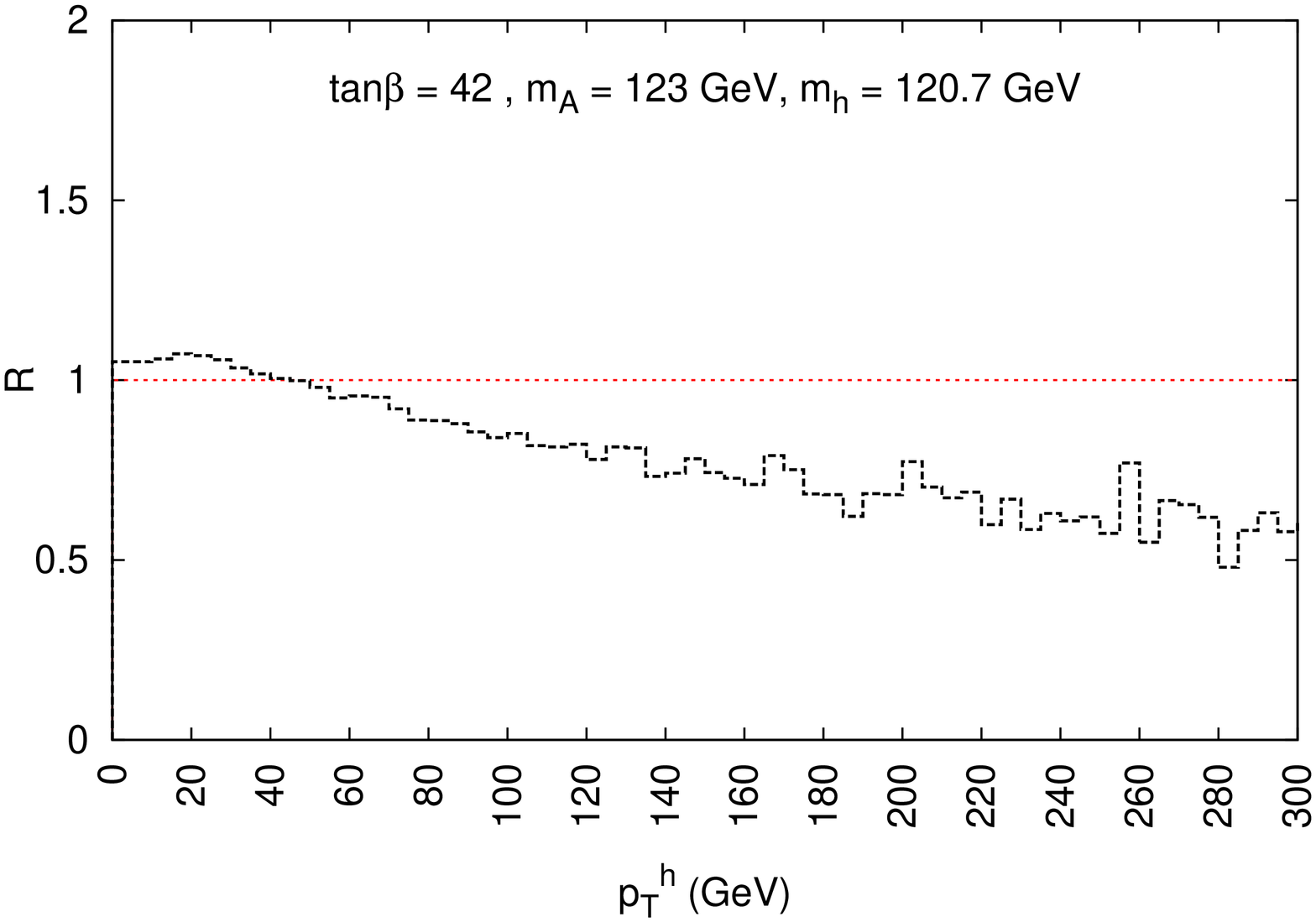}
~~~
\includegraphics[height=80mm,angle=270]{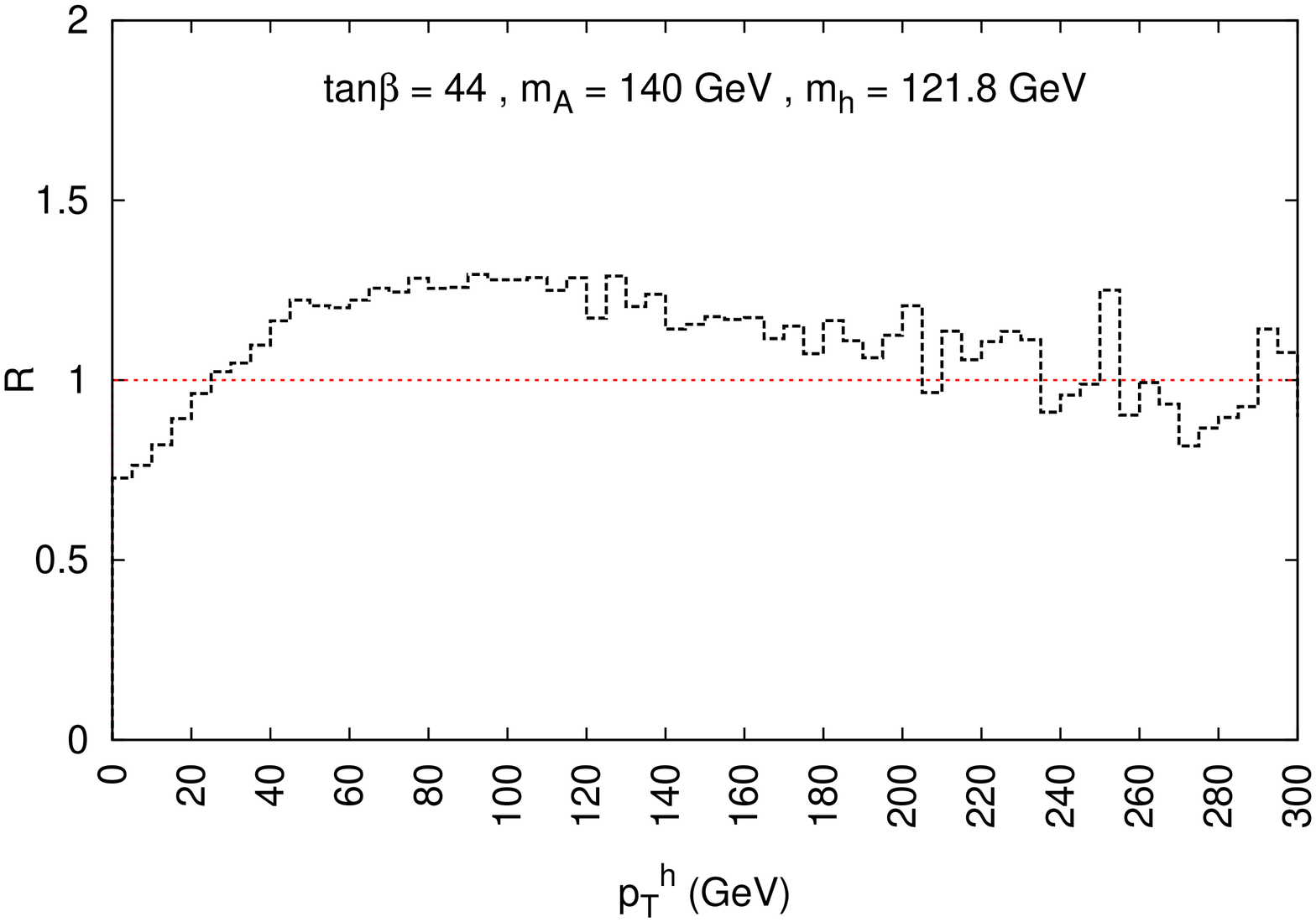}
\\~\\
\caption{\small Ratio of the transverse momentum distribution for the
  lightest scalar $h$ in the MSSM over the distribution for a SM Higgs
  with the same mass. The six plots correspond to different choices of
  $m_A$ and $\tan\beta$ for which the MSSM and SM predictions for the
  total cross section agree within $5\%$. The plots on the left are
  for $\mu>0$ while the plots on the right are for $\mu<0$.
\label{fig:MSSMvsSM-pt-ratio}
} 
\end{center}
\end{figure}

\begin{figure}[p]
\begin{center}
\includegraphics[height=80mm,angle=270]
{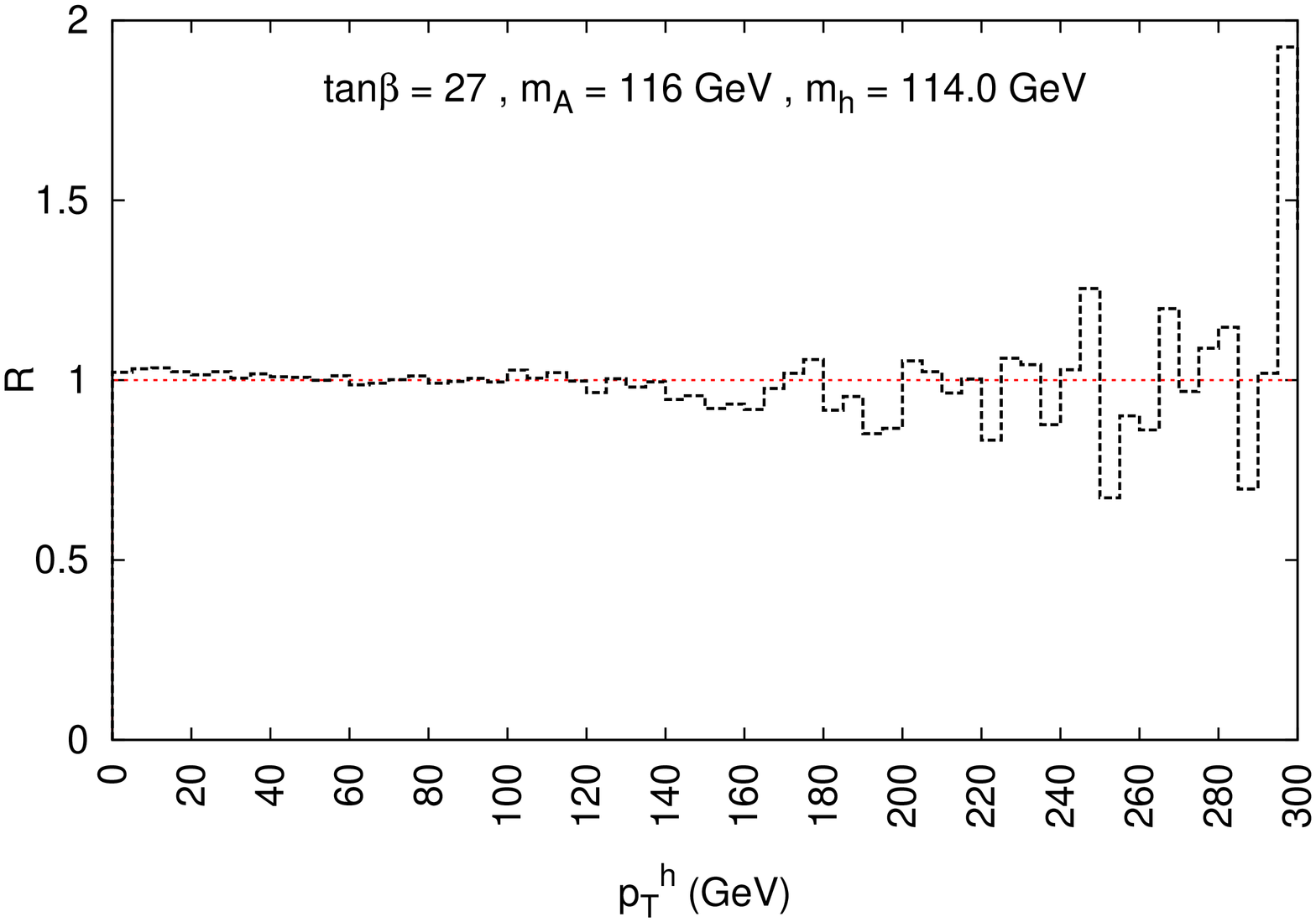}~~~
\includegraphics[height=80mm,angle=270]
{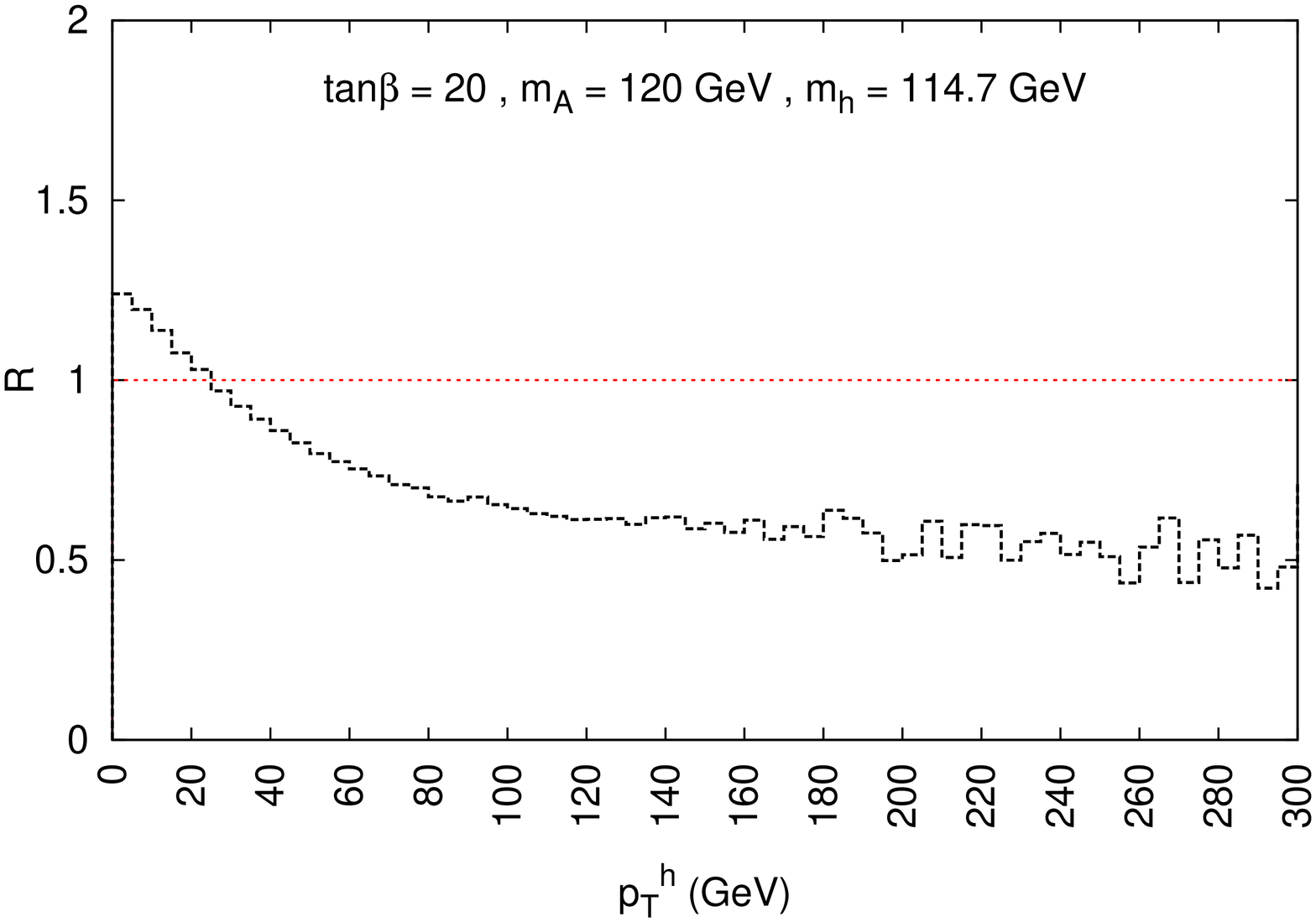}\\~\\
\includegraphics[height=80mm,angle=270]
{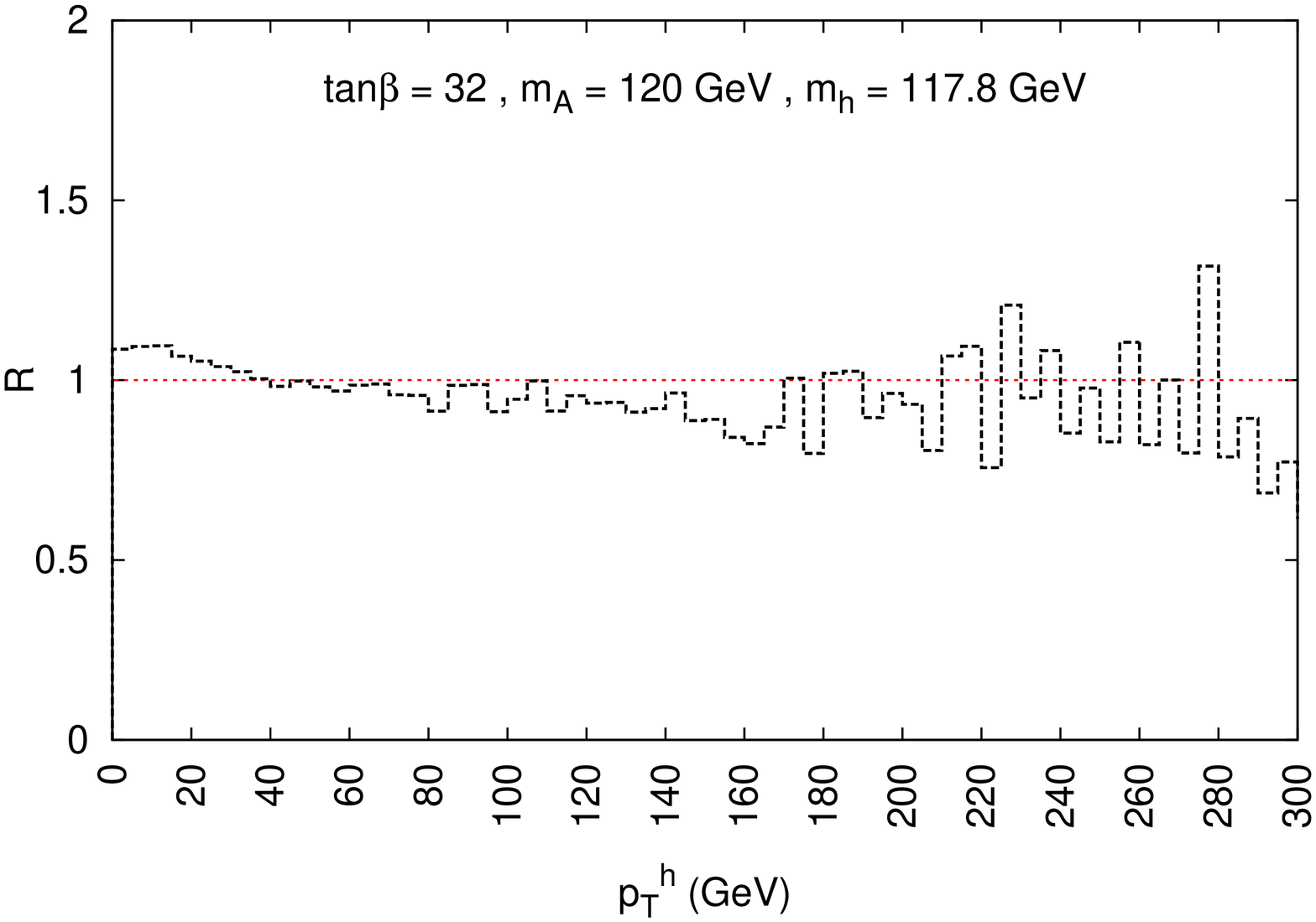}~~~
\includegraphics[height=80mm,angle=270]
{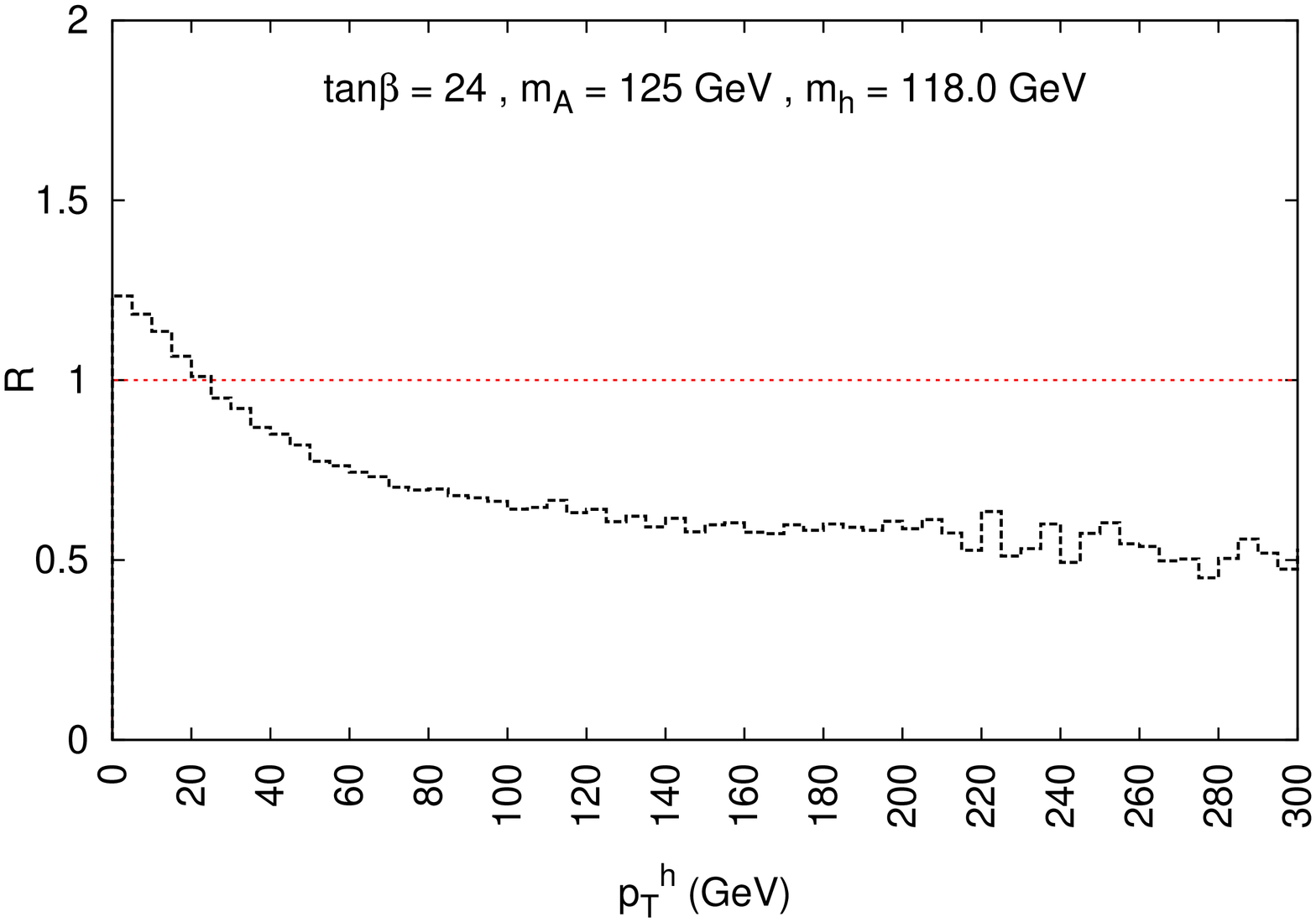}\\~\\
\includegraphics[height=80mm,angle=270]
{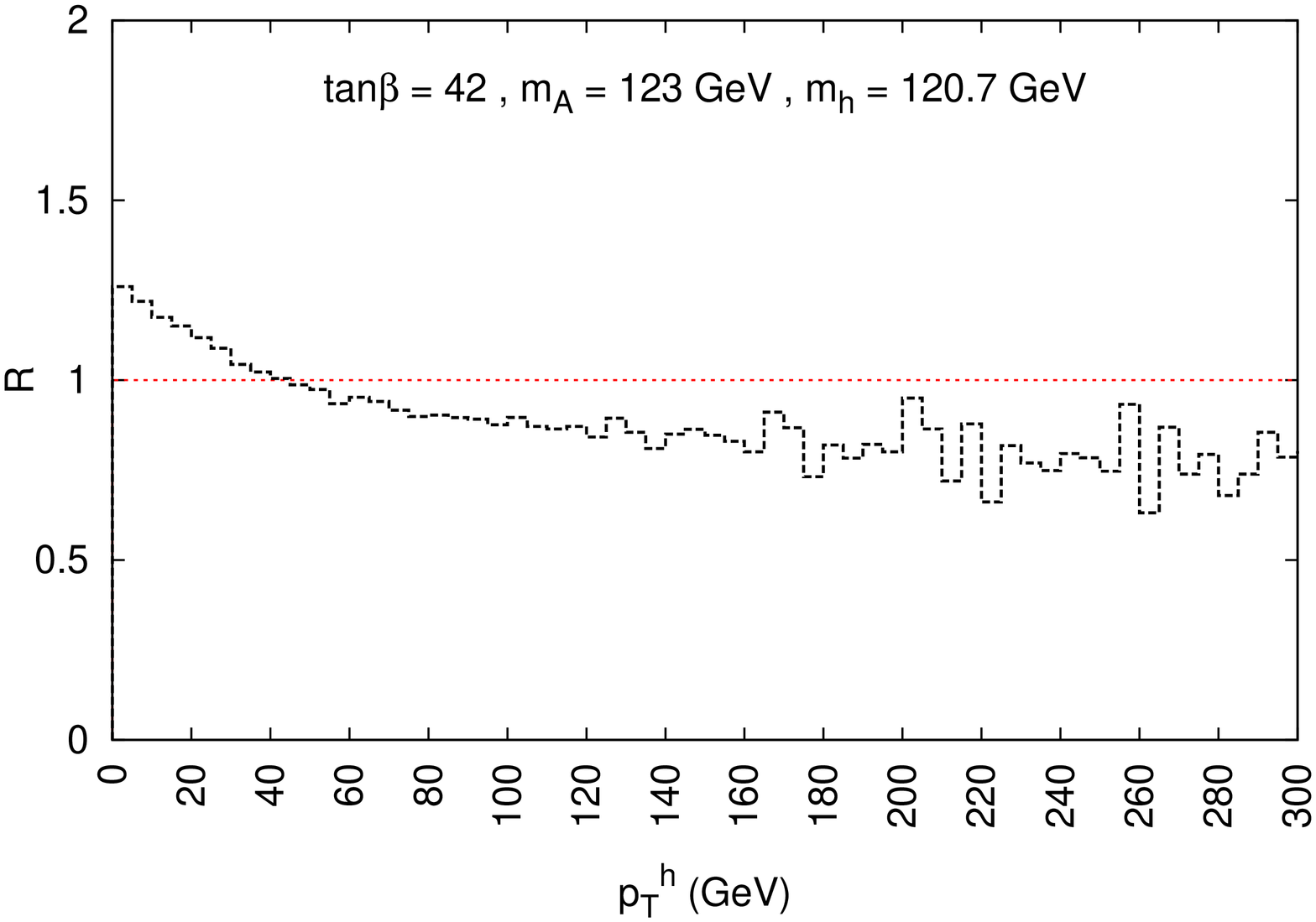}~~~
\includegraphics[height=80mm,angle=270]
{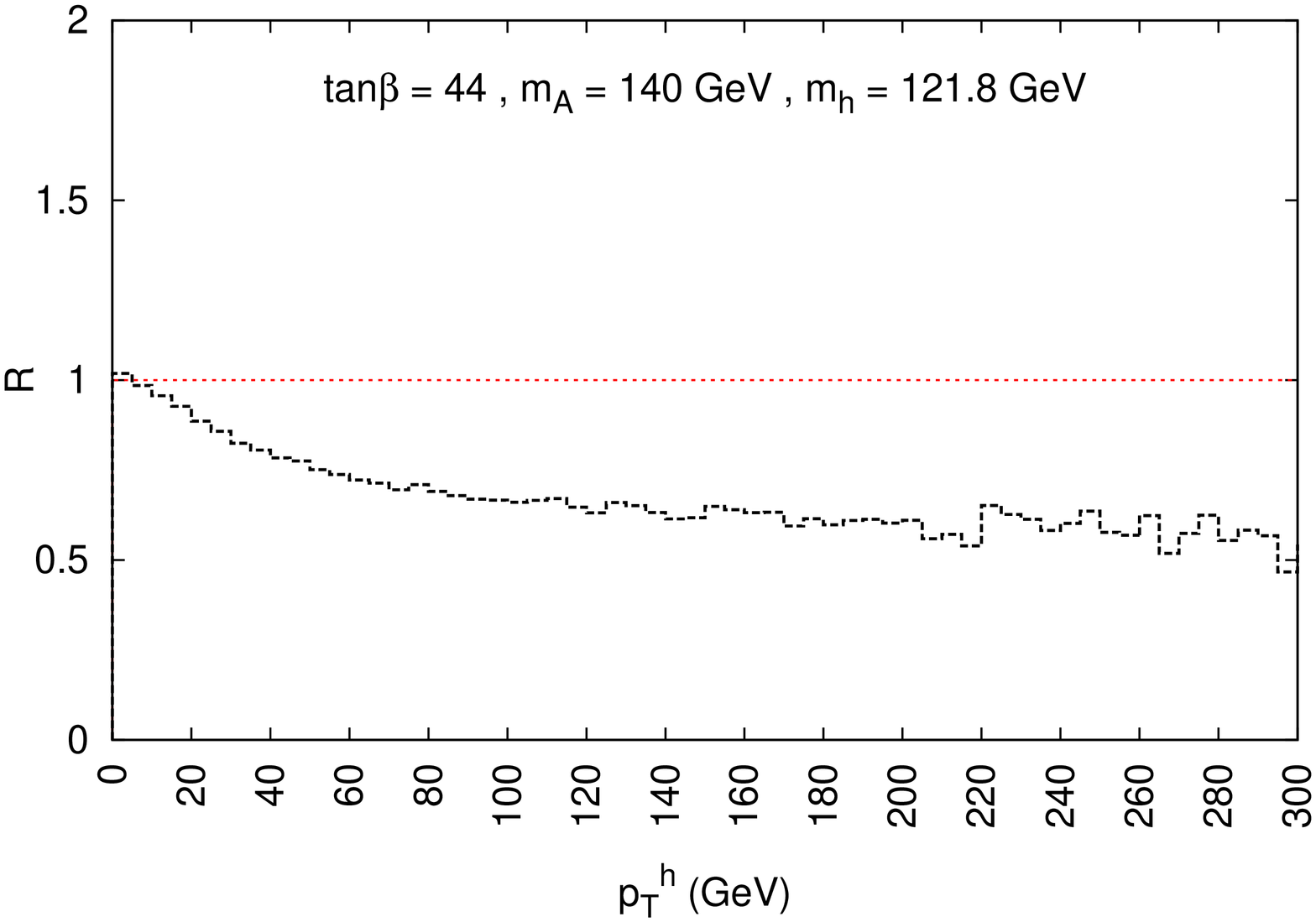}\\~\\
\caption{\small Ratio of the transverse momentum distribution for the
  lightest scalar $h$ in the MSSM over the approximate distribution
  computed with only quarks running in the loops.  The six plots
  correspond to different choices of $m_A$ and $\tan\beta$ for which
  the MSSM and SM predictions for the total cross section agree within
  $5\%$. The plots on the left are for $\mu>0$ while the plots on the
  right are for $\mu<0$.
\label{fig:MSSMvsONLYQUARKS-pt-ratio}
} 
\end{center}
\end{figure}

We then consider the opposite situation, namely when the coupling of
$h$ to the bottom quark is significantly enhanced. In this situation
two tree-level channels, i.e.~$b \bar{b} \to g h$ and $b g \to bh$,
can also contribute to the production mechanism and influence the
shape of the $p_T^h$ distribution~\cite{Brein}. Leaving a study of the
effects of those additional channels to a future analysis, we will now
illustrate how the kinematic distribution of the Higgs boson can help
discriminate between the SM and the MSSM.
The six plots in figure \ref{fig:MSSMvsSM-pt-ratio} correspond to
different points in the $(m_A,\tan\beta)$ plane characterized by the
fact that the MSSM and SM predictions for the total cross section
agree with each other within $5\%$ (therefore, we are effectively
comparing the shape of the transverse momentum distributions).  The
three plots on the left are obtained with $\mu>0$, while the three
plots on the right are obtained with $\mu<0$. The Higgs boson masses
corresponding to these points range between 114 and 122 GeV (i.e., a
SM Higgs with the same mass as $h$ would not yet be excluded by direct
searches).  Figure \ref{fig:MSSMvsSM-pt-ratio} shows that the region
at small $p_T^h$ receives a positive correction with respect to the SM
result for moderate values of $\tan\beta$.  The correction decreases
with increasing $\tan\beta$ and eventually becomes negative at large
$\tan\beta$ for $\mu<0$.  The region at large $p_T^h$ shows an
opposite behavior with respect to $\tan\beta$.

In figure \ref{fig:MSSMvsONLYQUARKS-pt-ratio} we show, for the same
six points in the $(m_A,\tan\beta)$ plane as in figure
\ref{fig:MSSMvsSM-pt-ratio}, the ratio of the $p_T^h$ distribution
over the approximate distribution computed with only quarks running in
the loops. Even though the light Higgs boson cannot resolve the squark
loops, we see that the squark contributions can modify the form of the
$p_T^h$ distribution, because of the interference with the bottom
contribution.  In particular, we observe that the squark contributions
may yield a positive correction on the distribution at small $p_T^h$,
which turns negative for larger values of the transverse momentum.
The negative correction becomes quite flat for $p_T^h>100$ GeV, and in
the $\mu<0$ case it reaches a $-50\%$ effect at very large $p_T^h$.

\section{Conclusions \label{sec:concl}}

We have presented a new implementation\footnote{A beta version of our
  \PH\ code for $gg \to \phi$ can be obtained upon request.} in the
\PH\ approach of the process of Higgs boson production via gluon
fusion in the SM and in the MSSM.  In the NLO-QCD contributions, we
have retained the exact dependence on all the particle masses in the
one-loop diagrams with real-parton emission and in the two-loop
diagrams with quarks and gluons, whereas we have employed the
approximation of vanishing Higgs mass in the two-loop diagrams
involving superpartners. We have also included the effects due to the
two-loop EW corrections.

The exact mass dependence at NLO QCD and its matching with the
multiple gluon emission produced by the {\tt PYTHIA} PS have important
effects on the total and differential cross sections of the Higgs
boson.  In the SM, the exact dependence on the bottom-quark mass
induces, for a light Higgs boson, a non-trivial distortion in the
shape of the transverse momentum distribution in the small-$p_T^H$
region. The effect is comparable in size with the current estimate of
the theoretical uncertainty on this observable, which was derived in
the HQET limit in ref.~\cite{deFlorian:2011xf}. In the case of a heavy
Higgs boson, the role of the bottom quark is negligible, but the exact
dependence on the top-quark mass yields, with respect to the HQET
results, an increase of the transverse momentum distribution at small
$p_T^H$, and a large negative correction at large $p_T^H$.

In the MSSM, our code allows for a systematic study of the parameter
space of the model in a realistic experimental setup.  As an
illustration, we considered representative choices in the MSSM
parameter space, modeled on the so-called $m_h^{\rm max}$ scenario.
We studied the role of the bottom diagrams and the impact of the
inclusion of diagrams involving superpartners at NLO QCD, both on the
total and on the differential cross sections.  In the
large-$\tan\beta$ regime, where the role of the bottom quark is very
relevant, the differential distributions can receive large
corrections, which cannot be described in the HQET approximation. A
detailed study of the Higgs kinematic distributions could help
discriminate between the SM and the MSSM, in case a scalar particle
with a cross section compatible with the SM prediction is observed.

\section*{Acknowledgments}
We thank P.~Nason, S.~Alioli and E.~Re for many clarifying discussions
about the original {\tt POWHEG} version of the code.
This work was partially supported by the Research Executive Agency
(REA) of the European Union under the Grant Agreement number
PITN-GA-2010-264564 (LHCPhenoNet).



\begin{thebibliography}{99}


\bibitem{TevatronExp}
The CDF and D0 Collaborations, 
  arXiv:1107.5518 [hep-ex].

\bibitem{LHCpapers}
  S.~Chatrchyan {\it et al.}  [CMS Collaboration],
  Phys.\ Lett.\  B {\bf 699} (2011) 25
  [arXiv:1102.5429 [hep-ex]];
 %
      G.~Aad {\it et al.}  [ATLAS Collaboration],
      Eur.\ Phys.\ J.\  C {\bf 71} (2011) 1728
      [arXiv:1106.2748 [hep-ex]],
%
  arXiv:1109.3615 [hep-ex],
%
  arXiv:1109.5945 [hep-ex].


\bibitem{Dittmaier:2011ti}
  S.~Dittmaier {\it et al.}  [LHC Higgs Cross Section Working Group],
  arXiv:1101.0593 [hep-ph].


\bibitem{H2gQCD0}
H.~M.~Georgi, S.~L.~Glashow, M.~E.~Machacek and D.~V.~Nanopoulos,
Phys.\ Rev.\ Lett.\  {\bf 40} (1978) 692.



\bibitem{H2gQCD1}
  S.~Dawson,
  Nucl.\ Phys.\ B {\bf 359} (1991) 283;
%
  A.~Djouadi, M.~Spira and P.~M.~Zerwas,
  Phys.\ Lett.\ B {\bf 264} (1991) 440.


\bibitem{SDGZ}
  M.~Spira, A.~Djouadi, D.~Graudenz and P.~M.~Zerwas,
  Nucl.\ Phys.\ B {\bf 453} (1995) 17
  [arXiv:hep-ph/9504378].

\bibitem{HK}
  R.~Harlander and P.~Kant,
  JHEP {\bf 0512} (2005) 015
  [arXiv:hep-ph/0509189].

\bibitem{babis1}
C.~Anastasiou, S.~Beerli, S.~Bucherer, A.~Daleo and Z.~Kunszt,
JHEP {\bf 0701} (2007) 082
[arXiv:hep-ph/0611236].

\bibitem{ABDV}
U.~Aglietti, R.~Bonciani, G.~Degrassi and A.~Vicini,
JHEP {\bf 0701} (2007) 021
[arXiv:hep-ph/0611266].

\bibitem{BDV} 
R.~Bonciani, G.~Degrassi and A.~Vicini,
  JHEP {\bf 0711} (2007) 095
  [arXiv:0709.4227 [hep-ph]].

\bibitem{H2gQCD2}
R.~V.~Harlander,
Phys.\ Lett.\ B {\bf 492} (2000) 74
[arXiv:hep-ph/0007289];
%
S.~Catani, D.~de Florian and M.~Grazzini,
JHEP {\bf 0105} (2001) 025
[arXiv:hep-ph/0102227];
%
R.~V.~Harlander and W.~B.~Kilgore,
Phys.\ Rev.\ D {\bf 64} (2001) 013015
[arXiv:hep-ph/0102241],
%
Phys.\ Rev.\ Lett.\  {\bf 88} (2002) 201801
[arXiv:hep-ph/0201206];
%
C.~Anastasiou and K.~Melnikov,
Nucl.\ Phys.\ B {\bf 646} (2002) 220
[arXiv:hep-ph/0207004];
%
V.~Ravindran, J.~Smith and W.~L.~van Neerven,
Nucl.\ Phys.\ B {\bf 665} (2003) 325
[arXiv:hep-ph/0302135].

\bibitem{H2gQCD3}
S.~Marzani, R.~D.~Ball, V.~Del Duca, S.~Forte and A.~Vicini,
  Nucl.\ Phys.\  B {\bf 800} (2008) 127
  [arXiv:0801.2544 [hep-ph]],
%
  Nucl.\ Phys.\ Proc.\ Suppl.\  {\bf 186} (2009) 98
  [arXiv:0809.4934 [hep-ph]];
%
R.~V.~Harlander and K.~J.~Ozeren,
  Phys.\ Lett.\  B {\bf 679} (2009) 467
  [arXiv:0907.2997 [hep-ph]], 
  JHEP {\bf 0911} (2009) 088
  [arXiv:0909.3420 [hep-ph]];
%
 A.~Pak, M.~Rogal and M.~Steinhauser,
  Phys.\ Lett.\  B {\bf 679} (2009) 473
  [arXiv:0907.2998 [hep-ph]],
  JHEP {\bf 1002} (2010) 025
  [arXiv:0911.4662 [hep-ph]];
%
 R.~V.~Harlander, H.~Mantler, S.~Marzani and K.~J.~Ozeren,
  Eur.\ Phys.\ J.\  C {\bf 66} (2010) 359
  [arXiv:0912.2104 [hep-ph]].


\bibitem{KLS}
M.~Kramer, E.~Laenen and M.~Spira,
Nucl.\ Phys.\ B {\bf 511} (1998) 523 
[arXiv:hep-ph/9611272].

\bibitem{H2gQCD4}
S.~Catani, D.~de Florian, M.~Grazzini and P.~Nason,
JHEP {\bf 0307} (2003) 028
[arXiv:hep-ph/0306211]. 


\bibitem{H2gQCD5}
S.~Moch and A.~Vogt,
Phys.\ Lett.\  B {\bf 631} (2005) 48
[arXiv:hep-ph/0508265];
%
  V.~Ravindran,
  Nucl.\ Phys.\  B {\bf 752} (2006) 173
  [arXiv:hep-ph/0603041].


\bibitem{H2gEW0}
A.~Djouadi and P.~Gambino,
Phys.\ Rev.\ Lett.\  {\bf 73} (1994) 2528
[arXiv:hep-ph/9406432];
%
A.~Djouadi, P.~Gambino and B.~A.~Kniehl,
Nucl.\ Phys.\ B {\bf 523} (1998) 17
[arXiv:hep-ph/9712330].




\bibitem{H2gEW1}
U.~Aglietti, R.~Bonciani, G.~Degrassi and A.~Vicini,
Phys.\ Lett.\ B {\bf 595} (2004) 432
[arXiv:hep-ph/0404071],
%
Phys.\ Lett.\ B {\bf 600} (2004) 57
[arXiv:hep-ph/0407162],
%
arXiv:hep-ph/0610033;
%
G.~Degrassi and F.~Maltoni,
Phys.\ Lett.\ B {\bf 600} (2004) 255
[arXiv:hep-ph/0407249].

\bibitem{APSU}
 S.~Actis, G.~Passarino, C.~Sturm and S.~Uccirati,
  Phys.\ Lett.\  B {\bf 670} (2008) 12
  [arXiv:0809.1301 [hep-ph]],
  Nucl.\ Phys.\  B {\bf 811} (2009) 182
  [arXiv:0809.3667 [hep-ph]].

\bibitem{BDVcpx}
  R.~Bonciani, G.~Degrassi, A.~Vicini,
  Comput.\ Phys.\ Commun.\  {\bf 182} (2011) 1253
  [arXiv:1007.1891 [hep-ph]].

\bibitem{H2gQCDEW}
  C.~Anastasiou, R.~Boughezal and F.~Petriello,
  JHEP {\bf 0904} (2009) 003
  [arXiv:0811.3458 [hep-ph]].


\bibitem{Demartin:2010er}
  F.~Demartin, S.~Forte, E.~Mariani, J.~Rojo and A.~Vicini,
  Phys.\ Rev.\  D {\bf 82} (2010) 014002
  [arXiv:1004.0962 [hep-ph]].

\bibitem{Alekhin:2011sk}
  S.~Alekhin {\it et al.},
  arXiv:1101.0536 [hep-ph].


\bibitem{Dawson:1996xz}
S.~Dawson, A.~Djouadi and M.~Spira,
Phys.\ Rev.\ Lett.\  {\bf 77} (1996) 16
[arXiv:hep-ph/9603423].

\bibitem{MS}
  M.~Muhlleitner and M.~Spira,
  Nucl.\ Phys.\  B {\bf 790} (2008) 1
  [arXiv:hep-ph/0612254].

\bibitem{HS}
R.~V.~Harlander and M.~Steinhauser,
Phys.\ Lett.\  B {\bf 574} (2003) 258
[arXiv:hep-ph/0307346],
%
Phys.\ Rev.\  D {\bf 68} (2003) 111701
[arXiv:hep-ph/0308210],
%
JHEP {\bf 0409} (2004) 066
[arXiv:hep-ph/0409010].

\bibitem{Franziska}
R.~V.~Harlander and F.~Hofmann,
JHEP {\bf 0603} (2006) 050
[arXiv:hep-ph/0507041].

\bibitem{DS1}
  G.~Degrassi and P.~Slavich,
  Nucl.\ Phys.\  B {\bf 805} (2008) 267
  [arXiv:0806.1495 [hep-ph]].

\bibitem{DDVS}
  G.~Degrassi, S.~Di Vita and P.~Slavich,
  JHEP {\bf 1108} (2011) 128
  [arXiv:1107.0914 [hep-ph]].

\bibitem{babis2}
  C.~Anastasiou, S.~Beerli and A.~Daleo,
  Phys.\ Rev.\ Lett.\  {\bf 100} (2008) 241806
  [arXiv:0803.3065 [hep-ph]].

\bibitem{spiraDb}
  M.~Muhlleitner, H.~Rzehak and M.~Spira,
  arXiv:1001.3214 [hep-ph].

\bibitem{DS2}
  G.~Degrassi and P.~Slavich,
  JHEP {\bf 1011} (2010) 044
  [arXiv:1007.3465 [hep-ph]].

\bibitem{hhm}
  R.~V.~Harlander, F.~Hofmann and H.~Mantler,
  JHEP {\bf 1102} (2011) 055
  [arXiv:1012.3361 [hep-ph]].



\bibitem{Hj0}
R.~K.~Ellis, I.~Hinchliffe, M.~Soldate and J.~J.~van der Bij,
Nucl.\ Phys.\ B {\bf 297} (1988) 221;
%
U.~Baur and E.~W.~N.~Glover,
Nucl.\ Phys.\ B {\bf 339} (1990) 38. 



\bibitem{Hj1}
S.~Catani, E.~D'Emilio and L.~Trentadue,
Phys.\ Lett.\ B {\bf 211} (1988) 335;
%
I.~Hinchliffe and S.~F.~Novaes,
Phys.\ Rev.\ D {\bf 38} (1988) 3475;
%
R.~P.~Kauffman,
Phys.\ Rev.\ D {\bf 44} (1991) 1415,
%
Phys.\ Rev.\ D {\bf 45} (1992) 1512;
%
  V.~Del Duca, W.~Kilgore, C.~Oleari, C.~Schmidt and D.~Zeppenfeld,
  Phys.\ Rev.\ Lett.\  {\bf 87} (2001) 122001
  [arXiv:hep-ph/0105129],
%
  Nucl.\ Phys.\  B {\bf 616} (2001) 367
  [arXiv:hep-ph/0108030];
%
A.~V.~Lipatov and N.~P.~Zotov,
Eur.\ Phys.\ J.\ C {\bf 44} (2005) 559
[arXiv:hep-ph/0501172]. 

\bibitem{DFGK}
D.~de Florian, M.~Grazzini and Z.~Kunszt,
Phys.\ Rev.\ Lett.\  {\bf 82} (1999) 5209
[arXiv:hep-ph/9902483];
%
D.~de Florian and M.~Grazzini,
Phys.\ Rev.\ Lett.\  {\bf 85} (2000) 4678
[arXiv:hep-ph/0008152],
%
Nucl.\ Phys.\ B {\bf 616} (2001) 247
[arXiv:hep-ph/0108273].

\bibitem{realtb}
  W.~Y.~Keung and F.~J.~Petriello,
  Phys.\ Rev.\  D {\bf 80} (2009) 013007
  [arXiv:0905.2775 [hep-ph]];
%
  C.~Anastasiou, S.~Bucherer and Z.~Kunszt,
  JHEP {\bf 0910} (2009) 068
  [arXiv:0907.2362 [hep-ph]];
%
  O.~Brein,
  Phys.\ Rev.\  D {\bf 81} (2010) 093006
  [arXiv:1003.4438 [hep-ph]].


\bibitem{Brein}
  O.~Brein and W.~Hollik,
  Phys.\ Rev.\  D {\bf 68} (2003) 095006
  [arXiv:hep-ph/0305321],
%
  Phys.\ Rev.\  D {\bf 76} (2007) 035002
  [arXiv:0705.2744 [hep-ph]].

\bibitem{MSSMH+j}
  B.~Field, J.~Smith, M.~E.~Tejeda-Yeomans and W.~L.~van Neerven,
  Phys.\ Lett.\  B {\bf 551} (2003) 137
  [arXiv:hep-ph/0210369];
%
  B.~Field, S.~Dawson and J.~Smith,
  Phys.\ Rev.\  D {\bf 69} (2004) 074013
  [arXiv:hep-ph/0311199];
%
  U.~Langenegger, M.~Spira, A.~Starodumov and P.~Trueb,
  JHEP {\bf 0606} (2006) 035
  [arXiv:hep-ph/0604156].

\bibitem{Hdiff1}
V.~Ravindran, J.~Smith and W.~L.~Van Neerven,
Nucl.\ Phys.\ B {\bf 634} (2002) 247
[arXiv:hep-ph/0201114];
%
C.~J.~Glosser and C.~R.~Schmidt,
JHEP {\bf 0212} (2002) 016
[arXiv:hep-ph/0209248];
%
J.~Smith and W.~L.~van Neerven,
Nucl.\ Phys.\ B {\bf 720} (2005) 182
[arXiv:hep-ph/0501098].

\bibitem{Hdiff2}
C.~Anastasiou, L.~J.~Dixon and K.~Melnikov,
Nucl.\ Phys.\ Proc.\ Suppl.\  {\bf 116}, 193 (2003)
[arXiv:hep-ph/0211141]; 
%
C.~Anastasiou, K.~Melnikov and F.~Petriello,
Phys.\ Rev.\ Lett.\  {\bf 93} (2004) 262002
[arXiv:hep-ph/0409088],
%
Nucl.\ Phys.\ B {\bf 724} (2005) 197
[arXiv:hep-ph/0501130];
%
S.~Catani and M.~Grazzini,
Phys.\ Rev.\ Lett.\  {\bf 98} (2007) 222002
[arXiv:hep-ph/0703012].

\bibitem{Balazs:2000wv}
  C.~Balazs and C.~P.~Yuan,
  Phys.\ Lett.\  B {\bf 478} (2000) 192
  [arXiv:hep-ph/0001103];
%
  C.~Balazs, J.~Huston and I.~Puljak,
  Phys.\ Rev.\  D {\bf 63} (2001) 014021
  [arXiv:hep-ph/0002032].
%
\bibitem{HqT}
G.~Bozzi, S.~Catani, D.~de Florian and M.~Grazzini,
Phys.\ Lett.\ B {\bf 564} (2003) 65
[arXiv:hep-ph/0302104],
%
Nucl.\ Phys.\ B {\bf 737} (2006) 73
[arXiv:hep-ph/0508068],
%
  Nucl.\ Phys.\  B {\bf 791} (2008) 1
  [arXiv:0705.3887 [hep-ph]].

\bibitem{deFlorian:2011xf}
  D.~de Florian, G.~Ferrera, M.~Grazzini and D.~Tommasini,
  arXiv:1109.2109 [hep-ph].


\bibitem{herwig}
  G.~Marchesini, B.~R.~Webber, G.~Abbiendi, I.~G.~Knowles, M.~H.~Seymour and L.~Stanco,
  Comput.\ Phys.\ Commun.\  {\bf 67} (1992) 465;
%
 G.~Corcella {\it et al.},
  arXiv:hep-ph/0210213.

\bibitem{pythia}
  T.~Sjostrand, S.~Mrenna and P.~Z.~Skands,
  JHEP {\bf 0605} (2006) 026
  [arXiv:hep-ph/0603175].


\bibitem{mcnlo}
  S.~Frixione and B.~R.~Webber,
  JHEP {\bf 0206} (2002) 029
  [arXiv:hep-ph/0204244].

\bibitem{Nason:2004rx}
  P.~Nason,
  JHEP {\bf 0411} (2004) 040
  [arXiv:hep-ph/0409146].

\bibitem{Frixione:2007vw}
  S.~Frixione, P.~Nason and C.~Oleari,
  JHEP {\bf 0711} (2007) 070
  [arXiv:0709.2092 [hep-ph]].

\bibitem{powhegggh}
  S.~Alioli, P.~Nason, C.~Oleari and E.~Re,
  JHEP {\bf 1006} (2010) 043
  [arXiv:1002.2581 [hep-ph]].

\bibitem{mcnloggh}
  S.~Frixione and B.~R.~Webber,
  arXiv:hep-ph/0309186.

\bibitem{Alioli:2008tz}
  S.~Alioli, P.~Nason, C.~Oleari and E.~Re,
  JHEP {\bf 0904} (2009) 002
  [arXiv:0812.0578 [hep-ph]].

\bibitem{herwig++ggh}
  M.~Bahr {\it et al.},
  arXiv:0812.0529 [hep-ph];
%
  S.~Frixione, F.~Stoeckli, P.~Torrielli and B.~R.~Webber,
  JHEP {\bf 1101} (2011) 053 
  [arXiv:1010.0568 [hep-ph]].

\bibitem{HIGLU}
 M.~Spira,
  arXiv:hep-ph/9510347,
%
  Nucl.\ Instrum.\ Meth.\  A {\bf 389} (1997) 357
  [arXiv:hep-ph/9610350].

\bibitem{iHixs}
  C.~Anastasiou, S.~Buhler, F.~Herzog and A.~Lazopoulos,
  arXiv:1107.0683 [hep-ph].

\bibitem{Alwall:2011cy}
  J.~Alwall, Q.~Li and F.~Maltoni,
  arXiv:1110.1728 [hep-ph].

\bibitem{Hamilton:2009za}
  K.~Hamilton, P.~Richardson and J.~Tully,
  JHEP {\bf 0904} (2009) 116
  [arXiv:0903.4345 [hep-ph]].

\bibitem{Frixione:1995ms}
  S.~Frixione, Z.~Kunszt and A.~Signer,
  Nucl.\ Phys.\  B {\bf 467} (1996) 399
  [arXiv:hep-ph/9512328].

\bibitem{Frixione:1997np}
  S.~Frixione,
  Nucl.\ Phys.\  B {\bf 507} (1997) 295
  [arXiv:hep-ph/9706545].

\bibitem{mstw2008}
  A.~D.~Martin, W.~J.~Stirling, R.~S.~Thorne and G.~Watt,
  Eur.\ Phys.\ J.\  C {\bf 63} (2009) 189
  [arXiv:0901.0002 [hep-ph]].



\bibitem{radcor}
  G.~Degrassi, S.~Heinemeyer, W.~Hollik, P.~Slavich and G.~Weiglein,
  Eur.\ Phys.\ J.\  C {\bf 28} (2003) 133
  [arXiv:hep-ph/0212020];
%
  B.~C.~Allanach, A.~Djouadi, J.~L.~Kneur, W.~Porod and P.~Slavich,
  JHEP {\bf 0409} (2004) 044
  [arXiv:hep-ph/0406166].

\bibitem{softsusy}
  B.~C.~Allanach,
  Comput.\ Phys.\ Commun.\  {\bf 143} (2002) 305
  [arXiv:hep-ph/0104145].

\bibitem{feynhiggs}
  S.~Heinemeyer, W.~Hollik and G.~Weiglein,
  Comput.\ Phys.\ Commun.\  {\bf 124} (2000) 76
  [arXiv:hep-ph/9812320];
  %
  M.~Frank, T.~Hahn, S.~Heinemeyer, W.~Hollik, H.~Rzehak and G.~Weiglein,
  JHEP {\bf 0702} (2007) 047
  [arXiv:hep-ph/0611326].

\bibitem{hrs}
  L.~J.~Hall, R.~Rattazzi and U.~Sarid,
  Phys.\ Rev.\  D {\bf 50} (1994) 7048
  [arXiv:hep-ph/9306309].

\bibitem{LHCsearches}
The ATLAS Collaboration, ATLAS-CONF-2011-132;
%
The CMS Collaboration, CMS-PAS-HIG-11-020.

\end{thebibliography}
\end{document}